\documentclass[prb,twocolumn,superscriptaddress,showpacs,floatfix]{revtex4}
\usepackage{mathrsfs}
\usepackage{amsmath,amsfonts,amssymb}
\usepackage{epsfig}
\usepackage{graphicx}
\usepackage{wasysym}
\usepackage{bbm}
\usepackage{psfrag}
\usepackage{color}
\usepackage{pstool}
\usepackage{braket}
\usepackage{lmodern}
\usepackage[dvips, bookmarks, colorlinks=true, plainpages = false, citecolor = blue, linkcolor = blue, urlcolor = blue, filecolor = blue]{hyperref}
\def\be{\begin{equation}}
\def\ee{\end{equation}}
\def\bea{\begin{eqnarray}}
\def\eea{\end{eqnarray}}

\begin{document}
\title{Engineering topological phases of any winding and Chern numbers in extended Su-Schrieffer-Heeger models}
\author{Rakesh Kumar Malakar}\email{rkmalakar75@gmail.com}
\author{Asim Kumar Ghosh}
 \email{asimk.ghosh@jadavpuruniversity.in}
\affiliation {Department of Physics, Jadavpur University, 
188 Raja Subodh Chandra Mallik Road, Kolkata 700032, India}
\begin{abstract}
  Simple route of engineering topological phases for any desired 
  value of winding and Chern numbers is found in 
  the Su-Schrieffer-Heeger (SSH) model by adding a further neighbor
  hopping term of varying distances. 
  It is known that the standard SSH model yields a
  single topological phase with winding number, $\nu=1$.
  In this study it is shown that how one can
  generate topological phases with any values 
   of winding numbers, for examples,
   $\nu=\pm 1,\pm 2,\pm 3,\cdots,$
   in the presence of a single further neighbor term
   which preserves inversion, particle-hole and chiral symmetries.
   Quench dynamics of the topological and trivial phases are
  studied in the presence of a specific nonlinear term. 
   Another version of SSH model with additional modulating
   nearest neighbor and next-nearest-neighbor hopping parameters 
   was introduced before which exhibit a single topological
   phase characterized by Chern number, $\mathcal C=\pm 1$. 
   Standard form of inversion, particle-hole and chiral symmetries are
   broken in this model.
Here this model has been studied in the presence of
   several types of parametrization
among which, for a special case the
system is found to yield a series of phases with
  Chern numbers, $\mathcal C=\pm 1,\pm 2,\pm 3,\cdots.$
  In another parametrization, multiple crossings within the edge states
  energy lines are found in both trivial and topological phases.  
  Topological phase diagrams are drawn for every case.
  Emergence of spurious topological phases is also reported.
  \vskip 1.5 cm
 Corresponding author: Asim Kumar Ghosh 
\end{abstract}
\maketitle
\section{INTRODUCTION}
Su-Schrieffer-Heeger (SSH) model is the most popular representative
of one-dimensional (1D) topological insulator which
paved the way for studying the
topological phases in the simplest manner\cite{SSH1,SSH3}. 
Both trivial and topological insulating phases have been realized 
by tuning the ratio of inter and intracell hopping amplitudes
in this staggered model composed by two-site unit cells.
Nontrivial phase is observed when this ratio exceeds 
unity and it is characterized by a nonzero
topological invariant known as
winding number ($\nu$) which is connected to the integral of
Berry curvature over the Brillouin zone (BZ) and known as
Pancharatnam-Berry (PB) phase or Zak phase.
This nontrivial phase is at the same time
associated with the emergence of symmetry protected
zero energy states which are found 
localized on both the edges of the open chain. 
Transition between those two phases with nonzero band gap is accompanied by
vanishing band gap found at the phase transition point. 

SSH model is connected with the 1D Kitaev model by
unitary transformation, which opens up a new field of
investigation known as topological superconductivity\cite{Kitaev}. 
Importance of topological matter lies in the fact that
additional topological robustness in the nontrivial
phase protects these systems from any 
kind of imperfections present in the materials.
This robustness enhances quantum correlations \cite{Chen} and
causes higher efficiency in electronic transport. 
As a result, topological materials
are expected to be more suitable in the development of
quantum processing devices \cite{DasSarma}.  

SSH model was introduced before in a totally different context,
as it was employed to understand the role of solitonic excitations
in conducting polymers, like polyacetylene, etc. 
The PB phase has been measured recently 
by mimicking the 1D periodic potential of polyacetylene
using system of ultracold atoms in optical lattices\cite{Atala}. 
Signature of topologically protected pair of bound states
is also detected by photonic quantum walk\cite{Kitagawa}. 
In addition, properties of tight-binding
SSH model have been experimentally
validated in photonic lattice composed of helical
waveguides \cite{Rechtsman} and
in phononic crystal composed of cylindrical
waveguides\cite{Li3}.

Existence of topological phase has been demonstrated in
various SSH-like dimerized models in numerous investigations.
For example, in a non-Hermitian SSH model,
where intracell hopping term is turned imaginary
keeping the intercell hopping real, 
the same type of topological
behaviour is obtained\cite{Ziwei}.
The same topological phase appears  
again in another dimerized model constituted by
bigger unit cell comprising of four lattice points
\cite{Maffei,Xie}. In another study, existence of anomalous Floquet
topological $\pi$ mode is successfully demonstrated in 
periodically driven SSH model\cite{Cheng}. 
Topological properties of a hybrid system comprised of
SSH and Kitaev models are studied in order to
find the role of particle-hole symmetry embedded in the
individual models\cite{Wakatsuki}. Another type of
SSH-like staggered model, where particle number is not
conserved is employed before in order
to study its quantum phase transition
along with to explain the nontrivial quench during the 
transition \cite{You,Jafari}. However, most
of these models incorporate no further neighbor
hopping term. At the same time it is also true that
no topological phase with $\nu>1$ appears without further
neighbor terms. 

The topological phase in two-band SSH model is
defined uniquely by $\nu=1$ for each band.
Besides, search of new topological phases,
preferably with higher values of $\nu$ continues afterwards
by adding further neighbor
hopping terms. Collectively they are called
extended SSH (eSSH) models.
Emergence of a new phase with $\nu=-1$
has been demonstrated before by adding a single further neighbor
hopping term\cite{Li}. In another investigation, additional phase
with $\nu=2$ has been obtained simply by adding a pair of
staggered further neighbour terms \cite{Maffei}. 
By invoking multiple further neighbor
hopping terms new phases with $\nu=2,3,4$
have been generated later\cite{Platero,Ghosh}.
PB phase of eSSH model is determined with the Wannier
functions by taking into account the different postions for
two sites within the unit cell\cite{Hetenyi}.
Emergence of multiple topological phases in Kitaev chain with
long range couplings is reported before \cite{Kartik}. 
In this work it is shown that the eSSH model
is capable to host indefinite number of topological phases
with a series of different winding numbers as one wishes.
And remarkably, in this series of eSSH models only a
single extra further neighbor
hopping term is sufficient for their realization. 

Interestingly, demonstration of topological phases in two-dimensional (2D)
system has been started, long before, with the discovery of
integer quantum Hall effect\cite{TKNN,Hasan}.
Subsequently, this phenomenon is observed in other systems as well,  
when Haldane found its realization on a tight-binding model
with complex further neighbor hopping terms
formulated on honeycomb lattice\cite{Haldane}.
This finding gives birth to new area of research known as 
quantum anomalous Hall (QAH) effect 
where the magnetic field is replaced by phase dependent 
hoppings. This state of matter was experimentally realized in periodically
modulated optical honeycomb lattice\cite{Jotzu}. 
For the 2D systems, Chern number
${\mathcal C}$, is treated as the topological invariant.
In the two-band Haldane model, topological phase is defined by
${\mathcal C}=\pm 1$, values of opposite signs 
for the two different energy bands.
Realization of topological phase for higher values of 
${\mathcal C}$s continues thereafter by either invoking
further neighbour hopping terms \cite{Moumita1,Moumita2,Moumita3}
or imposing periodic drive\cite{Arghya,Agarwal}, etc. 
Experimental realization of QAH phases tunable up to ${\mathcal C}=\pm 5$
has been reported recently\cite{Zhao}.

In another development, finding of QAH effect breaks its
dimensional barrier, as the realization of this  phase
is possible in 1D eSSH model, where nearest neighbor (NN) and 
next-nearest-neighbor (NNN) hopping amplitudes
are modulated by two independent cyclic variables\cite{Li2}.
Remarkably, in this case, one of the
cyclic variable can be treated like an additional synthetic
dimension. So as a whole, this 1D model behaves like an
effective 2D model in the reciprocal space and at the same
time, hosts nontrivial topological phases. 

Again, in this investigation, the eSSH models are studied in
2D reciprocal space by introducing different kind of parametrization
in terms of those two cyclic parameters. 
And again, it is shown that these models are capable to host
indefinite number of topological phases
with a series of different Chern numbers.
Properties of these new phases with higher values
of ${\mathcal C}$s have been
characterized in details. Article has been organized in the
following manner.

Structure of these eSSH models are described in
the section \ref{model}. Topological phases of eSSH models are
characterized in Sec. \ref{phase-winding-number}.
Four different eSSH models are introduced here, whose
topological properties are studied in details in terms
of winding numbers, edge states, and quench dynamics.
Models for phases of higher values of $\nu$
will be generalized at the end of this section. 
Topological phases in terms of Chern numbers are 
studied in Sec. \ref{phase-Chern-number}. 
Several types of parametrization
are introduced and their topological properties are
characterized. Spurious topological phases are identified. 
Topological phase diagrams have been drawn in very case
and the symmetries of the Hamiltonian are explained. 
A discussion based on these results is available 
in Sec \ref{Discussion}.
\section{SSH models with further neighbor terms}
\label{model}
The standard SSH model \cite{SSH1} is defined on a 1D bipartite lattice where one primitive
cell contains two different sites, A and B. The corresponding Hamiltonian
is described as
\be
H_{vw}=\sum_{j=1}^N\left(v\,c^\dag_{{\rm A},j}c_{{\rm B},j}+w\,c^\dag_{{\rm A},j+1}c_{{\rm B},j} \right)\!+{\rm h.c.},
 \label{ham-ssh}
\ee
where $c_{{\rm A},j}$ and $c_{{\rm B},j}$ stand for the annihilation operators of
electron on sublattices A and B, respectively, in the $j$th primitive cell. 
$N$ is the total number of primitive cells where $v$ and $w$ are the
intracell and intercell hopping amplitudes, respectively. These
terms permit hopping only between the adjacent sites. 
Energy spectrum of $H_{vw}$ is gapless when $w=v$, while there is
a band gap when $w \ne v$. Between the two gapful regions around the
gapless point, one is topologically trivial ($\nu=0$) when $w<v$, and remarkably 
as long as $w>v$, this simple model hosts a single
nontrivial topological phase with $\nu=1$.

\begin{figure}[h]
\psfrag{A}{\large A}
\psfrag{B}{\large B}
\psfrag{v}{\large $v$}
\psfrag{w}{\large $w$}
\psfrag{z}{\large $z$}
\includegraphics[width=230pt]{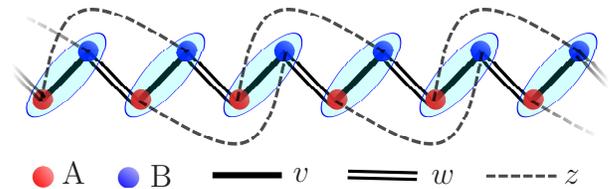}
\caption{Extended SSH model describing the hopping for the
total Hamiltonian, $H$, in Eq. \ref{H-z}.}
\label{Extended-SSH-1}
\end{figure}

In 2019, Li and Miroshnichenko \cite{Li} showed that a new topological
phase with $\nu=-1$ appears on introducing
additional terms which allow hopping between sites of A
sublattice and nonadjacent sites of B sublattice but
only among the NN primitive cells as shown in Fig. \ref{Extended-SSH-1}. 
A single pair of topological edge states is found to
appear associated with this new phase.

The chiral symmetry of the resultant system is preserved by this
specific choice of sites between which the hopping is allowed. 
If $z$ be the amplitude of this additional hopping,
total Hamiltonian can be expressed as
\bea
H&=& H_{vw}+H_z,\nonumber\\ [0.4em]
H_z&=&\sum_{j=1}^Nz\,c^\dag_{{\rm A},j}c_{{\rm B},j+1}+{\rm h.c.},
\label{H-z}
\eea
and distribution of winding numbers in the parameter space
of the system is given by
\be
\nu=\left\{\begin{array}{ll}
    0,& |w+z|<v,\\[0.3em]
    1,&|w+z|>v \;{\rm and}\;w>z,\\[0.3em]
    -1,&|w+z|>v \;{\rm and}\;w<z.
  \end{array}\right.
\ee
In another study, P\'erez-Gonz\'alez {\em et al} showed that
an additional topological phase with $\nu=2$ emerges
in the presence of more than one further neighbor
hopping terms \cite{Platero}. Two distinct pairs of
topological edge states are found to
appear. In the presence of multiple further neighbor
hopping terms, topological phases with
higher winding numbers, say, up to $\nu=4$ have been
reported so far \cite{Ghosh}. 

In this study we are going to show that a single additional hopping
term is sufficient to produce the topological
phases with any value of winding number as one wishes.
Topological phases with higher values of winding numbers
can be generated by systematically increasing the separation 
between the sites over which hopping is taken into account.
Multiple pairs of edge states, consistent with the
value of $\nu$, are found to appear. 
\section{Topological phases in terms of winding numbers}
\label{phase-winding-number}
In order to generate the topological phases with any values of
winding numbers in the most
simple way, two different types of eSSH models are 
introduced, however, both of them include a single further neighbor
hopping term.
Two different types of Hamiltonians are termed as
`A-B' and `B-A' depending on the
ordering of the sublattice sites and they are noted as
$H^{\rm {A\mbox{-}B}}_{z,n}$ and $H^{\rm {B\mbox{-}A}}_{z,n}$, respectively,
where $(n-1)$ is the number of intermediate
primitive cells being covered under the hopping distance
and $z$ is the amplitude of the further neighbour hopping.
In this nomenclature, Hamiltonian $H_z$ in Eq. \ref{H-z}
can be specified as $H^{\rm {A\mbox{-}B}}_{z,1}$.
However, hopping only between different sublattices
is allowed in this case.
This type of hopping term 
preserves the particle-hole and inversion symmetries.
Conservation of these symmetries means the
preservation of chiral symmetry in addition.
Now the topological properties of four different eSSH models
will be studied in great details. Among them, 
two are of type `B-A' and the remaining two are of type `A-B'.
\subsection{Topological phases for $H=H_{vw}+H^{\rm {B\mbox{-}A}}_{z,2}$}
\begin{figure}[h]
\psfrag{A}{\large A}
\psfrag{B}{\large B}
\psfrag{v}{\large $v$}
\psfrag{w}{\large $w$}
\psfrag{z}{\large $z$}
\includegraphics[width=230pt]{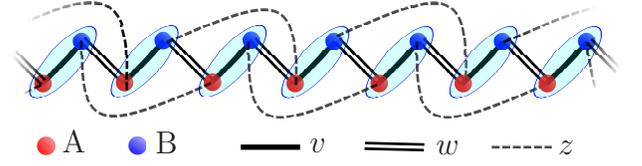}
\caption{Extended SSH model describing the hopping in $H^{\rm {B\mbox{-}A}}_{z,2}$.}
\label{Extended-SSH-2}
\end{figure}

Total Hamiltonian in this case is expressed as
\bea
H&=& H_{vw}+H^{\rm {B\mbox{-}A}}_{z,2},\nonumber\\ [0.4em]
H^{\rm {B\mbox{-}A}}_{z,2}&=&\sum_{j=1}^Nz\,c^\dag_{{\rm B},j}c_{{\rm A},j+2}+{\rm h.c.},
\label{HBA-z-2}
\eea
where the hopping term extends over one intermediate primitive cell,
which is shown in Fig. \ref{Extended-SSH-2}.
Under the Fourier transformations, 
\bea c_{{\rm A},j}&=&\frac{1}{\sqrt N}\sum_{\rm k\in {\rm BZ}}a_{\rm k}\,e^{i\rm kj},\nonumber\\ [0.4em]
c_{{\rm B},j}&=&\frac{1}{\sqrt N}\sum_{\rm k\in {\rm BZ}}b_{\rm k}\,e^{i\rm kj},\nonumber
\eea
where the summation extends over BZ,
and assuming periodic boundary condition (PBC),
the Hamiltonian in the k-space becomes
\[H=\sum_{\rm k\in {\rm BZ}}\Psi^\dag_{\rm k}H({\rm k})\psi_{\rm k},\]
for which $\Psi^\dag_{\rm k}=[a^\dag_{\rm k}\;b^\dag_{\rm k}],$ and
$H(\rm k)=\boldsymbol g(\rm k)\cdot \boldsymbol \sigma$. 
Here, $\boldsymbol \sigma=(\sigma_x,\sigma_y,\sigma_z)$,
are the Pauli matrices, 
and assuming the unit lattice parameter, ($a=1$), 
\[\boldsymbol g(\rm k)\equiv\left\{\begin{array}{l}
g_x=v+w\cos{\!(\rm k)}+z\cos{\!(2\rm k)},\\[0.3em]
g_y=w\sin{\!(\rm k)}+z\sin{\!(2\rm k)},\\[0.3em]
g_z=0.
\end{array}\right.\]
It can be shown that $H({\rm k})$ satisfies the
following transformation relations under the three
different operators:
\[\left\{\begin{array}{l}
\mathcal T H({\rm k}) \mathcal T^{-1}=H(-{\rm k}),\\ [0.4em]
\mathcal P H({\rm k}) \mathcal P^{-1}=-H(-{\rm k}),\\ [0.4em]
 \sigma_z H({\rm k}) \sigma_z=-H({\rm k}),\end{array}\right. \]
where $\mathcal T = \mathcal K$, $\mathcal P = \mathcal K \sigma_z$ and
$\mathcal K$ is the complex conjugation operator. 
These relations correspond to the conservation of
time-reversal, particle-hole and chiral symmetries.
As a consequence, inversion symmetry is preserved as 
$\sigma_x H({\rm k}) \sigma_x=H(-{\rm k})$.
 
\begin{figure}[h]
\psfrag{x}{\large $w/|v+z|$}
\psfrag{y}{\large k}
\psfrag{ep}{\large $E_+({\rm k})$}
\psfrag{a}{\large (a)}
\psfrag{b}{\large (b)}
\psfrag{v}{\large $v=1$}
\psfrag{za}{\large $z=1/2$}
\psfrag{vb}{\large $v=3/4$}
\psfrag{zb}{\large $z=1$}
\psfrag{5}{\large 5}
\psfrag{4}{\large 4}
\psfrag{3}{\large 3}
\psfrag{2}{\large 2}
\psfrag{1}{\large 1}
\psfrag{0}{\large 0}
\psfrag{0.0}{\large 0}
\psfrag{-2}{\large $-$2}
\psfrag{-1}{\large $-$1}
\psfrag{-3.1}{\large $-\pi$}
\psfrag{-1.6}{\large $-\pi/2$}
\psfrag{3.1}{\large $\pi$}
\psfrag{1.6}{\large $\pi/2$}
  \includegraphics[width=230pt]{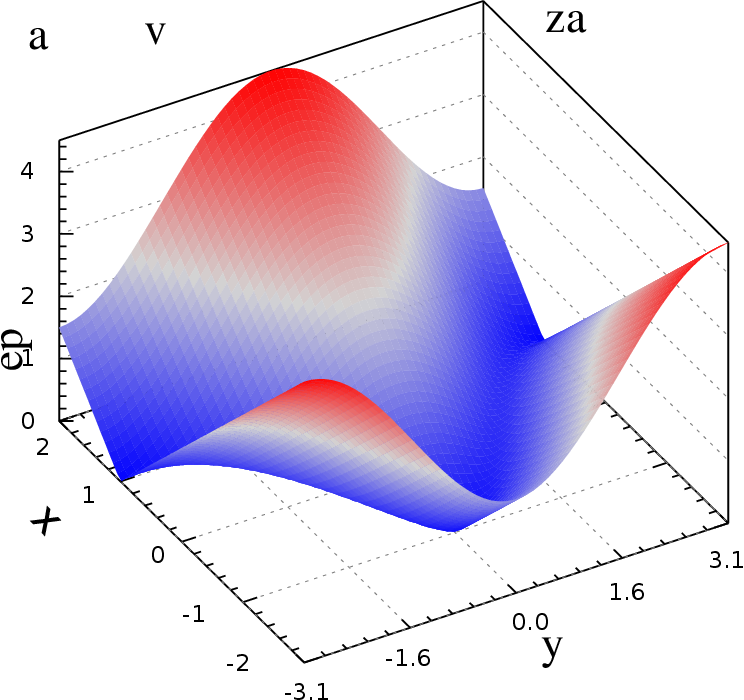}
  \vskip 0.6cm
  \includegraphics[width=230pt]{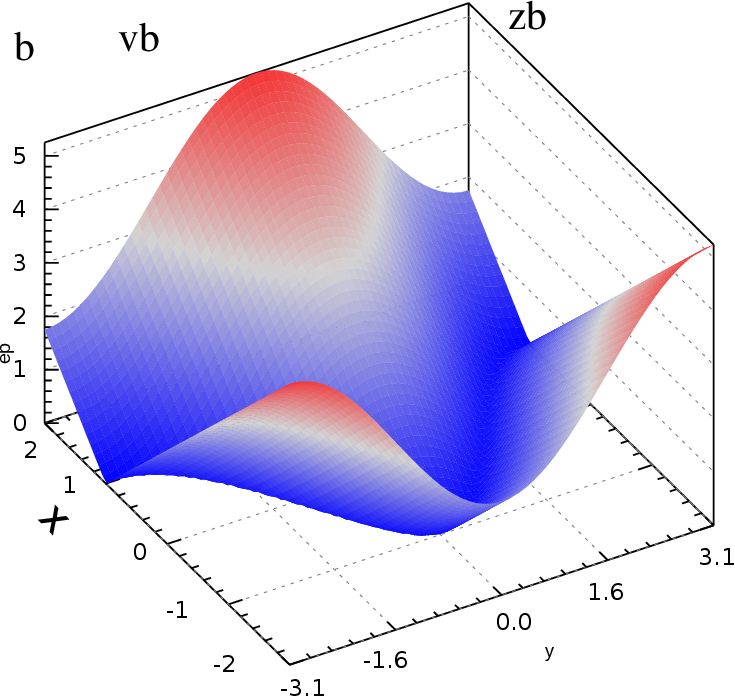}
\caption{Dispersion relation for $H=H_{vw}+H^{\rm {A\mbox{-}B}}_{z,2}$
  when $v/z>1$, (a) and $v/z<1$, (b).}
\label{Dispersion-SSH-2}
\end{figure}

$\boldsymbol g(\rm k)$ can be spanned as
a vector in the $g_x\mbox{-}g_y$ complex plane,
due to the conservation of chiral symmetry. 
As a result, the dispersion relation can be expressed as
$E_\pm({\rm k})=\pm|g({\rm k})|$, or,
$E_\pm({\rm k})=\pm \sqrt{v^2+w^2+z^2+2[vw\cos{\!(\rm k)}
    +vz\cos{\!(2\rm k)}+wz\cos{\!(\rm k)}]}$.
Dispersions are symmetric around the energy, $E=0$,
since the Hamiltonian preserves particle-hole symmetry.  
Variation of dispersion relation, $E_+({\rm k})$,
with $w/|v+z|$ for $v=1$, $z=1/2$ and $v=3/4$, $z=1$ are shown in
Fig. \ref{Dispersion-SSH-2} (a) and (b), respectively.
The lower band, $E_-({\rm k})$ is not drawn. 
The figures in (a) and (b) are serving as prototype figures for
$v/z>1$ and $v/z<1$, respectively.
Dispersions comprise of one broad peak when
$w/|v+z|\le 1$ for both the cases $v/z>1$ and $v/z<1$. 
Band gap vanishes at the BZ boundaries, ${\rm k}=\pm \pi$ and
${\rm k}=0$, when $w=|v+z|$. 
As a result, $\nu$ is undefined at the point
when $w=|v+z|$.

\subsection{Winding number, $\nu$}
Tip of the vector $\boldsymbol g(\rm k)$ traces out closed loops
in the $g_x\mbox{-}g_y$ plane 
if k runs from $-\pi$ to $\pi$ on the BZ. Winding number is defined
to enumerate the number of closed loops around the origin of the plane.
Mathematically it is expressed as
\[\nu=\frac{1}{2\pi}\int_{-\pi}^\pi \left[\boldsymbol {\hat g}(\rm k)\times
  \frac{d}{d\rm k}\,\boldsymbol {\hat g}(\rm k)\right]_z\!d{\rm k},\]
where $\boldsymbol {\hat g}(\rm k)
=\boldsymbol g(\rm k)/|\boldsymbol g(\rm k)|.$
Two distinct topological phases, with $\nu=1,2$
are found for this case in the parameter space as
\be
\nu=\left\{\begin{array}{ll}
    1,& w>|z+v|,\\[0.3em]
    0,&w<|z+v|,\;{\rm and}\;v>z,\\[0.3em]
    2,&w<|z+v|, \;{\rm and}\;v<z,
  \end{array}\right.
\ee
and these are associated with a number of topological phase
transitions.

For examples, when $v>z$, a transition takes place
at $w=|v+z|$, separating trivial phase, $\nu=0$ 
for $w<|v+z|$ and topological phase, $\nu=1$
for $w>|v+z|$. Whereas, transition occurs at the
same point between two topological phases when $v<z$.
In this case, the phase for $w>|v+z|$ is marked by
$\nu=1$, while that for $w<|v+z|$ is identified by
$\nu=2$. In all cases transition takes place 
between the phases with energy gap, and
obviously, gap closes at the transition point, 
$w=|v+z|$. 

\begin{figure}[h]
\psfrag{gxx}{\large $g_x$}
\psfrag{gyy}{\large $g_y$}
\psfrag{a}{\large (a)}
\psfrag{b}{\large (b)}
\psfrag{c}{\large (c)}
\psfrag{d}{\large (d)}
\includegraphics[width=230pt]{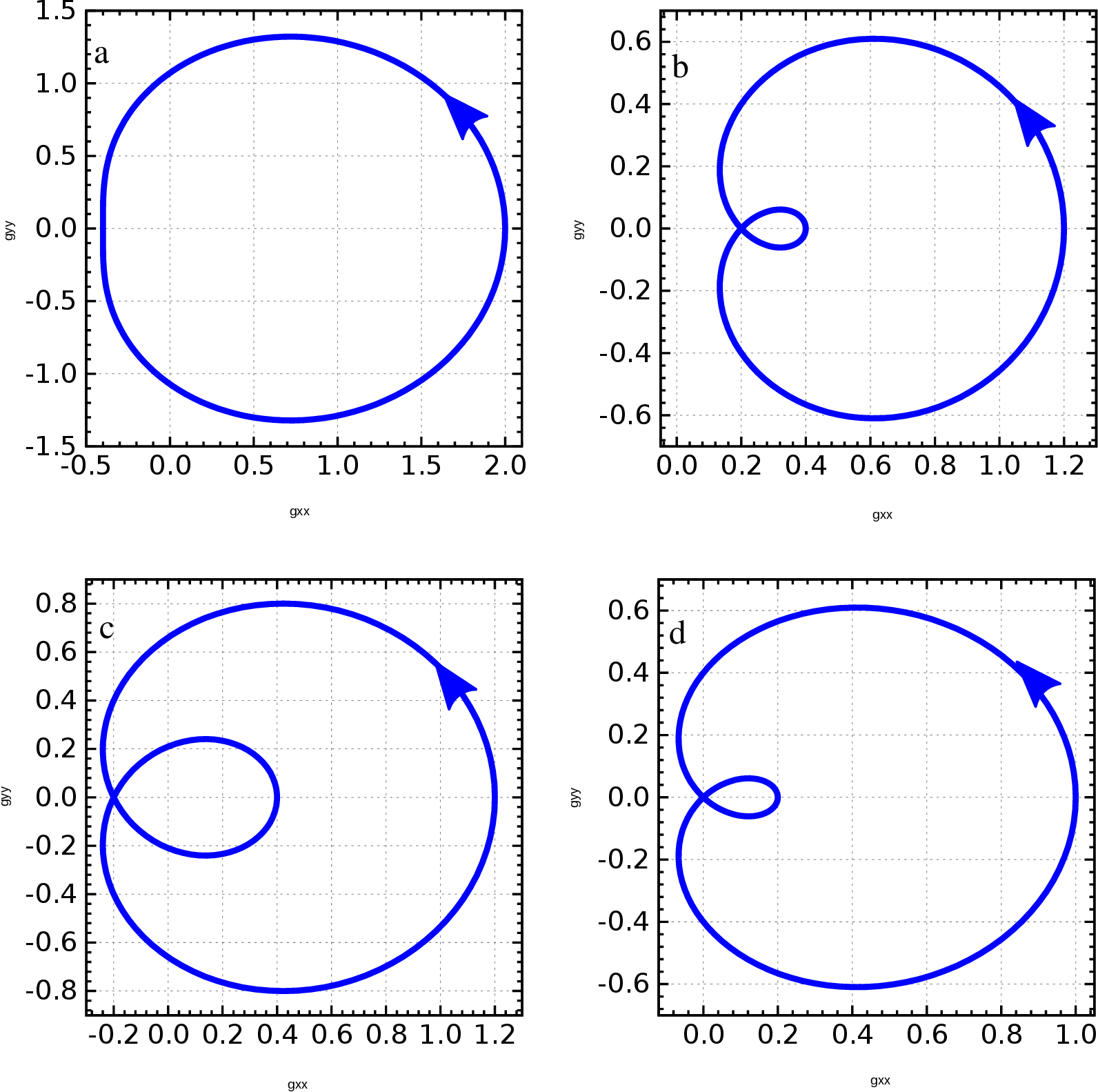}
\caption{Parametric winding diagrams in the
$g_x\mbox{-}g_y$ plane for the Hamiltonian,
  $H^{\rm {B\mbox{-}A}}_{z,2}$. Four figures are drawn for
  (a) $v=0.5$, $w=1.2$, $z=0.3$, (b) $v=0.5$, $w=0.4$, $z=0.3$,
  (c) $v=0.3$, $w=0.4$, $z=0.5$, and (c) $v=0.3$, $w=0.4$, $z=0.3$.}
\label{parametric-windings-SSH-2}
\end{figure}

The parametric plot of winding diagrams in the
$g_x\mbox{-}g_y$ complex plane are shown in
Fig. \ref{parametric-windings-SSH-2}. 
Four figures are drawn for
  (a) $v=0.5$, $w=1.2$, $z=0.3$, (b) $v=0.5$, $w=0.4$, $z=0.3$,
(c) $v=0.3$, $w=0.4$, $z=0.5$, and (c) $v=0.3$, $w=0.4$, $z=0.3$.
Arrow head indicates the direction of move of the
the vector $\boldsymbol g({\rm k})$ for an infinitesimal
increment of ${\rm k}$. 
Tip of $\boldsymbol g({\rm k})$ moves
in counterclockwise direction over all the closed contours. 
The contours in (a) and (c) enclose the the origin, while
that in (b) does not. On the other hand, contour in (d)
passes over the origin. The curve passes around the
origin once in (a) and twice in (c). 
Those figures serve as the prototype windings for the
four different regions, 
$w>|v+z|$, $v>z$ for $\nu=1$, $w<|v+z|$, $v>z$
for $\nu=0$, $w<|v+z|$, $v<z$, for $\nu=2$, and $w<|v+z|$, $v=z$.
There is gap for the first three cases
while the spectrum is gapless for the last. 

\begin{figure}[h]
  \psfrag{w}{\large $w/|v\!+\!z|$}
  \psfrag{3}{\hskip -0.15 cm 3}
\psfrag{2}{\hskip -0.15 cm 2}
\psfrag{1}{\hskip -0.15 cm 1}
\psfrag{0}{\hskip -0.15 cm 0}
\psfrag{-3}{\hskip -0.15 cm -3}
\psfrag{-2}{\hskip -0.15 cm -2}
\psfrag{-1}{\hskip -0.15 cm -1}
\psfrag{y}{\large k}
\psfrag{E}{\hskip -0.5 cm Energy}
\psfrag{a}{\large (a)}
\psfrag{b}{\large (b)}
\psfrag{v}{\large $v=1$}
\psfrag{za}{\large $z=1/2$}
\psfrag{vb}{\large $v=3/4$}
\psfrag{zb}{\large $z=1$}
\hskip -3.0 cm
\begin{minipage}{0.20\textwidth}
  \includegraphics[width=130pt,angle=-90]{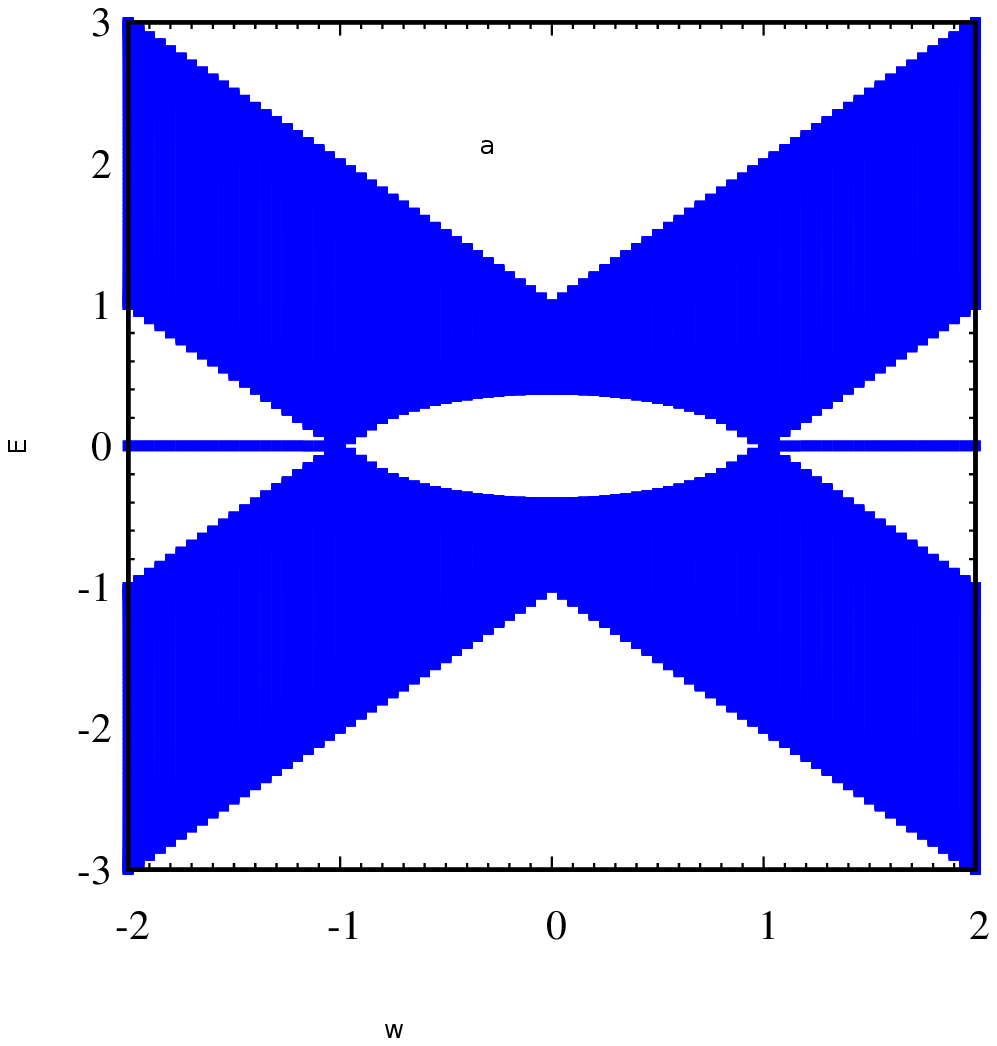}
  \end{minipage}\hskip 0.6cm
  \begin{minipage}{0.2\textwidth}
  \includegraphics[width=130pt,angle=-90]{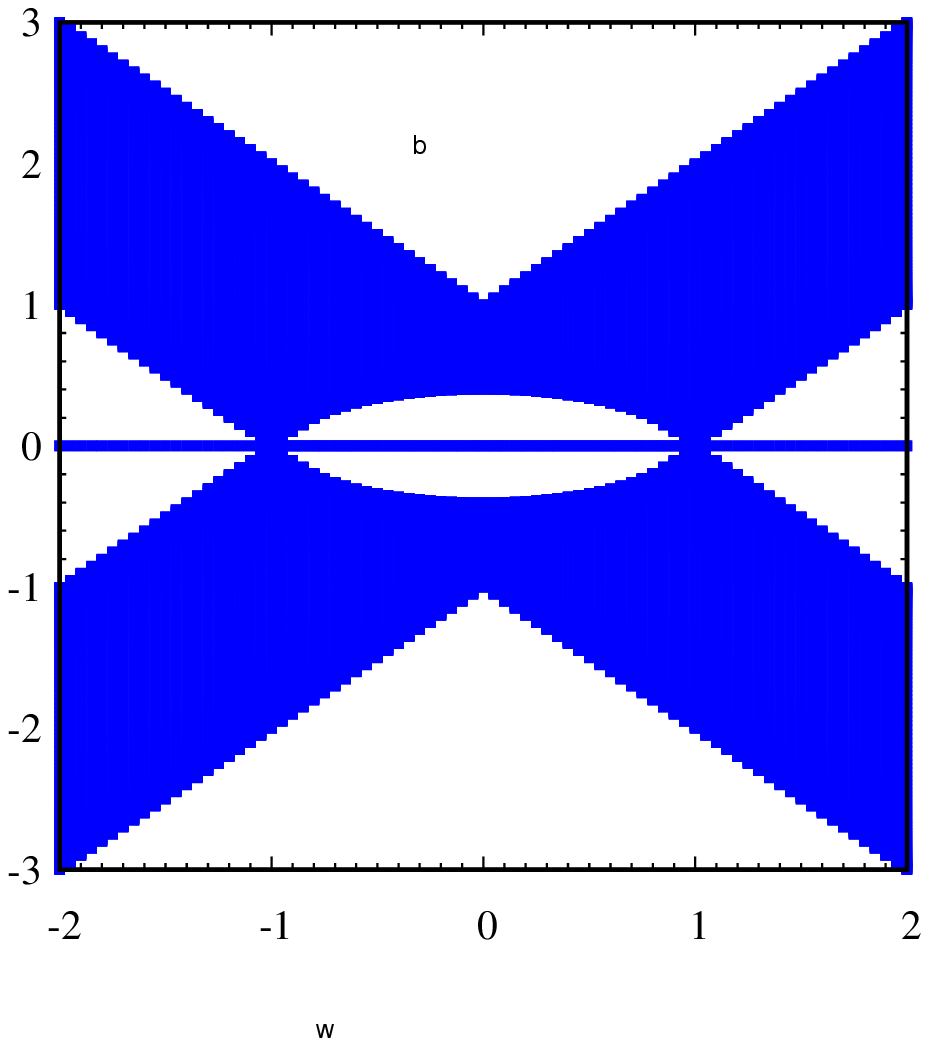}
  \end{minipage}
\caption{Bulk and edge state energies for $H=H_{vw}+H^{\rm {B\mbox{-}A}}_{z,2}$
  when $v/z>1$, (a) for $v=0.7$, $z=0.3$, and $v/z<1$, (b)
  for $v=0.3$, $z=0.7$. Variation of energy with
$w/|v\!+\!z|$ are shown for the lattice of 200 sites.}
\label{Edge-states-SSH-2}
\end{figure}
\begin{figure}[]
\psfrag{mps}{\hskip 0.1 cm $|\psi|^2$}
\psfrag{a}{ (a)}
\psfrag{b}{ (b)}
\psfrag{n}{\hskip -0.2 cm sites}
\psfrag{p}{$v=0.25,\,w=2.5,\,z=0.25$}
\psfrag{q}{$v=0.25,\,w=0.25,\,z=2.5$}
\psfrag{140}{140}
\psfrag{120}{120}
\psfrag{100}{100}
\psfrag{80}{80}
\psfrag{60}{60}
\psfrag{40}{40}
\psfrag{20}{20}
\psfrag{0.5}{0.5}
\psfrag{1.0}{1.0}
\psfrag{0} {0}
\psfrag{0.0}{0.0}
\includegraphics[width=230pt]{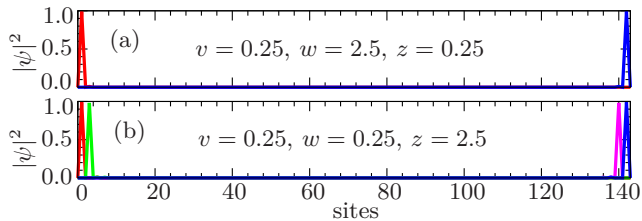}
\caption{Probability density of edge states for $H^{\rm {B\mbox{-}A}}_{z,2}$: 
  (a) for $v=0.25$, $w=2.5$, $z=0.25$, one pair of edge states, 
  (b) for $v=0.25$, $w=0.25$, $z=2.5$, two pairs of edge states.
Figures are drawn for the lattice of 142 sites.}
\label{Edge-state-probability-for-SSH-2}
\end{figure}

Variation of bulk-edge state energies with respect to
$w/|v+z|$ is shown in Fig. \ref{Edge-states-SSH-2}
as long as $w/|v+z|\le \pm 2$. 
A single pair of zero energy edge states
survives when $w>|v\!+\!z|$ as shown in (a). No edge state is there
in this system when $w<|v\!+\!z|$ and and $v>z$. In contrast,
zero energy edge states are always there
when $w>|v\!+\!z|$ and $v<z$ which is shown in
Fig. \ref{Edge-states-SSH-2} (b).
Actually, a single pair of zero energy edge states
survives when $w>|v\!+\!z|$, and 
two pairs of edge states are there when
$w<|v\!+\!z|$ and $v<z$.
The figures are drawn for lattice of sites 200. 
All these results are consistent with the bulk-boundary correspondence rule.

In order to confirm the presence of zero energy edge states,
probability density of those states are drawn in Fig. 
\ref{Edge-state-probability-for-SSH-2} for the lattice of
142 sites.
Two figures are drawn for two distinct topological phases.
In the upper panel (a), probability densities of two
distinct edge states with $E=0$ are shown when
$v=0.25$, $w=2.5$, $z=0.25$, as these values confirm to the
conditions, $w>|v\!+\!z|$. Probability density
of one edge state exhibits sharp peak at site $m=1$
and another one at site $m=142$. This corresponds to
the topological phase with $\nu=1$.
On the other hand, for $w<|v\!+\!z|$ 
and $v<z$, probability densities of four 
distinct edge states with $E=0$ are shown
in the lower panel (b) when
$v=0.25$, $w=0.25$, $z=2.5$,
as these values are in accordance to 
the last conditions. 
Probability density 
of four orthogonal edge states exhibit
sharp peak at sites $m=1$, $m=3$, $m=140$,
and $m=142$. It indicates that zero energy states
near the left edge are localized on the A sublattice,
while those close to the right edge are localized
on the B sublattice. 
This result is in accordance to
the topological phase with $\nu=2$.

\begin{figure}[h]
  \psfrag{nu}{\large $\nu$}
   \psfrag{nu0}{\hskip -.5cm \color{white} \Large $\nu=0$}
   \psfrag{nu2}{\hskip -.5cm \color{white} \Large $\nu=2$}
   \psfrag{nu1}{\hskip -.0cm \color{blue} \Large $\nu\!=\!1$}
\psfrag{v}{\large $v$}
\psfrag{zv}{\large $v+z=2$}
\psfrag{w}{\hskip -0.4 cm \large $w/|v+z|$}
\psfrag{0.5}{\large 0.5}
\psfrag{0.0}{\large 0.0}
\psfrag{1.0}{\large 1.0}
\psfrag{1.5}{\large 1.5}
\psfrag{2.0}{\large 2.0}
\psfrag{2}{\large 2}
\psfrag{1}{\large 1}
\psfrag{0}{\large 0}
\psfrag{-0.5}{\large -0.5}
\psfrag{-1.0}{\large -1.0}
\psfrag{-1.5}{\large -1.5}
\includegraphics[width=230pt,angle=-90]{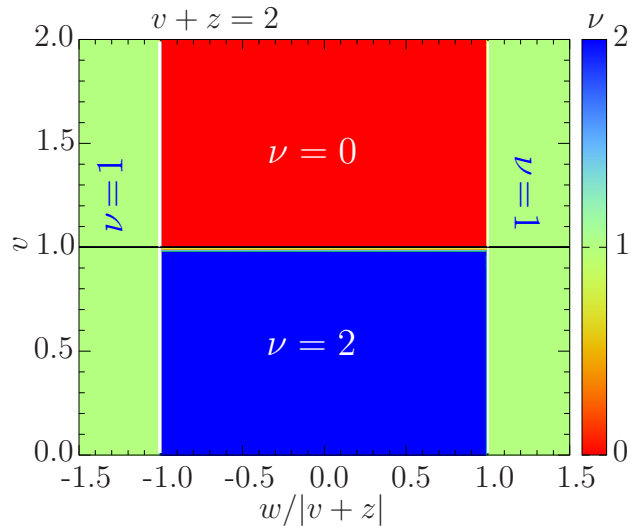}
\caption{Topological phase diagram for the Hamiltonian,
  $H^{\rm {B\mbox{-}A}}_{z,2}$. Trivial phase is shown by red
  ($\nu=0$)  while two distinct topological
  phases are shown by green ($\nu=1$) and blue ($\nu=2$).
  This diagram is drawn for $v+z=2.$
The horizontal line 
indicates the value $v=1$ or $v/z=1.$}
\label{topological-phase-windings-SSH-2}
\end{figure}

An extensive phase diagram of the total Hamiltonian 
including $H^{\rm {B\mbox{-}A}}_{z,2}$ is shown in Fig
\ref{topological-phase-windings-SSH-2}. Here contour plot
for $\nu$ is drawn in the $v$-$w/|v+z|$ space.
Presence of two distinct topological phases,
$\nu=1$ and 2 along with the trivial phase, $\nu=0$ are
shown in green, blue and red, respectively.
The horizontal line is drawn at 
$v=1$ or $v/z=1$ since this diagram is drawn for
$v+z=2$. The line segment within the points $w/(v+z)=\pm 1$ 
separates the trivial phase from 
the topological phase with $\nu=2$.
Hence phase transition occurs around this segment. 
Another topological phase with $\nu=1$
appears beyond the two vertical lines
drawn at $w/(v+z)=\pm 1$. They separate
topological phases with $\nu=1$ and 2
when $v/z<1$ and topological ($\nu=1$) and trivial phase
when $v/z>1$. So the system undergoes phase transition
around those straight lines. Band gap vanishes over those
lines as well as on the line segment.

\subsection{Quenched dynamics in the presence of nonlinear terms}
Now the effect of nonlinearity on the topological phase will be
studied following the method developed by
Ezawa \cite{Ezawa1}.
Schr\"odinger equation for a Hamiltonian matrix, $M$, spanned
on a lattice composed of $L$ sites can be written
as ($\hbar =1$),
\be
i\,\frac{\partial \psi_l}{\partial t}+\sum_{m=1}^LM_{lm}\,\psi_l=0.
\label{Schrodinger-1}
\ee
It actually comprises $L$ coupled linear equations 
and governs the time evolution of the system
where $M_{lm}$ is recognized as the element of   
hopping matrix in case of tight-binding model. 
This system hosts the topological as well as trivial phases
for different parameter regime. 

\begin{figure*}[t]
\psfrag{a}{\hskip 0.2 cm \large \color{white} (a)}
\psfrag{c}{\hskip 0.4 cm \large \color{white} (c)}
\psfrag{b}{\hskip 0.2 cm \large \color{white}  (b)}
\psfrag{sites}{\hskip -0.1 cm sites}
\psfrag{t}{ time}
\hskip -2.4 cm
\begin{minipage}{0.277\textwidth}
  \includegraphics[width=177pt,angle=-90]{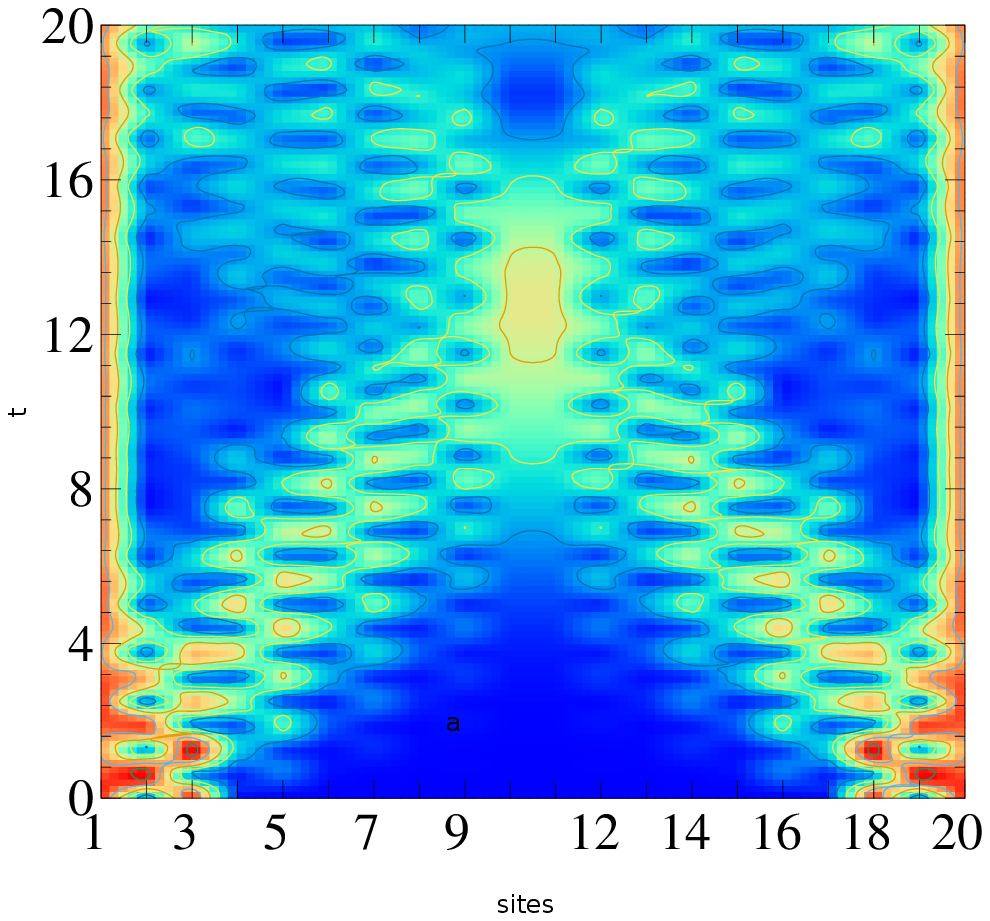}
    \end{minipage}\hskip 0.3cm
  \begin{minipage}{0.277\textwidth}
  \includegraphics[width=178pt,angle=-90]{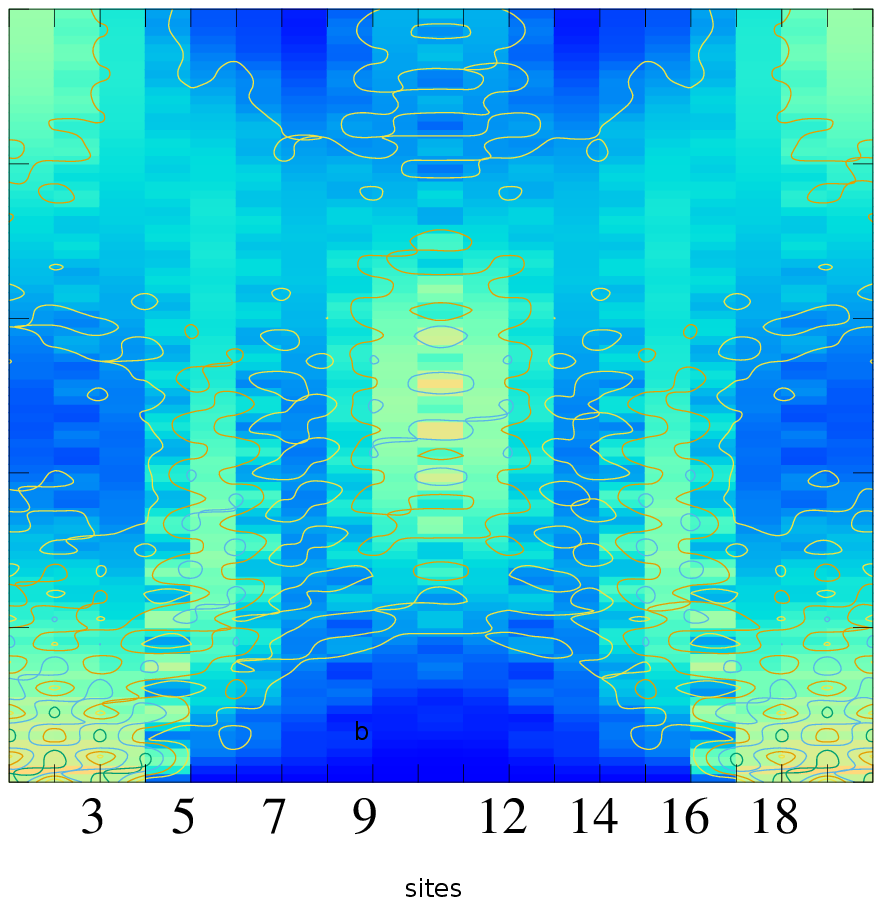}
  \end{minipage}\hskip 0.3cm
  \begin{minipage}{0.28\textwidth}
  \includegraphics[width=180pt,angle=-90]{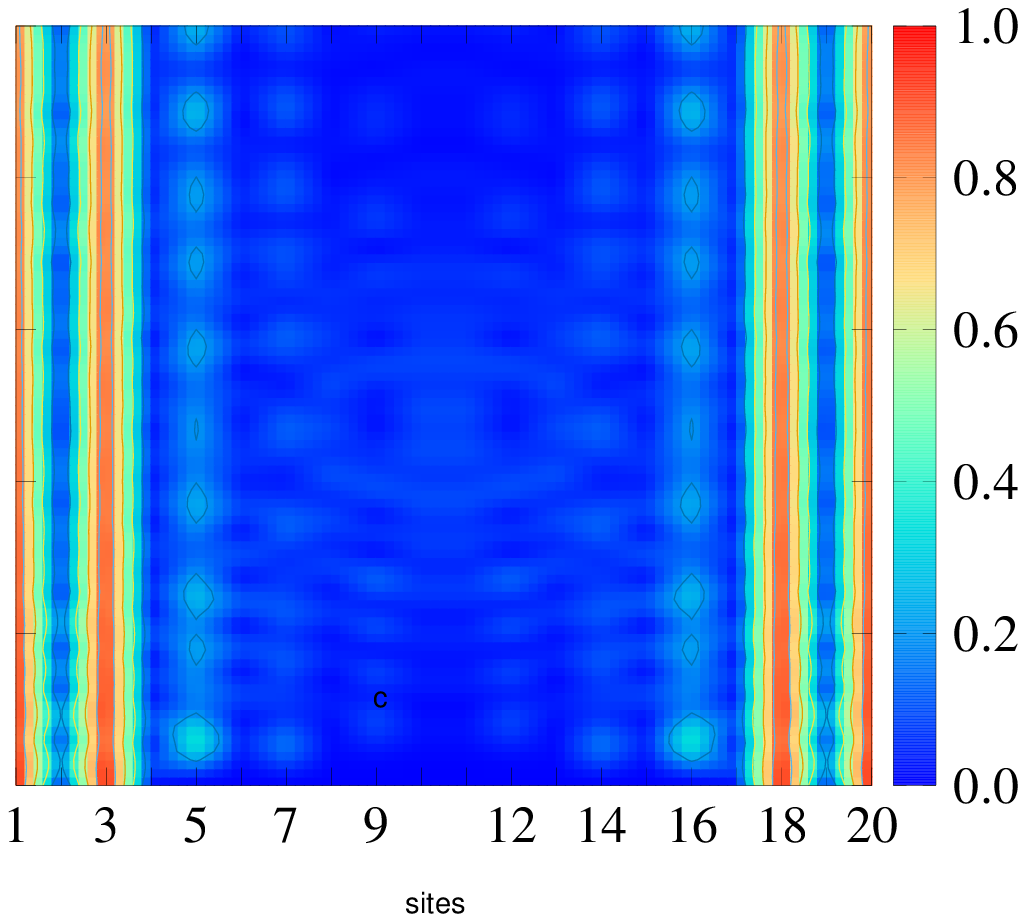}
  \end{minipage}
\caption{Quench dynamics for $H^{\rm {B\mbox{-}A}}_{z,2}$
  when $\zeta=0.5$, (a) for $v=0.25$, $w=2.5$, $z=0.25$, 
   (b) for $v=2.5$, $w=0.25$, $z=0.25$,
  (c) for $v=0.25$, $w=0.25$, $z=2.5$.
Figures are drawn for lattice with 20 sites. }
\label{Quence-dynamics-SSH-2}
\end{figure*}

The eigenvalue equation for the hopping matrix,
$M$ is written as
\be
M\bar \phi_q=E_q\bar \phi_q,\;1\le q\le L,
\label{eigen-system}
\ee
where $q$ serves as the quantum index.
Hence, time evolution of the model
is governed by the 
solution of Eq. \ref{Schrodinger-1} as 
\be
\bar \phi_q(t) = e^{-itE_q}\,\bar \phi_q(t),
\label{time-evolution}
\ee
since the Schr\"odinger equation eventually turns into a
set of decoupled equations
\be
i\,\frac{\partial \bar \phi_q}{\partial t}+\sum_{m=1}^LM_{lm}\,\bar \phi_q=0.
\ee
The variation of energies $E_q$ with $w/|v+z|$
for two different topological phases
have been shown in the Fig. \ref{Edge-states-SSH-2}, when
the hopping matrix is constituted for the Hamiltonian
defined in Eq. \ref{HBA-z-2} 
for the lattice of sites $L=200$. 
Topological phases are always protected by the
zero energy edge states by virtue of 
particle-hole symmetry of the system.
As a result, no time evolution of those localized
states is permissible according to the Eq. \ref{time-evolution}.
In light of this fact, time evolution of the
edge states in the presence of additional
nonlinear term will be studied. 

The Schr\"odinger equation in the presence of
nonlinear term for the one-dimensional tight-binding model
of hopping matrix $M_{lm}$ is defined by 
\be
i\frac{\partial \psi_l}{\partial t}+\sum_{m}^LM_{lm}\psi_l
+\zeta |\psi_l|^2\psi_l=0,
\label{Schrodinger-2}
\ee
where the effect of nonlinearity is controlled
by the parameter $\zeta$. 
Explicit form of the set of coupled nonlinear first order differential
equation for finite chain of $L$ sites and for the Hamiltonian
defined in Eq. \ref{HBA-z-2} with open boundary condition (OBC)
is given by 
\bea
i\,\frac{\partial \psi_{2j-1}}{\partial t}&=&v(\psi_{2j}-\psi_{2j-1})+
w(\psi_{2j-2}-\psi_{2j-1})\nonumber \\[0.0 em]
&&+z(\psi_{2j-4}-\psi_{2j-1})-\zeta |\psi_{2j-1}|^2\psi_{2j-1},\nonumber \\[0.2 em]
\vdots\quad\; &=&\qquad \qquad \vdots \label{coupled-equations-1} \\[0.3 em]
i\,\frac{\partial \psi_{2j}}{\partial t}&=&w(\psi_{2j+1}-\psi_{2j})+
v(\psi_{2j-1}-\psi_{2j})\nonumber \\[0.0 em]
&&+z(\psi_{2j+3}-\psi_{2j})-\zeta |\psi_{2j}|^2\psi_{2j},\nonumber
\eea
where $j$ denotes the cell index which ultimately
generates $L$ number of coupled equations each one for every
site. So, $j=1,2,3,\cdots,N/2$.
Differential equations for odd and even sites
are different since the translational symmetry
of one lattice unit is broken. 

 The fate of the
topological state when $\zeta \ne 0$
will be studied in terms of the time evolution
of the nonlinear system by imposing
an initial condition,
\[\psi_l(t)=\delta_{l,m}\quad {\rm when}\;\;t=0.\]
It means a delta-function like pulse at the $m$-th site is
given initially.  Henceforth dynamics of the resulting 
nonlinear system will be examined by maintaining 
the conservation rule imposed by the equation,
\be
\sum_{l=1}^L|\psi_l(t)|^2={\rm constant}. 
\ee
Value of the constant may be fixed depending on 
the choice of the initial conditions. The initial conditions
in turn depend on the value of winding number
for a particular topological phase. 
It is shown that topological phase defined in the
linear system is robust against the introduction
of the nonlinear term as long as $\zeta<1$, 
as a result, quenching of the edge states are
observed.

As the general solution of the Eq. \ref{Schrodinger-2}
can be expanded as
\[\psi_l(t)=\sum_q c_q(t) \bar \phi_q(t),\]
the initial state can be expressed as
\be
\psi_l(0)=\delta_{l,m}=\sum_q c_q \bar \phi_q(0).
\label{initial-condition-1}
\ee
The topological phase of the linear system is
always protected by the presence of zero energy edge
(localized) states. So keeping in
mind the position of edge states, initial condition is
imposed either by $l=1$ or $l=L$, when $\nu=1$.
Here  $l=1$ ($l=L$) denotes the leftmost (rightmost)
site of the lattice. 
Now for $l=1$, initial condition turns out as
$\psi_l(0)=\delta_{l,1}$.
Right hand side of the Eq. \ref{initial-condition-1}
may be simplified by labeling the
zero energy state by $\bar \phi_1$ with $E_1=0$
in Eq. \ref{eigen-system}. So, at $t=0$,
\be
\psi_1(0)= c_1 \bar \phi_1(0).
\ee

As $E_1=0$, $\bar \phi_q(t) = \bar \phi_q(0)$,
which leads to the fact that
\[\psi_1(t)= c_1 \bar \phi_1(0).\]
It means no time evolution of the edge states
is there or in other words a non-zero probability
amplitude at the edge site remains at any time. 
It corresponds to the quenching of the edge states.
No such quenching is possible for the bulk states
by virtue of their non-zero energy, ($E_q \ne 0$).

In contrast, zero-energy localized states are
absent in the trivial phase, and all the states are
found to extend within the bulk. As a result,
quenching dynamics of edge states may serve as
an alternative numerical tool to distinguish the
topological and trivial phases by investigating
the effect of nonlinear component on the initial
condition. At the same time,
investigation of quench dynamics for systems under
PBC is meaningless, since no edge state is there. 

Quenching of edge states for the
nonlinear system is shown in 
Fig. \ref{Quence-dynamics-SSH-2}, 
by solving the set of Eq. \ref{coupled-equations-1},
for $L=20$ when $\zeta=0.5$.
Contour plot for the time evolution of the absolute value 
of complex amplitude, $|\psi_l(t)|$, is drawn for
every site, $l=1,2,3,\cdots,20$, 
which is shown along the horizontal axis. 
Three contour plots are shown 
(a) for $v=0.25$, $w=2.5$, $z=0.25$, 
   (b) for $v=2.5$, $w=0.25$, $z=0.25$,
(c) for $v=0.25$, $w=0.25$, $z=2.5$,
where (b) indicates trivial phase while
(a) and (c) for the topological phases of $\nu=1$
and $\nu=2$. Initial condition is set by 
$\psi_l(0)=\delta_{l,m}$, where $m=1,3,18,20$. 
Which means the initial pulse is given
only at those sites.
As a result, conservation
rule follows the relation,
$\sum_{l=1}^L|\psi_l(t)|^2=4$.

Time evolution is explored
for the span, $0\le t\le 20$, which is plotted along the vertical
axis. The diagram clearly indicates that
probability amplitudes for $l=1,20$,  {\em i. e.}, $|\psi_1(t)|$
and $|\psi_{20}(t)|$ survive with time in (a). 
So the edge states bound to the topological phase
with $\nu=1$ exhibit their quenching. No such quenching
is found for the trivial phase as shown in (b).
Quenching of four edge states, $|\psi_l(t)|$, 
$l=1,3,18,20$ are found in (c) which correspond to the
topological phase with $\nu=2$.
So for the lattice with $L$ sites,
quenching are found for the
amplitude with sites $l=1,3,L\!-\!2,L$. 
It is true that the diagram exhibiting the quenching of edge states
will be different if the initial conditions are made different
from this set. However, this particular choice of
initial conditions is considered from the previous
knowledge of locations of the peaks of
probability density of edge states
as shown in Fig. \ref{Edge-state-probability-for-SSH-2}.
Hence the quench
dynamics provide another route for distinguishing
topological and trivial phases for a system. 

\subsection{Topological phases for $H=H_{vw}+H^{\rm {A\mbox{-}B}}_{z,2}$}
\begin{figure}[h]
\psfrag{A}{\large A}
\psfrag{B}{\large B}
\psfrag{v}{\large $v$}
\psfrag{w}{\large $w$}
\psfrag{z}{\large $z$}
\includegraphics[width=230pt]{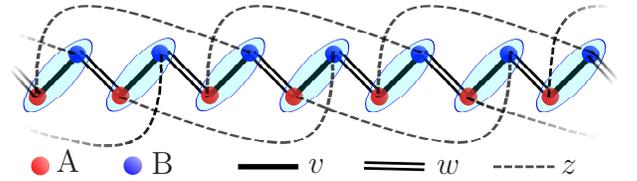}
\caption{Extended SSH model describing the hopping in $H^{\rm {A\mbox{-}B}}_{z,2}$.}
\label{Extended-SSH-3}
\end{figure}

\begin{figure}[h]
\psfrag{x}{\large $w/|v+z|$}
\psfrag{y}{\large k}
\psfrag{ep}{\large $E_+({\rm k})$}
\psfrag{a}{\large (a)}
\psfrag{b}{\large (b)}
\psfrag{v}{\large $v=1$}
\psfrag{za}{\large $z=1/2$}
\psfrag{vbb}{\large $v=1/2$}
\psfrag{zb}{\large $z=1$}
\psfrag{5}{\large 5}
\psfrag{4}{\large 4}
\psfrag{3}{\large 3}
\psfrag{2}{\large 2}
\psfrag{1}{\large 1}
\psfrag{0}{\large 0}
\psfrag{0.0}{\large 0}
\psfrag{-2}{\large $-$2}
\psfrag{-1}{\large $-$1}
\psfrag{-3.1}{\large $-\pi$}
\psfrag{-1.6}{\large $-\pi/2$}
\psfrag{3.1}{\large $\pi$}
\psfrag{1.6}{\large $\pi/2$}
  \includegraphics[width=230pt]{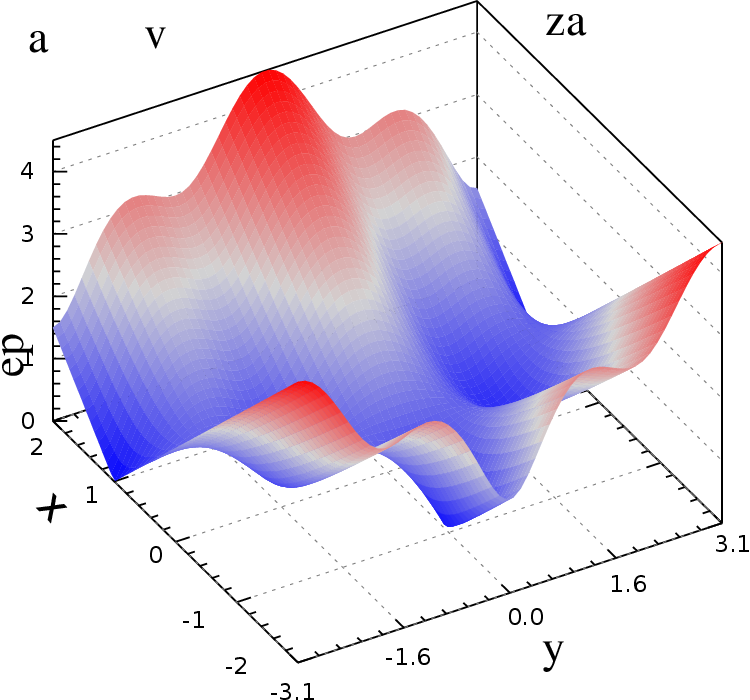}
  \vskip 0.6cm
  \includegraphics[width=230pt]{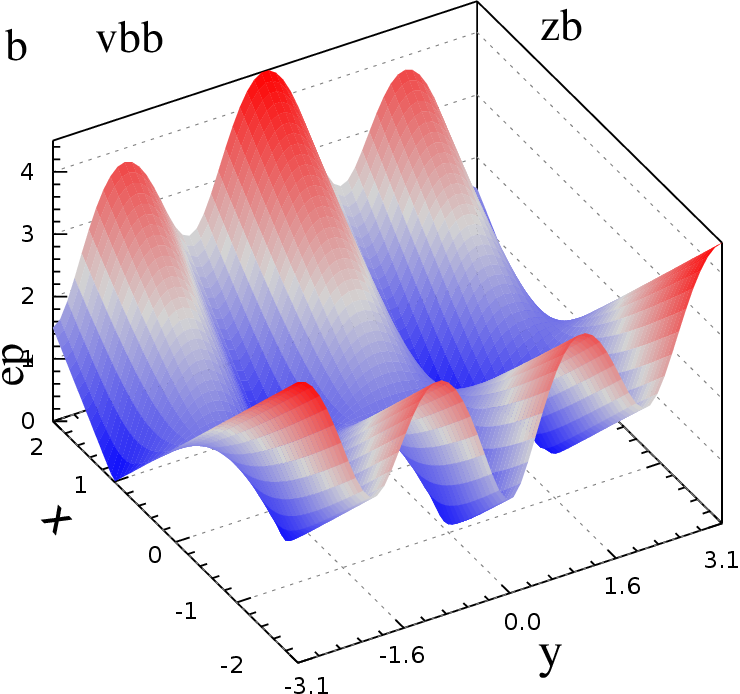}
\caption{Dispersion relation for $H^{\rm {A\mbox{-}B}}_{z,2}$
  when $v/z>1$, (a) and $v/z<1$, (b).}
\label{Dispersion-SSH-3}
\end{figure}

Total Hamiltonian in this case is 
\bea
H&=& H_{vw}+H^{\rm {A\mbox{-}B}}_{z,2},\nonumber\\ [0.4em]
H^{\rm {A\mbox{-}B}}_{z,2}&=&\sum_{j=1}^Nz\,c^\dag_{{\rm A},j}c_{{\rm B},j+2}+{\rm h.c.},\label{HAB-z-2}
\eea
where the hopping term once again extends over
one intermediate primitive cell,
which is shown in Fig. \ref{Extended-SSH-3}.
As a result, 
\[\boldsymbol g(\rm k)\equiv\left\{\begin{array}{l}
g_x=v+w\cos{\!(\rm k)}+z\cos{\!(2\rm k)},\\[0.3em]
g_y=w\sin{\!(\rm k)}-z\sin{\!(2\rm k)},\\[0.3em]
g_z=0.\end{array}\right.\]
Dispersion relation in this case is
$E_\pm({\rm k})=\pm \sqrt{v^2+w^2+z^2+2[vw\cos{\!(\rm k)}
    +vz\cos{\!(2\rm k)}+wz\cos{\!(3\rm k)}]}$.
Variation of dispersion relation, $E_+({\rm k})$,
with $w/|v+z|$ for $v=1$, $z=1/2$ and $v=1/2$, $z=1$ are shown in
Fig. \ref{Dispersion-SSH-3} (a) and (b), respectively.
Those are serving as prototype figures for
$v/z>1$ and $v/z<1$, respectively.
Dispersions comprise of three peaks for any values of
the parameters, $v$, $w$ and $z$. 
Like the previous case, band gap vanishes at ${\rm k}=\pm \pi$,
and ${\rm k}=0$, when $w=|v+z|$, for both the cases $v/z>1$ and $v/z<1$. 
As a result, $\nu$ is undefined again at the point
when $w=|v+z|$. But in the region, $w<|v+z|$, for $v<z$,
the system undergoes an additional phase
transition at the point defined by
the set of equations, $E_\pm({\rm k})=0$, and 
$\frac{dE_\pm({\rm k})}{d{\rm k}}=0$,
which will be discussed later.

Also, in this case, two different topological phases appear
in the parameter space as given below which are separated by
phase transition lines. 
\be
\nu=\left\{\begin{array}{ll}
    1,& w>|z+v|,\\[0.3em]
    0,&w<|z+v|,\;{\rm and}\;v>z,\\[0.3em]
    -2,0&w<|z+v|, \;{\rm and}\;v<z.
  \end{array}\right.
\ee
Topological phase with $\nu=1$ exists as along as
the relation $w>|z+v|$ holds irrespective of individual values of
$v$ and $z$. Another nontrivial phase with
$\nu=-2$ appears in a limited region for $w<|z+v|$
and $v<z$, separated by trivial phase.
The equation of phase transition line 
can be obtained by satisfying the
conditions, $E_\pm({\rm k})=0$, and
$\frac{dE_\pm({\rm k})}{d{\rm k}}=0$.
Anyway, this model hosts the new topological phase with $\nu=-2$.

\begin{figure}[h]
\psfrag{gxx}{\large $g_x$}
\psfrag{gyy}{\large $g_y$}
\psfrag{a}{\large (a)}
\psfrag{b}{\large (b)}
\psfrag{c}{\large (c)}
\psfrag{d}{\large (d)}
\includegraphics[width=230pt]{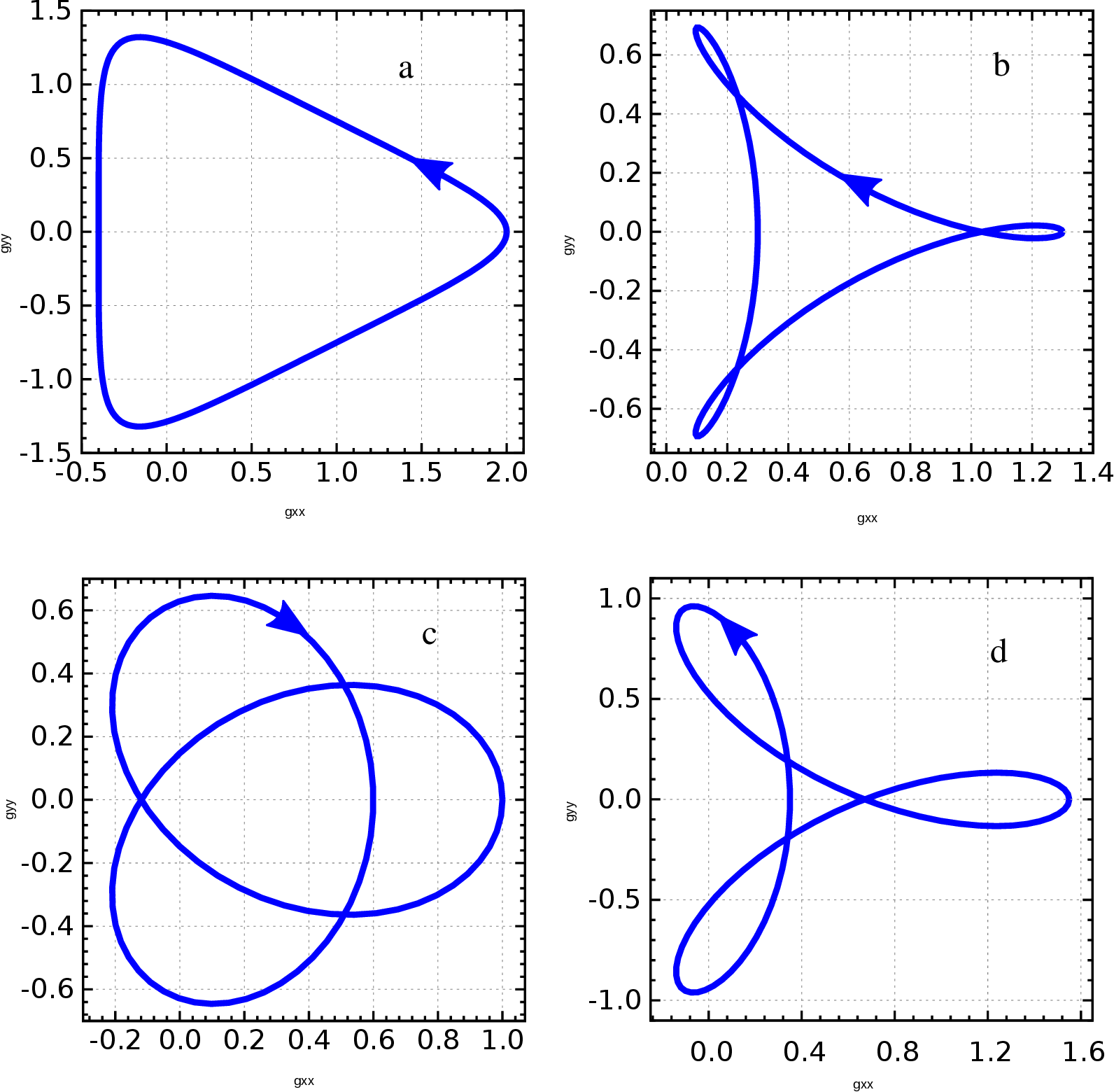}
\caption{Parametric winding diagrams in the
$g_x\mbox{-}g_y$ plane for the Hamiltonian,
  $H^{\rm {A\mbox{-}B}}_{z,2}$. Four figures are drawn for
  (a) $v=0.5$, $w=1.2$, $z=0.3$, (b) $v=0.5$, $w=0.5$, $z=0.3$,
  (c) $v=0.3$, $w=0.2$, $z=0.5$, and (c) $v=0.45$, $w=0.6$, $z=0.5$.}
\label{parametric-windings-SSH-3}
\end{figure}
The parametric plot of winding by the tip of
the vector, $\boldsymbol g(\rm k)$ 
in the $g_x\mbox{-}g_y$ complex plane is shown in
Fig. \ref{parametric-windings-SSH-3}. 
Four figures are drawn for
  (a) $v=0.5$, $w=1.2$, $z=0.3$, (b) $v=0.5$, $w=0.5$, $z=0.3$,
(c) $v=0.3$, $w=0.2$, $z=0.5$, and (d) $v=0.45$, $w=0.6$, $z=0.5$.
Those figures serve as the prototype contours for the
four different regions, 
$w>|v+z|$, for $\nu=1$, $w<|v+z|$, $v>z$
for $\nu=0$, $w<|v+z|$, $v<z$, for $\nu=-2$ and $\nu=0$.
$\boldsymbol g({\rm k})$ traces the closed contour
in counter clockwise direction for (a) and (b) while it is
clockwise for (c) and (d).
Curve encloses the origin once in (a) and
twice in (c) but in opposite direction
which corresponds to winding numbers of opposite sign. 
Nonzero band gap is there for all the cases.

\begin{figure}[h]
  \psfrag{w}{\large $w/|v\!+\!z|$}
  \psfrag{3}{\hskip -0.15 cm 3}
\psfrag{2}{\hskip -0.15 cm 2}
\psfrag{1}{\hskip -0.15 cm 1}
\psfrag{0}{\hskip -0.15 cm 0}
\psfrag{-3}{\hskip -0.15 cm -3}
\psfrag{-2}{\hskip -0.15 cm -2}
\psfrag{-1}{\hskip -0.15 cm -1}
\psfrag{y}{\large k}
\psfrag{E}{\hskip -0.5 cm Energy}
\psfrag{a}{\large (a)}
\psfrag{b}{\large (b)}
\psfrag{v}{\large $v=1$}
\psfrag{za}{\large $z=1/2$}
\psfrag{vb}{\large $v=3/4$}
\psfrag{zb}{\large $z=1$}
\hskip -3.0 cm
\begin{minipage}{0.20\textwidth}
  \includegraphics[width=130pt,angle=-90]{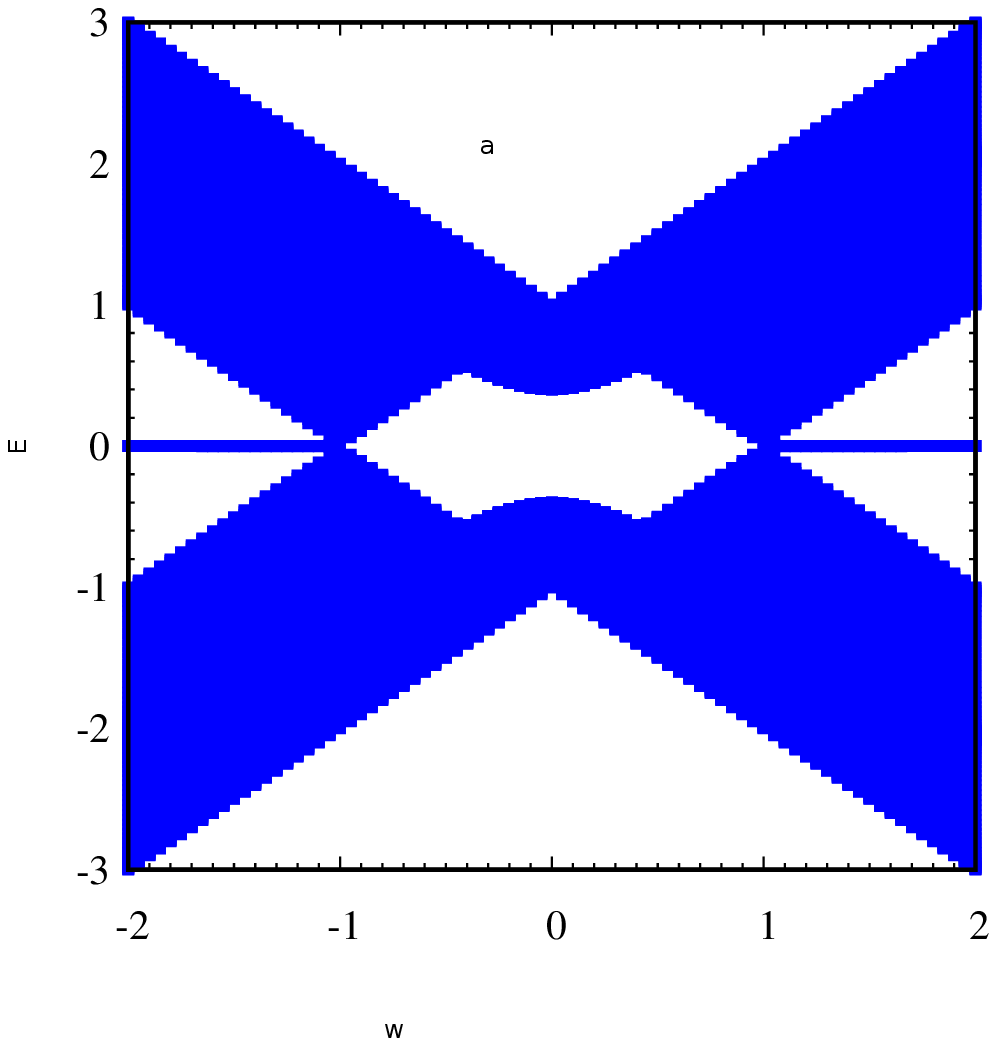}
  \end{minipage}\hskip 0.6cm
  \begin{minipage}{0.2\textwidth}
  \includegraphics[width=130pt,angle=-90]{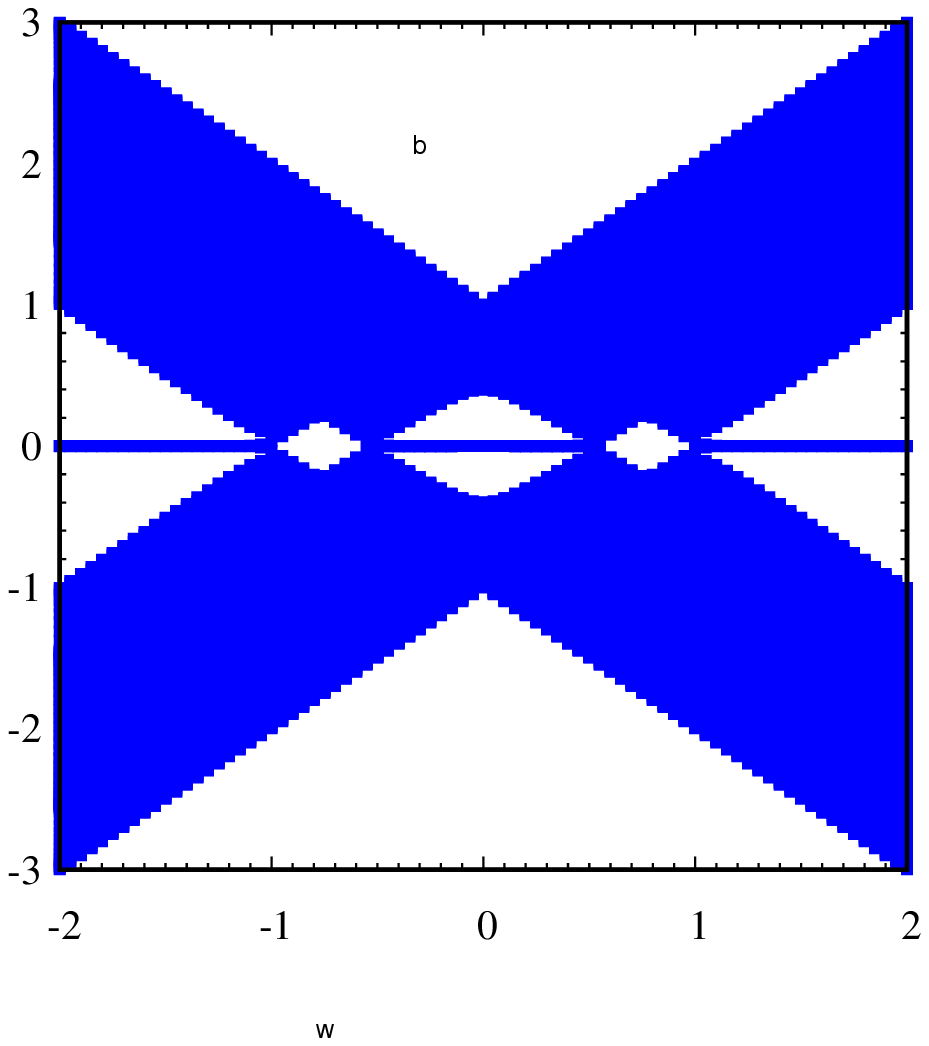}
  \end{minipage}
\caption{Bulk and edge state energies for $H=H_{vw}+H^{\rm {A\mbox{-}B}}_{z,2}$
  when $v/z>1$, (a) for $v=0.7$, $z=0.3$, and $v/z<1$, (b)
  for $v=0.3$, $z=0.7$.}
\label{Edge-states-SSH-3}
\end{figure}

Variation of bulk-edge state energies with respect to
$w/|v+z|$ is shown in Fig. \ref{Edge-states-SSH-3}
for the regime $-2\le (w/|v+z|)\le 2$. 
A single pair of zero energy edge states
is there when $w>|v+z|$ as shown in (a). No edge state is there
in this system when $w<|v+z|$ and and $v>z$. However, 
two pairs of zero energy edge states appear
in a region around the point  $w/|v+z|=0$ 
when $w<|v+z|$ and $v<z$ which is shown in
Fig. \ref{Edge-states-SSH-3} (b). This particular
region is surrounded by a trivial phase as long as
$-1\le (w/|v+z|)\le 1$.
The figures are drawn for lattice of sites 200, and
the results confirm the existence of
edge states in the topological phases.

\begin{figure}[h]
\psfrag{mps}{\hskip -0.1 cm $|\psi|^2$}
\psfrag{a}{ (a)}
\psfrag{b}{ (b)}
\psfrag{n}{sites}
\psfrag{p}{$v=0.25,\,w=2.5,\,z=0.25$}
\psfrag{q}{$v=0.25,\,w=0.25,\,z=2.5$}
\psfrag{150}{150}
\psfrag{125}{125}
\psfrag{100}{100}
\psfrag{75}{75}
\psfrag{50}{50}
\psfrag{25}{25}
\psfrag{0.5}{0.5}
\psfrag{1.0}{1.0}
\psfrag{0} {0}
\psfrag{0.0}{0.0}
\includegraphics[width=230pt]{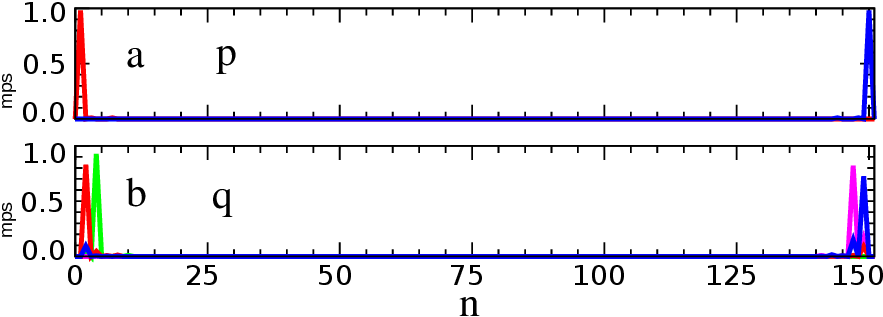}
\caption{Probability density of edge states for $H=H_{vw}+H^{\rm {A\mbox{-}B}}_{z,2}$: 
  (a) for $v=0.25$, $w=2.5$, $z=0.25$, one pair of edge state, 
  (b) for $v=0.25$, $w=0.25$, $z=2.5$, two pairs of edge states.
Figures are drawn for the lattice of 150 sites. }
\label{Edge-state-probability-for-SSH-3}
\end{figure}

To make sure the presence of zero energy edge states,
probability densities of those states are drawn in Fig. 
\ref{Edge-state-probability-for-SSH-3} for the lattice of
150 sites.
Two figures are drawn for two distinct topological phases.
In the upper panel (a), probability densities of two
distinct edge states with $E=0$ are shown when
$v=0.25$, $w=2.5$, $z=0.25$. Those values are
selected for satisfying the 
conditions, $w>|v+z|$. Probability density
of one edge state exhibits sharp peak at site $m=1$
and another one at site $m=150$. This corresponds to
the topological phase with $\nu=1$.
On the other hand, for $w<|v+z|$ 
and $v<z$, probability density of four 
distinct zero energy edge states are shown
in the lower panel (b) when
$v=0.25$, $w=0.25$, $z=2.5$. 
Probability density 
of four orthogonal edge states exhibits 
sharp peak at sites $m=2$, $m=4$, $m=147$,
and $m=149$. In this case zero energy states
close to the left edge are localized on the B sublattice,
while those close to the right edge are localized
on the A sublattice. The difference on localization
with respect to the previous case attributes to the change in the
sign of the winding number, as the new
topological phase of $\nu=-2$, appears
with opposite sign with respect to previous case. 

\begin{figure}[h]
  \psfrag{nu}{\large $\nu$}
  \psfrag{nu0}{\hskip -.3cm \color{blue} \Large $\nu=0$}
  \psfrag{nu-2}{\hskip -.5cm \color{white} \Large $\nu=-2$}
   \psfrag{nu1}{\hskip -.0cm \color{white} \Large $\nu\!=\!1$}
\psfrag{v}{\large $v$}
\psfrag{zv}{\large $v+z=2$}
\psfrag{w}{\hskip -0.4 cm \large $w/|v+z|$}
\psfrag{0.5}{\large 0.5}
\psfrag{0.0}{\large 0.0}
\psfrag{1.0}{\large 1.0}
\psfrag{1.5}{\large 1.5}
\psfrag{2.0}{\large 2.0}
\psfrag{ 1}{\large 1}
\psfrag{ 0}{\large 0}
\psfrag{-1}{\large -1}
\psfrag{-2}{\large -2}
\psfrag{2}{\large 2}
\psfrag{1}{\large 1}
\psfrag{0}{\large 0}
\psfrag{-0.5}{\large -0.5}
\psfrag{-1.0}{\large -1.0}
\psfrag{-1.5}{\large -1.5}
\psfrag{-2.0}{\large -2.0}
\includegraphics[width=230pt,angle=-90]{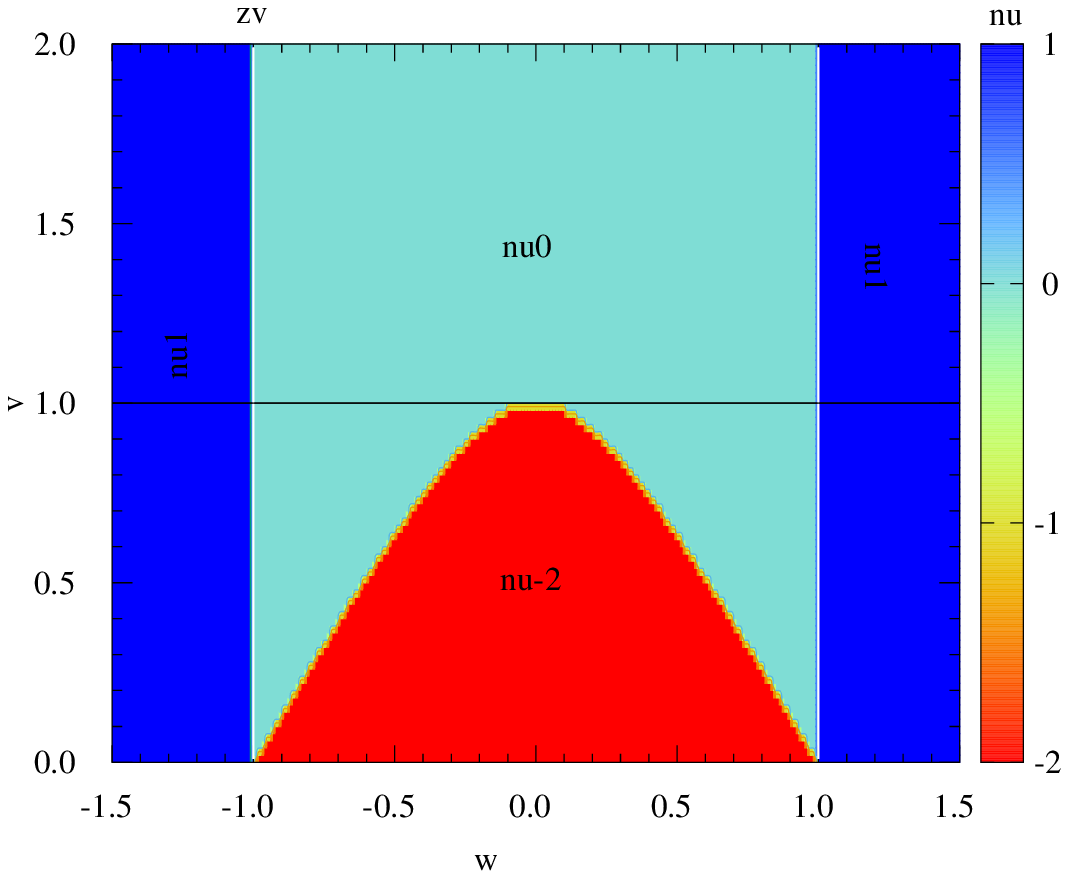}
\caption{Topological phase diagram for the Hamiltonian,
  $H=H_{vw}+H^{\rm {A\mbox{-}B}}_{z,2}$. Two distinct topological
  phases are shown by blue ($\nu=1$) and red  ($\nu=-2$).
  The remaining portion is trivial ($\nu=0$). 
  This diagram is drawn for $v+z=2.$ The horizontal line 
indicates the value $v=1$ or $v/z=1.$}
\label{topological-phase-windings-SSH-3}
\end{figure}

A comprehensive phase diagram for this model is shown in Fig
\ref{topological-phase-windings-SSH-3} where contour plot
for $\nu$ is drawn in the $v$-$w/|v+z|$ space.
Variation of the parameters is made by maintaining
the constraint $v+z=2$.
Existence of two different topological phases,
$\nu=1$ and $-2$, along with the trivial phase, $\nu=0$ are
shown in three different colours. 
The horizontal line is drawn at $v/z=1$,
above which topological phase with $\nu=-2$
does not survive.
This phase exists over the line segment,
$-1 \le w/(v+z)\le +1$, when $v=0$. However the length
of this segment reduces symmetrically around
$w/(v+z)=0$ and vanishes at the point $v=z$.
The boundary lines of these
phases can be obtained by simultaneously solving the Eqs. 
$E_\pm({\rm k})=0$, and $\frac{dE_\pm({\rm k})}{d{\rm k}}=0$.
As a result, the transition lines are given by the
two solutions of quadratic equation, 
$v^2+w^2+z^2+2\{vwp+vz(2p^2-1)+wzp(4p^2-3)\}=0$,  
where $p=\cos^{-1}{\left(\frac{-vz \pm \sqrt{v^2z^2-3w^2z(v-3z)}}{6wz}\right)}$,
along with the constraint, $v+z=2$. 
These curved lines are symmetric
around the straight line $w/(v+z)=0$ and meet
at the point, $w/(v+z)=0$, $v=1$.
Another topological phase with $\nu=1$
appears beyond the two vertical lines
drawn at $w/(v+z)=\pm 1$. They separate
topological phase with $\nu=1$ from the trivial phase. 
So the system undergoes phase transition
around those straight lines. 

As the quenching of edge states provides their exact location 
more clearly, dynamics of the edge states in the presence of
nonlinear terms for the topological phases
of this model will be discussed. 
The set of coupled nonlinear first order differential
equation for finite chain of $L$ sites and for the Hamiltonian
defined in Eq. \ref{HAB-z-2} with OBC is explicitly 
given by 
\bea
i\,\frac{\partial \psi_{2j-1}}{\partial t}&=&v(\psi_{2j}-\psi_{2j-1})+
w(\psi_{2j-2}-\psi_{2j-1})\nonumber \\[0.0 em]
&&+z(\psi_{2j+4}-\psi_{2j-1})-\zeta |\psi_{2j-1}|^2\psi_{2j-1},\nonumber \\[0.2 em]
\vdots\quad\; &=&\qquad \qquad \vdots \label{coupled-equations-2}\\[0.3 em]
i\,\frac{\partial \psi_{2j}}{\partial t}&=&w(\psi_{2j+1}-\psi_{2j})+
v(\psi_{2j-1}-\psi_{2j})\nonumber \\[0.0 em]
&&+z(\psi_{2j-5}-\psi_{2j})-\zeta |\psi_{2j}|^2\psi_{2j}.\nonumber
\eea

\begin{figure*}[t]
\psfrag{a}{\hskip 0.2 cm \large \color{white} (a)}
\psfrag{c}{\hskip -0.0 cm \large \color{white} (c)}
\psfrag{b}{\hskip 0.2 cm \large \color{white}  (b)}
\psfrag{sites}{\hskip -0.1 cm sites}
\psfrag{t}{time}
\hskip -2.2 cm
\begin{minipage}{0.28\textwidth}
  \includegraphics[width=177pt,angle=-90]{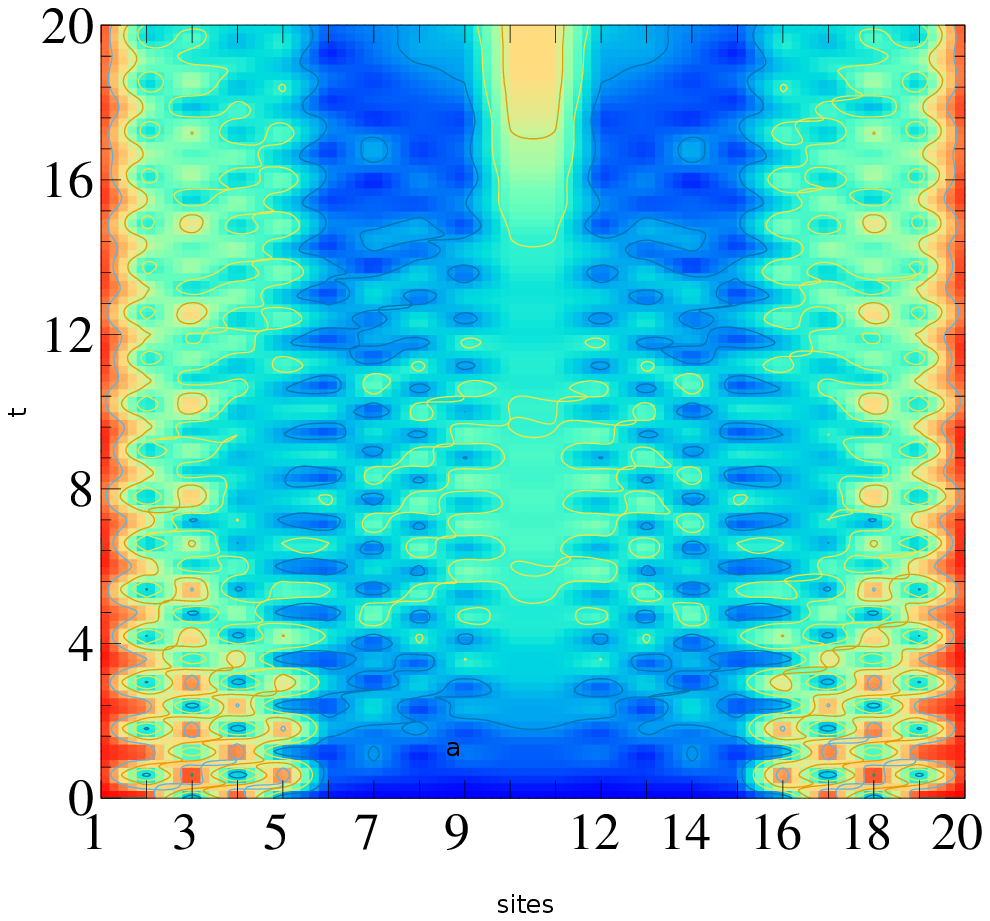}
    \end{minipage}\hskip 0.3cm
  \begin{minipage}{0.28\textwidth}
  \includegraphics[width=177pt,angle=-90]{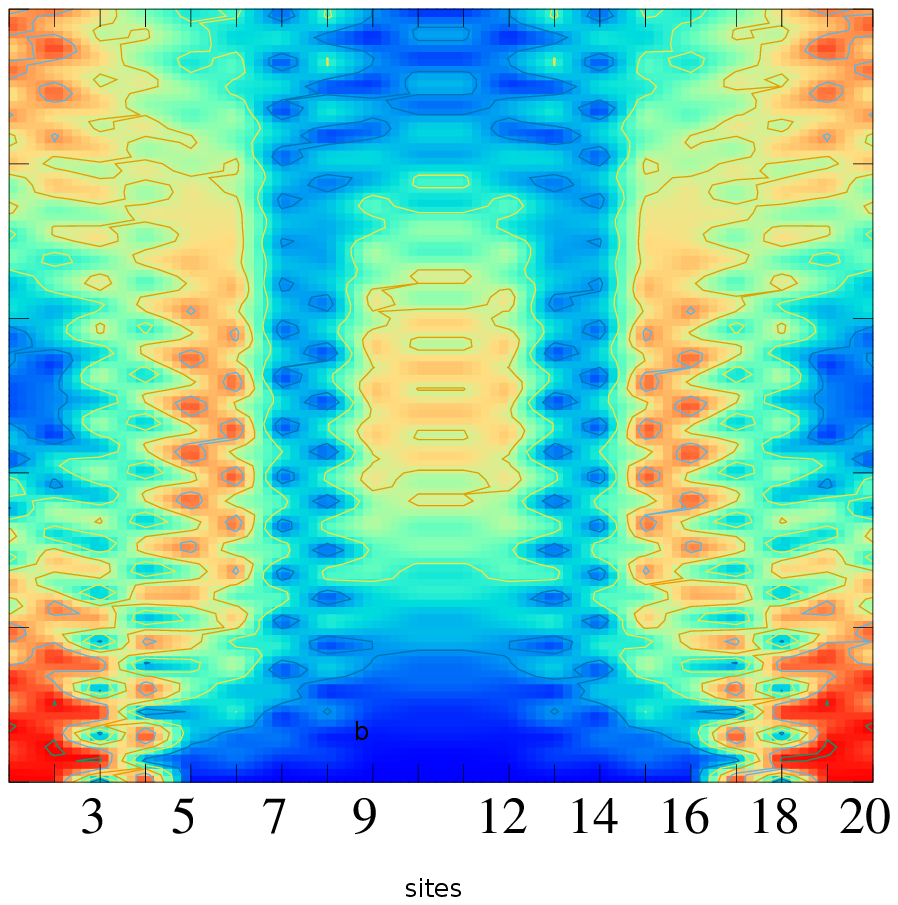}
  \end{minipage}\hskip 0.3cm
  \begin{minipage}{0.28\textwidth}
  \includegraphics[width=180pt,angle=-90]{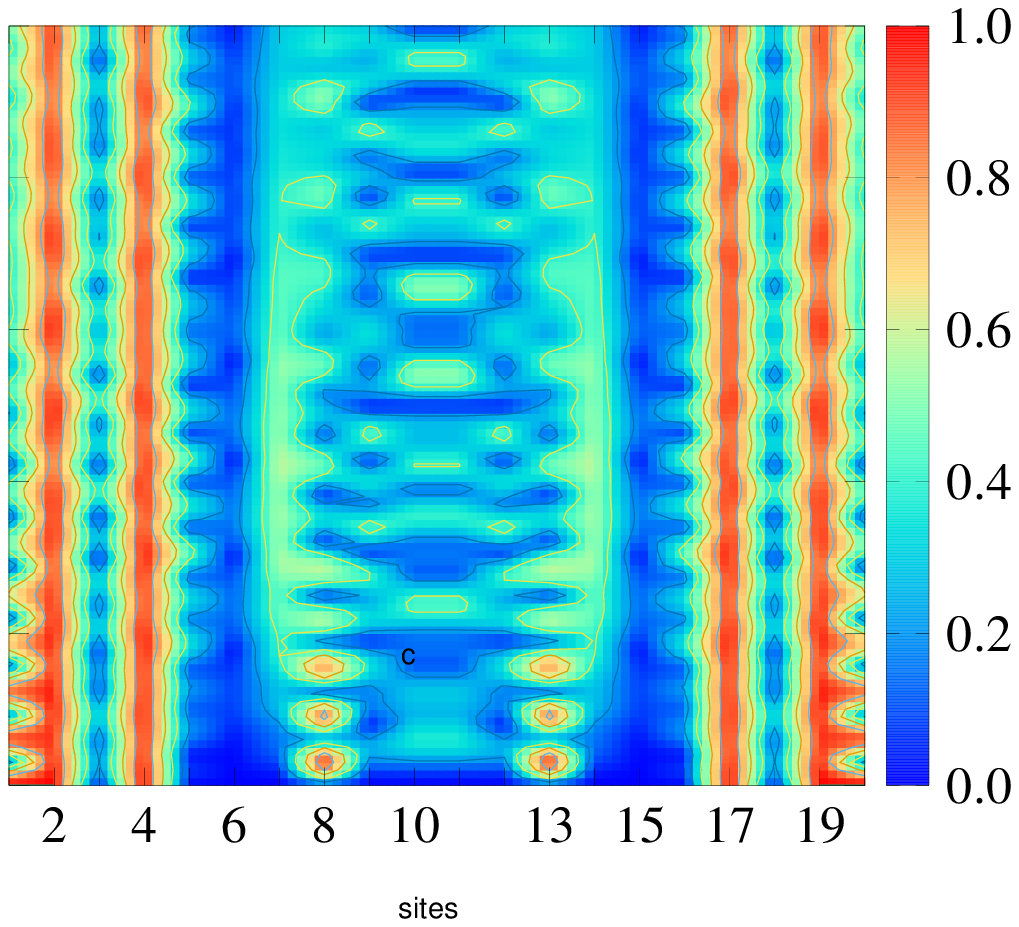}
  \end{minipage}
\caption{Quench dynamics for $H=H_{vw}+H^{\rm {A\mbox{-}B}}_{z,2}$
  when $\zeta=0.5$, (a) for $v=0.25$, $w=2.5$, $z=0.25$, 
   (b) for $v=2.5$, $w=0.25$, $z=0.25$,
  (c) for $v=0.25$, $w=0.25$, $z=2.5$.
Figures are drawn for lattice with 20 sites. }
\label{Quench-dynamics-SSH-3}
\end{figure*}

Quenching of edge states for the
nonlinear system is shown in 
Fig. \ref{Quench-dynamics-SSH-3}, 
by solving the set of Eq. \ref{coupled-equations-2},
for $L=20$, when $\zeta=0.5$.
Contour plot for the time evolution of
$|\psi_l(t)|$, is drawn for every site 
which is shown along the horizontal axis. 
Three contour plots are shown 
(a) for $v=0.25$, $w=2.5$, $z=0.25$, 
   (b) for $v=2.5$, $w=0.25$, $z=0.25$,
(c) for $v=0.25$, $w=0.25$, $z=2.5$,
where (b) indicates trivial phase as before while
(a) and (c) for the topological phases of $\nu=1$
and $\nu=-2$, respectively. In this case,
initial condition is set by 
$\psi_l(0)=\delta_{l,m}$, where $m=2,4,17,19$. 
As a result, conservation
rule follows the same equation as before,
$\sum_{l=1}^L|\psi_l(t)|^2=4$.

Evolution of the system is explored
for the time span, $0\le t\le 20$, where
the time is plotted along the vertical
axis. The diagram in (a) clearly indicates that
probability amplitudes for $l=1,20$,  {\em i. e.}, $|\psi_1(t)|$
and $|\psi_{20}(t)|$ survive with time. 
So the edge states bound to the topological phase
with $\nu=1$ exhibit their quenching. No such quenching
is found for the trivial phase as shown in (b).
Those results are similar to the previous case, although
the respective figures are qualitatively different. 
Quenching of four edge states, $|\psi_l(t)|$, when 
$l=2,4,17,19$ are found in (c) which correspond to the
topological phase with $\nu=-2$. In contrast to
this result, quenching of four edge states,
for $l=1,3,18,20$ are found when $\nu=2$,
as discussed in the previous model.
It means quenching over A and B sublattices
interchange their edges with the change in sign of $\nu$. 
Thus, quenching are found for the
amplitude on sites, $l=2,4,L\!-\!3,L\!-\!1$, 
for any arbitrary length of lattice. 

\subsection{Topological phases for $H=H_{vw}+H^{\rm {B\mbox{-}A}}_{z,3}$}
\begin{figure}[h]
\psfrag{A}{\large A}
\psfrag{B}{\large B}
\psfrag{v}{\large $v$}
\psfrag{w}{\large $w$}
\psfrag{z}{\large $z$}
\includegraphics[width=230pt]{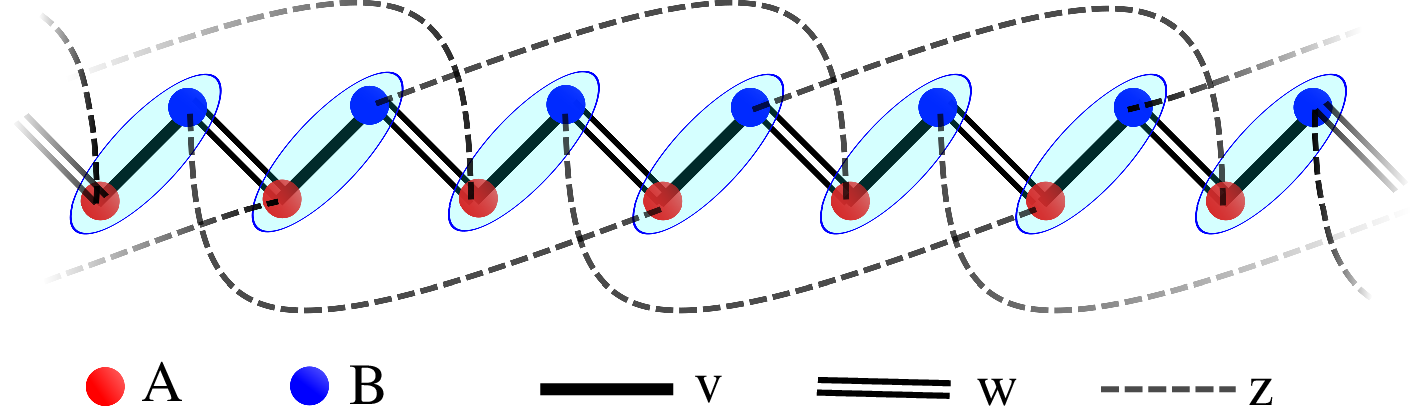}
\caption{Extended SSH model describing the hopping in $H^{\rm {B\mbox{-}A}}_{z,3}$.}
\label{Extended-SSH-4}
\end{figure}

\begin{figure}[h]
\psfrag{x}{\large $v/|w+z|$}
\psfrag{y}{\large k}
\psfrag{ep}{\large $E_+({\rm k})$}
\psfrag{a}{\large (a)}
\psfrag{b}{\large (b)}
\psfrag{v}{\large $w=1$}
\psfrag{za}{\large $z=1/2$}
\psfrag{vbb}{\large $w=1/2$}
\psfrag{zb}{\large $z=1$}
\psfrag{5}{\large 5}
\psfrag{4}{\large 4}
\psfrag{3}{\large 3}
\psfrag{2}{\large 2}
\psfrag{1}{\large 1}
\psfrag{0}{\large 0}
\psfrag{0.0}{\large 0}
\psfrag{-2}{\large $-$2}
\psfrag{-1}{\large $-$1}
\psfrag{-3.1}{\large $-\pi$}
\psfrag{-1.6}{\large $-\pi/2$}
\psfrag{3.1}{\large $\pi$}
\psfrag{1.6}{\large $\pi/2$}
  \includegraphics[width=230pt]{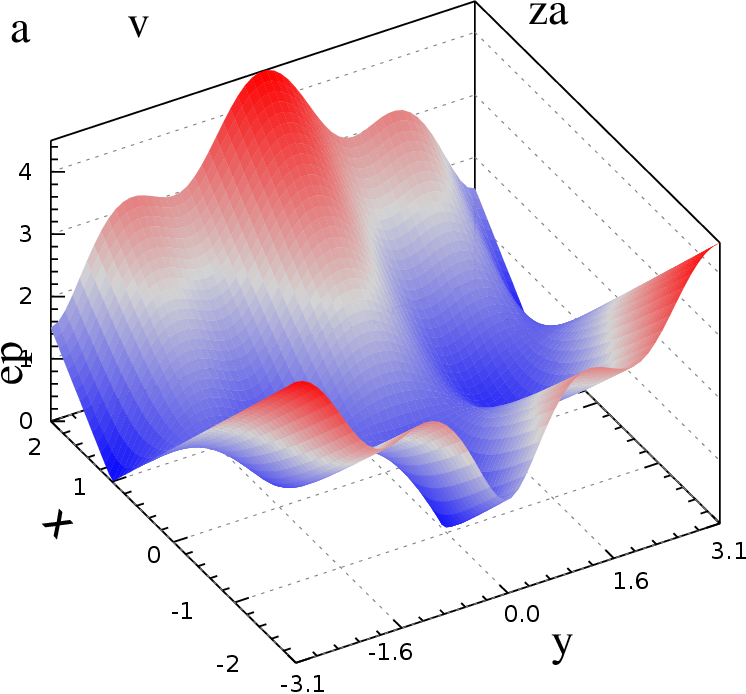}
  \vskip 0.6cm
  \includegraphics[width=230pt]{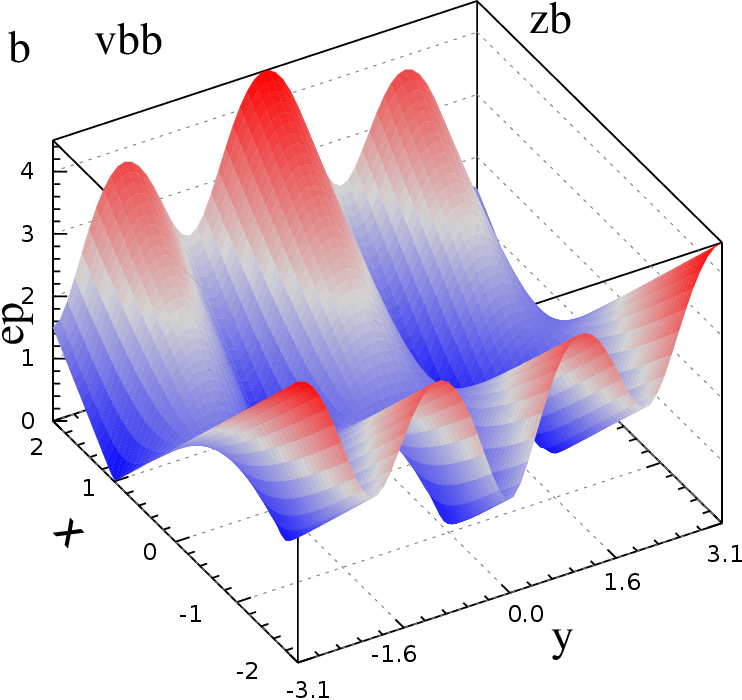}
\caption{Dispersion relation for $H^{\rm {B\mbox{-}A}}_{z,3}$
  when $w/z>1$, (a) and $w/z<1$, (b).}
\label{Dispersion-SSH-4}
\end{figure}

Now the total Hamiltonian is 
\bea
H&=& H_{vw}+H^{\rm {B\mbox{-}A}}_{z,3},\nonumber\\ [0.4em]
H^{\rm {B\mbox{-}A}}_{z,3}&=&\sum_{j=1}^Nz\,c^\dag_{{\rm B},j}c_{{\rm A},j+3}+{\rm h.c.},
\label{HBA-z-3}
\eea
where the hopping term extends over two intermediate primitive cell,
which is shown in Fig. \ref{Extended-SSH-4}.
In this model every cell is connected to the third NN
cell by the hopping parameter $z$. The
$\boldsymbol g(\rm k)$ vector assumes the form, 
\[\boldsymbol g(\rm k)\equiv\left\{\begin{array}{l}
g_x=v+w\cos{\!(\rm k)}+z\cos{\!(3\rm k)},\\[0.3em]
g_y=w\sin{\!(\rm k)}+z\sin{\!(3\rm k)},\\[0.3em]
g_z=0.
\end{array}\right.\]
Dispersion relation in this case is
$E_\pm({\rm k})=\pm \sqrt{v^2+w^2+z^2+2[vw\cos{\!(\rm k)}
    +vz\cos{\!(3\rm k)}+wz\cos{\!(2\rm k)}]}$.
  Variation of dispersion relation, $E_+({\rm k})$,
with $v/|w+z|$ for $w=1$, $z=1/2$ and $w=1/2$, $z=1$ are shown in
Fig. \ref{Dispersion-SSH-4} (a) and (b), respectively.
Those are serving as prototype figures for
$w>z$ and $w<z$, respectively.
Dispersions comprise of three broad peaks when
$v/|w+z|\le 1$, for both the cases $w/z>1$ and $w/z<1$. 
Band gap vanishes at the BZ boundaries, ${\rm k}=\pm \pi$,
and ${\rm k}=0$ when $v=|w+z|$. 
As a result, $\nu$ is undefined at the point
when $v=|w+z|$. The dispersions plotted in Figs.
\ref{Dispersion-SSH-3} and \ref{Dispersion-SSH-4}
look alike although they are different in a sense that
they are plotted with respect to different parameters,
say, $w/|v+z|$ in Fig. \ref{Dispersion-SSH-3} and
$v/|w+z|$ in Fig. \ref{Dispersion-SSH-4}.
This similarity attributes to the fact that dispersions for
the Hamiltonians in Eqs. \ref{HAB-z-2} and \ref{HBA-z-3}
are interchangeable upon interchange of $v$ and $w$. 

In this case also two different types of topological phases
with $\nu=1$ and 3 appear
in the parameter space as given below and they are separated by
phase transition lines. 
\be
\nu=\left\{\begin{array}{ll}
    0,&  v>|w+z|,\\[0.3em]
    1,& v<|w+z|,\;{\rm and}\;w>z,\\[0.3em]
    3,1,&v<|w+z|, \;{\rm and}\;w<z.
  \end{array}\right.
\ee
The system is trivial as long as $v>|w+z|$,
irrespective of the values of $w$ and $z$. 
Topological phase with $\nu=1$ exists when
the relations $v<|w+z|$ and $w>z$ do hold.
Another nontrivial phase with
$\nu=3$ appears in a limited region for $v<|w+z|$
and $w<z$, separated by the topological phase with $\nu=1$.
It means the phase with  $\nu=1$ emerges
for $v<|w+z|$ for both $w>z$ and $w<z$. 
The equation of phase transition lines 
can be obtained by satisfying the
conditions, $E_\pm({\rm k})=0$,
and $\frac{dE_\pm({\rm k})}{d{\rm k}}=0$. 
So, this model hosts the new topological phase with $\nu=3$.

\begin{figure}[h]
\psfrag{gxx}{\large $g_x$}
\psfrag{gyy}{\large $g_y$}
\psfrag{a}{\large (a)}
\psfrag{b}{\large (b)}
\psfrag{c}{\large (c)}
\psfrag{d}{\large (d)}
\includegraphics[width=230pt]{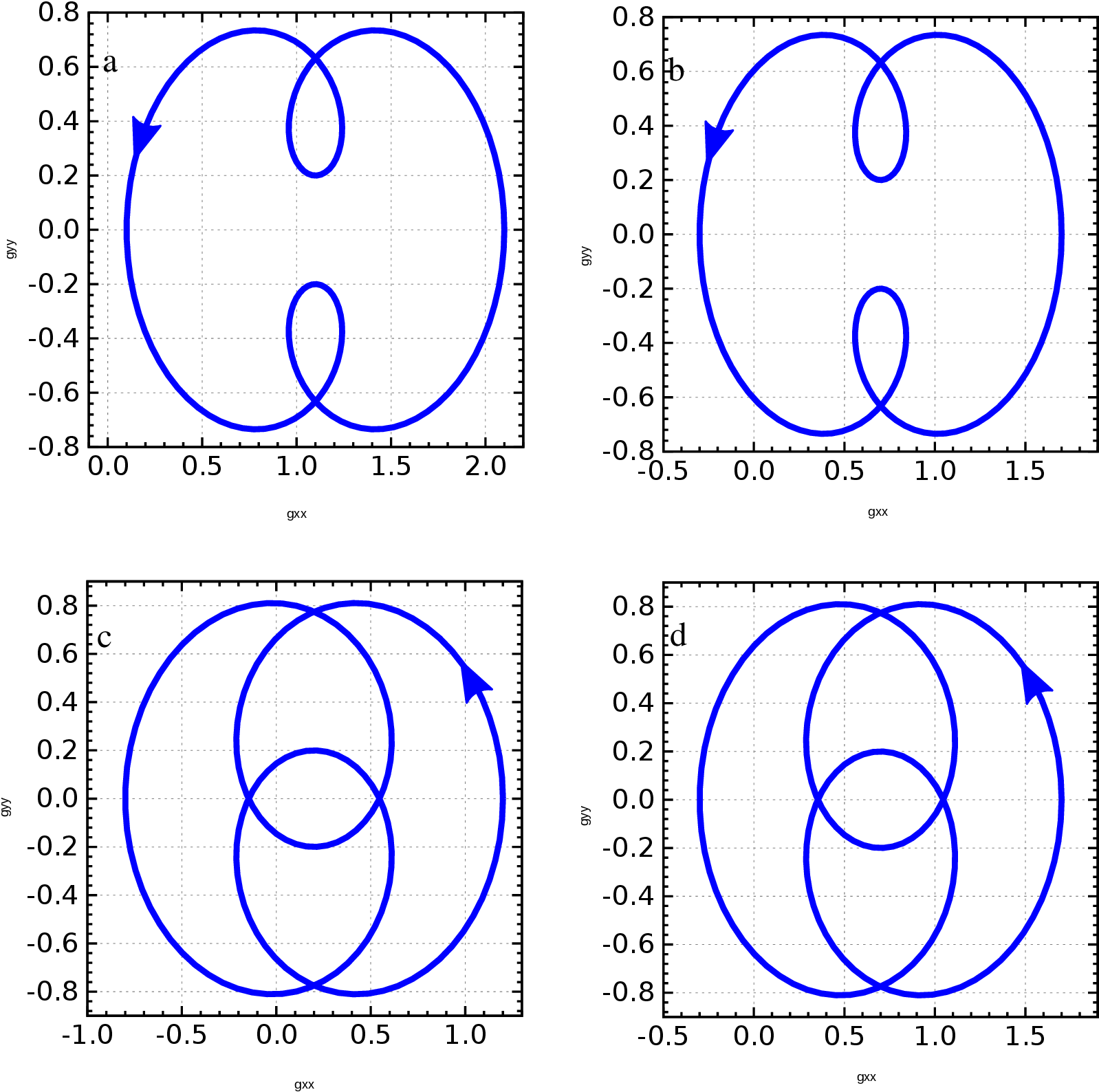}
\caption{Parametric winding diagrams in the
$g_x\mbox{-}g_y$ plane for the Hamiltonian,
  $H^{\rm {B\mbox{-}A}}_{z,3}$. Four figures are drawn for
  (a) $v=1.1$, $w=0.6$, $z=0.4$, (b) $v=0.7$, $w=0.6$, $z=0.4$,
  (c) $v=0.2$, $w=0.4$, $z=0.6$, and (c) $v=0.7$, $w=0.4$, $z=0.6$.}
\label{parametric-windings-SSH-4}
\end{figure}

The parametric plot of winding by the tip of
the vector, $\boldsymbol g(\rm k)$ 
in the $g_x\mbox{-}g_y$ complex plane are shown in
Fig. \ref{parametric-windings-SSH-4}. 
Four figures are drawn for
  (a) $v=1.1$, $w=0.6$, $z=0.4$, (b) $v=0.7$, $w=0.6$, $z=0.4$,
(c) $v=0.2$, $w=0.4$, $z=0.6$, and (d) $v=0.7$, $w=0.4$, $z=0.6$.
All the curves traverse in the counter clockwise direction,
as a result of which, all the winding numbers are positive. 
Those figures serve as the prototype contours for the
four different regions, 
$v>|w+z|$, for $\nu=0$, $v<|w+z|$, $w>z$
for $\nu=1$, $v<|w+z|$, $w<z$, for $\nu=3$ and $\nu=1$.
$\boldsymbol g({\rm k})$. 
Nonzero band gap is there for all the cases.

\begin{figure}[h]
  \psfrag{v}{\large $v/|w\!+\!z|$}
  \psfrag{3}{\hskip -0.15 cm 3}
\psfrag{2}{\hskip -0.15 cm 2}
\psfrag{1}{\hskip -0.15 cm 1}
\psfrag{0}{\hskip -0.15 cm 0}
\psfrag{-3}{\hskip -0.15 cm -3}
\psfrag{-2}{\hskip -0.15 cm -2}
\psfrag{-1}{\hskip -0.15 cm -1}
\psfrag{y}{\large k}
\psfrag{E}{\hskip -0.5 cm Energy}
\psfrag{a}{\large (a)}
\psfrag{b}{\large (b)}
\hskip -3.0 cm
\begin{minipage}{0.20\textwidth}
  \includegraphics[width=130pt,angle=-90]{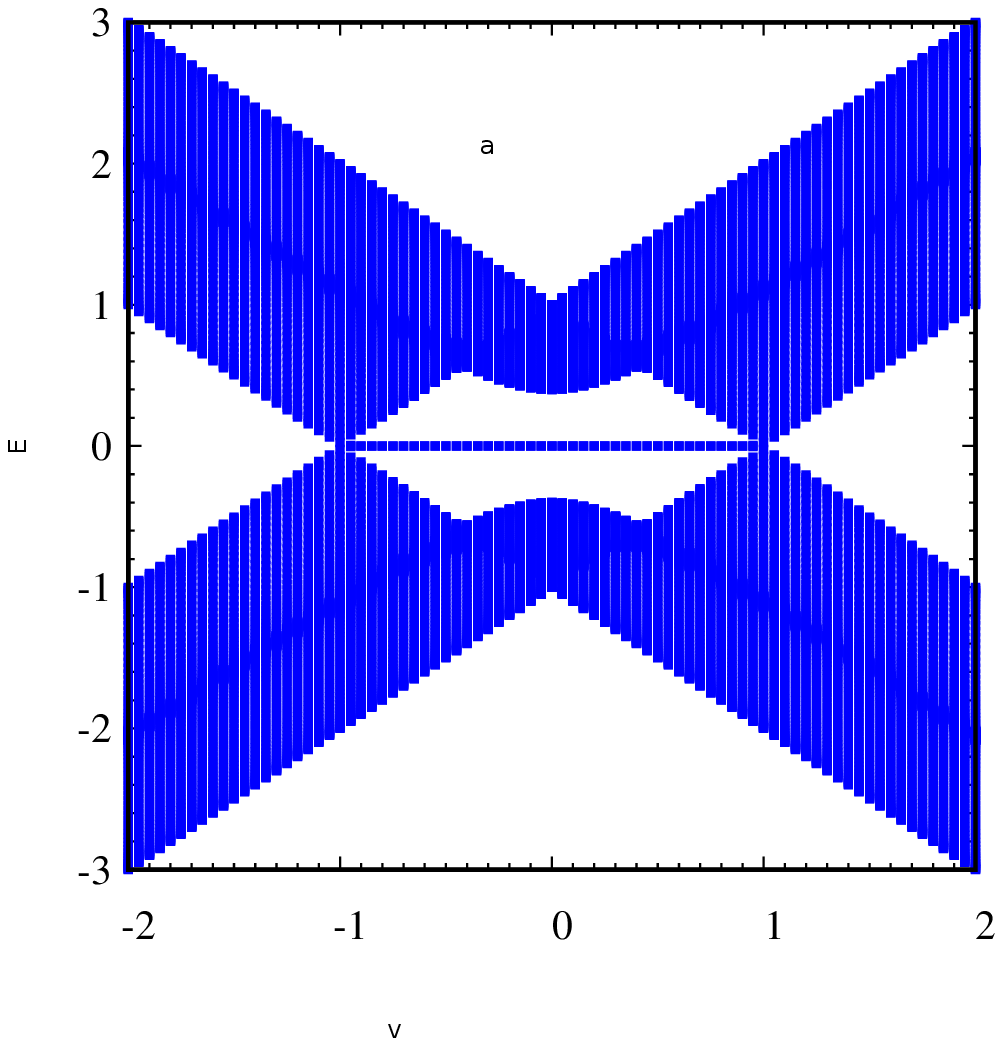}
  \end{minipage}\hskip 0.6cm
  \begin{minipage}{0.2\textwidth}
  \includegraphics[width=130pt,angle=-90]{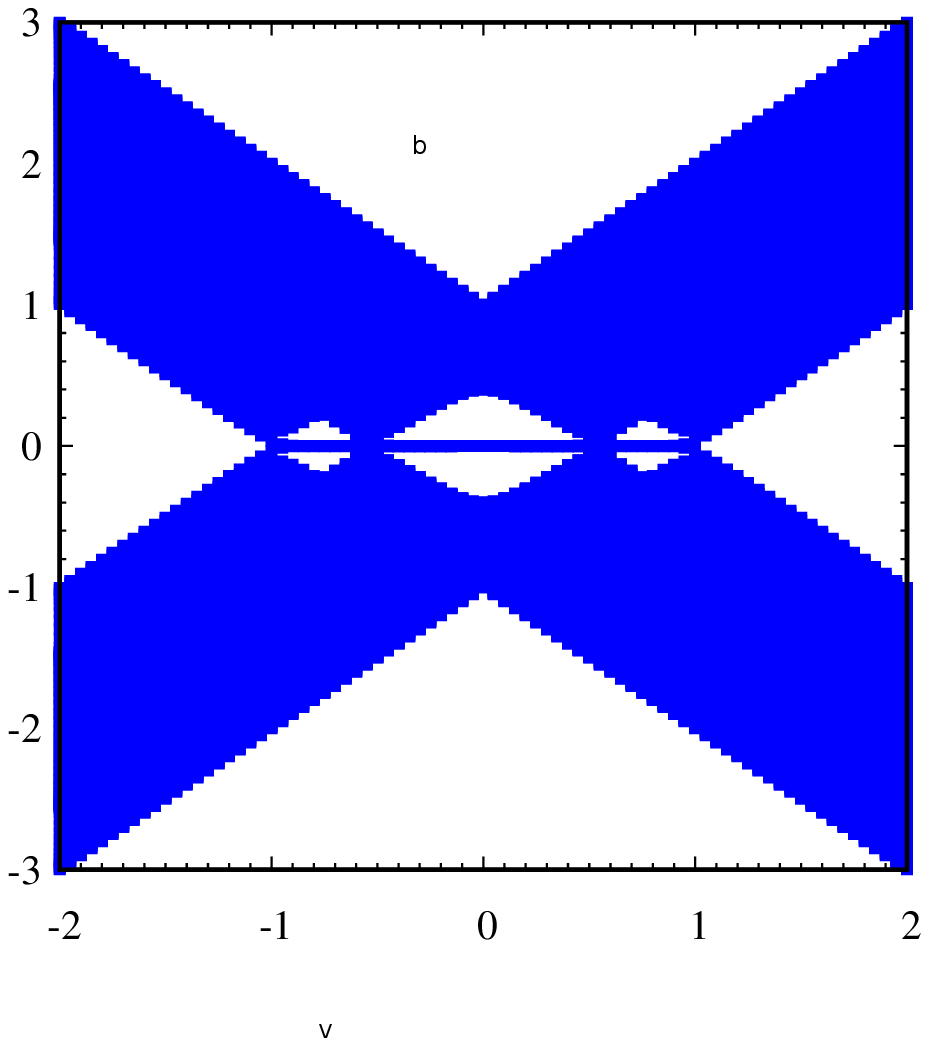}
  \end{minipage}
\caption{Edge states for $H=H_{vw}+H^{\rm {B\mbox{-}A}}_{z,3}$
  when $w/z>1$, (a) for $w=0.7$, $z=0.3$, and $w/z<1$, (b)
  for $w=0.3$, $z=0.7$.}
\label{Edge-states-SSH-4}
\end{figure}

Variation of bulk-edge state energies with respect to
$v/|w+z|$ is shown in Fig. \ref{Edge-states-SSH-4}
for the regime $-2\le (v/|w+z|)\le 2$. 
No zero energy edge states
is there when $v>|w+z|$ as shown in (a). Single pair of edge state is there
in this system when $v<|w+z|$ and and $w>z$. However, 
three pairs of zero energy edge states appear
in a region around the point  $v/|w+z|=0$ 
when $v<|w+z|$ and $w<z$ which is shown in
Fig. \ref{Edge-states-SSH-4} (b). This particular
region is surrounded by a single pair of edge states as long as
$-1\le (v/|w+z|)\le 1$.
The figures are drawn for lattice of sites 200, and
the results confirm the existence of
edge states in the topological phases.

\begin{figure}[h]
\psfrag{mps}{\hskip -0.1 cm  $|\psi|^2$}
\psfrag{a}{ (a)}
\psfrag{b}{ (b)}
\psfrag{n}{sites}
\psfrag{p}{$v=0.25,\,w=2.5,\,z=0.25$}
\psfrag{q}{$v=0.25,\,w=0.25,\,z=2.5$}
\psfrag{200}{200}
\psfrag{180}{180}
\psfrag{160}{160}
\psfrag{140}{140}
\psfrag{120}{120}
\psfrag{100}{100}
\psfrag{80}{80}
\psfrag{60}{60}
\psfrag{40}{40}
\psfrag{20}{20}
\psfrag{0.5}{0.5}
\psfrag{1.0}{1.0}
\psfrag{0} {0}
\psfrag{0.0}{0.0}
\includegraphics[width=230pt]{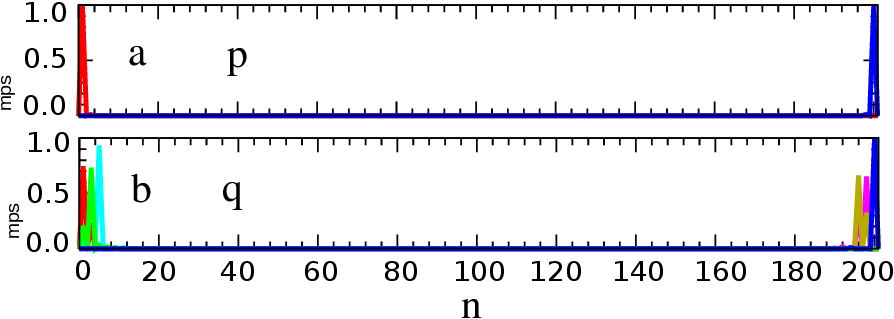}
\caption{Probability density of edge states for $H=H_{vw}+H^{\rm {B\mbox{-}A}}_{z,3}$: 
  (a) for $v=0.25$, $w=2.5$, $z=0.25$, one pair of edge states, 
  (b) for $v=0.25$, $w=0.25$, $z=2.5$, three pairs of edge states.
Figures are drawn for 200 sites. }
\label{Edge-state-probability-for-SSH-4}
\end{figure}

In order to confirm the existence of zero energy edge states,
probability densities of those states are drawn in Fig. 
\ref{Edge-state-probability-for-SSH-4} for the lattice of
200 sites.
Two figures are drawn for two distinct topological phases.
In the upper panel (a), probability densities of two
distinct edge states with $E=0$ are shown when
$v=0.25$, $w=2.5$, $z=0.25$. Those values are
selected for satisfying the 
conditions, $v<|w+z|$ and $w>z$. Probability density
of one edge state exhibits sharp peak at site $m=1$
and another one at site $m=200$. This corresponds to
the topological phase with $\nu=1$.
On the other hand, for $v<|w+z|$ 
and $w<z$, probability density of four 
distinct zero energy edge states are shown
in the lower panel (b) when
$v=0.25$, $w=0.25$, $z=2.5$. 
Probability density 
of six orthonormal edge states exhibit
sharp peak at sites $m=1,3,5,196,198$ and 200. 
This result is in accordance to
the topological phase of $\nu=3$.
Here localization of zero energy states are found on
A sublattice near left edge and B sublattice near right edge. 
\begin{figure}[h]
  \psfrag{nu}{\large $\nu$}
   \psfrag{nu0}{\hskip -.0cm \color{white} \Large $\nu\!=\!0$}
   \psfrag{nu3}{\hskip -.5cm \color{white} \Large $\nu=3$}
   \psfrag{nu-1}{\hskip -.0cm \color{white} \Large $\nu=-1$}
   \psfrag{nu1}{\hskip -.30cm \color{blue} \Large $\nu=1$}
\psfrag{w}{\large $w$}
\psfrag{zw}{\large $w+z=2$}
\psfrag{v}{\hskip -0.4 cm \large $v/|w+z|$}
\psfrag{0.5}{\large 0.5}
\psfrag{0.0}{\large 0.0}
\psfrag{1.0}{\large 1.0}
\psfrag{ 1}{\large 1}
\psfrag{ 0}{\large 0}
\psfrag{-1}{\large -1}
\psfrag{ 2}{\large 2}
\psfrag{ 3}{\large 3}
\psfrag{1.5}{\large 1.5}
\psfrag{2.0}{\large 2.0}
\psfrag{2.5}{\large 2.5}
\psfrag{3.0}{\large 3.0}
\psfrag{-0.5}{\large -0.5}
\psfrag{-1.0}{\large -1.0}
\psfrag{-1.5}{\large -1.5}
\includegraphics[width=230pt,angle=-90]{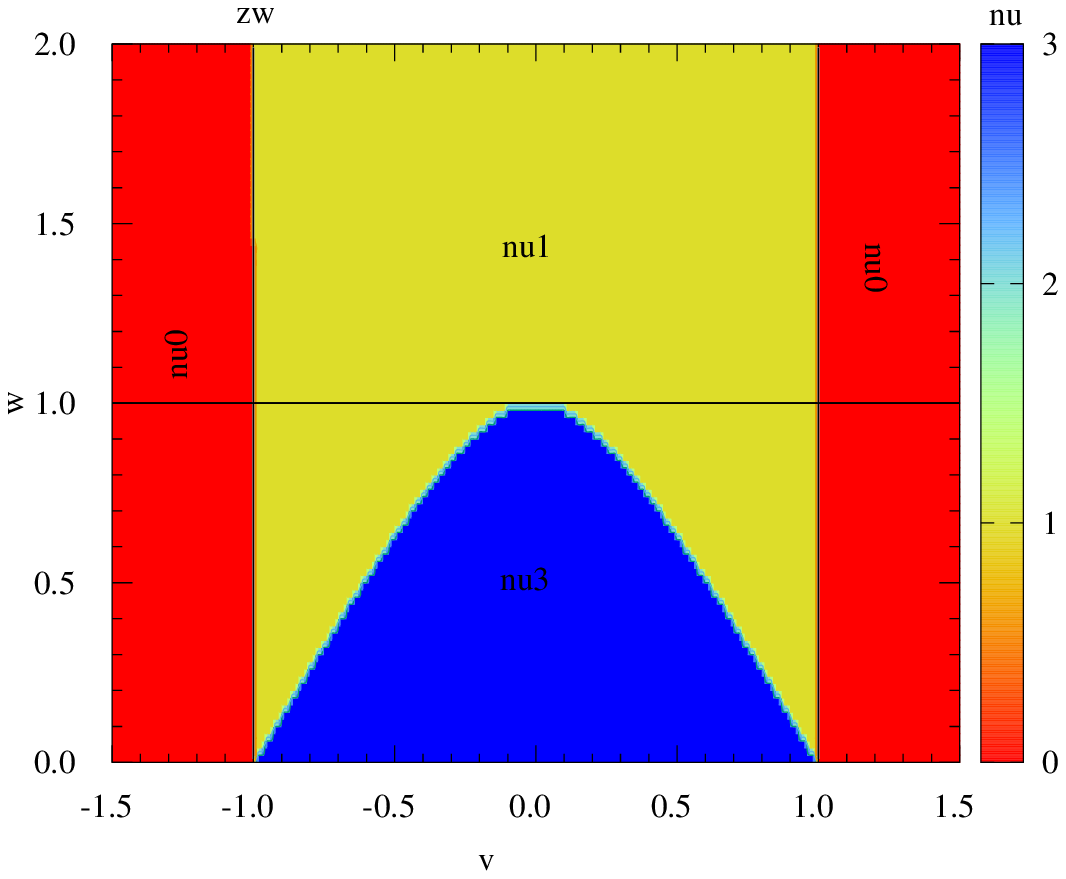}
\caption{Topological phase diagram for the Hamiltonian,
  $H^{\rm {B\mbox{-}A}}_{z,3}$. Trivial phase is shown by red
  ($\nu=0$)  while two distinct topological
  phases are shown by yellow ($\nu=1$) and blue  ($\nu=3$).
  This diagram is drawn for $w+z=2.$
The horizontal line 
indicates the value $w=1$ or $w/z=1.$}
\label{topological-phase-windings-SSH-4}
\end{figure}

A rigorous phase diagram for this model is shown in Fig
\ref{topological-phase-windings-SSH-4} where contour plot
for $\nu$ is drawn in the $w$-$v/|w+z|$ space.
Variation of the parameters is made by maintaining
the constraint $w+z=2$.
Existence of two different topological phases,
$\nu=1$ and $3$ along with the trivial phase, $\nu=0$ are
shown in yellow, blue and red. 
The horizontal line is drawn at $w/z=1$,
above which topological phase with $\nu=3$
does not survive.
This phase exists over the line segment,
$-1 \le v/(w+z)\le +1$, when $w=0$. However the length
of this segment reduces symmetrically around
$v/(w+z)=0$ and vanishes at the point $w=z$.
The boundary lines of separation of those
phases can be obtained as before by solving the Eqs. 
$E_\pm({\rm k})=0$, and  $\frac{dE_\pm({\rm k})}{d{\rm k}}=0$.
Combination of those two equations leads to a quadratic
equation,   
$v^2+w^2+z^2+2\{vwp+wz(2p^2-1)+vzp(4p^2-3)\}=0$,
where $p=\cos^{-1}{\left(\frac{-wz \pm \sqrt{w^2z^2-3v^2z(w-3z)}}{6vz}\right)}$.
Two solutions of this equation along with the
constraint, $w+z=2$, yield the equation of phase transition lines. 
Those curved lines are symmetric
around the straight line $v/(w+z)=0$.

Trivial phase ($\nu=0$) 
appears beyond the two vertical lines
drawn at $v/(w+z)=\pm 1$. They separate
topological phases with $\nu=1$ and 3 from the trivial phase. 
So the system undergoes phase transition
around those straight lines.
The structure of this phase diagram looks similar to
that shown in Fig. \ref{topological-phase-windings-SSH-3}.
However, a closer scrutiny will reveal that
the positions of topological phases are different.
At the same time parameters plotted along the two
axes are also different.
Topological phase with $\nu=-2$ is replaced by that 
of $\nu=3$ and the trivial phase ($\nu=0$)
and another topological phase with $\nu=1$
interchange their positions.

According to the formalism for quenching
of edge states as discussed before,
dynamics of the edge states in the presence of
the same nonlinear terms for the topological phases
of this model has been studied.  
The set of coupled nonlinear 
equation for chain of $L/2$ unit cells and for the Hamiltonian
defined in Eq. \ref{HBA-z-3} with OBC is 
given by 
\bea
i\,\frac{\partial \psi_{2j-1}}{\partial t}&=&v(\psi_{2j}-\psi_{2j-1})+
w(\psi_{2j-2}-\psi_{2j-1})\nonumber \\[0.0 em]
&&+z(\psi_{2j-6}-\psi_{2j-1})-\zeta |\psi_{2j-1}|^2\psi_{2j-1},\nonumber \\[0.2 em]
\vdots\quad\; &=&\qquad \qquad \vdots \label{coupled-equations-3}\\[0.3 em]
i\,\frac{\partial \psi_{2j}}{\partial t}&=&w(\psi_{2j+1}-\psi_{2j})+
v(\psi_{2j-1}-\psi_{2j})\nonumber \\[0.0 em]
&&+z(\psi_{2j+5}-\psi_{2j})-\zeta |\psi_{2j}|^2\psi_{2j}.\nonumber
\eea
\begin{figure*}[t]
 \psfrag{a}{\hskip 0.2 cm \large \color{white} (a)}
\psfrag{c}{\hskip 0.0 cm \large \color{white} (c)}
\psfrag{b}{\hskip 0.2 cm \large \color{white}  (b)}
\psfrag{sites}{\hskip -0.1 cm sites}
\psfrag{t}{time}
\hskip -2.2 cm
\begin{minipage}{0.277\textwidth}
  \includegraphics[width=177pt,angle=-90]{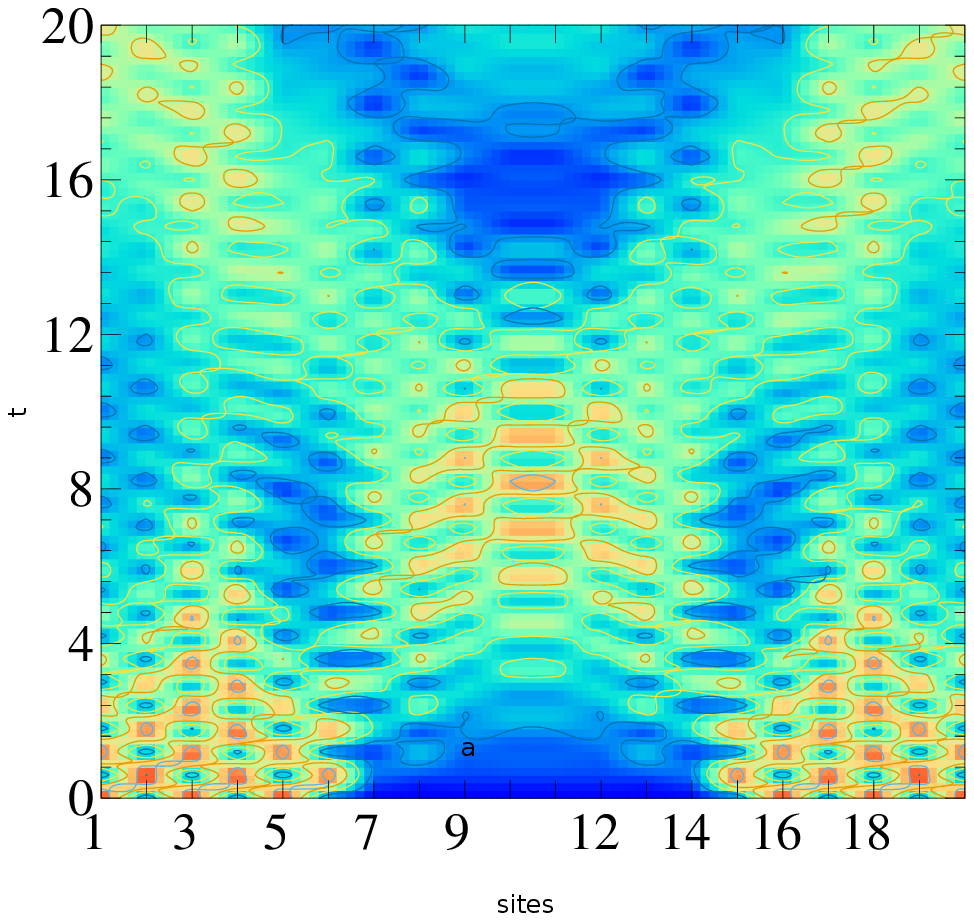}
    \end{minipage}\hskip 0.3cm
  \begin{minipage}{0.277\textwidth}
  \includegraphics[width=177pt,angle=-90]{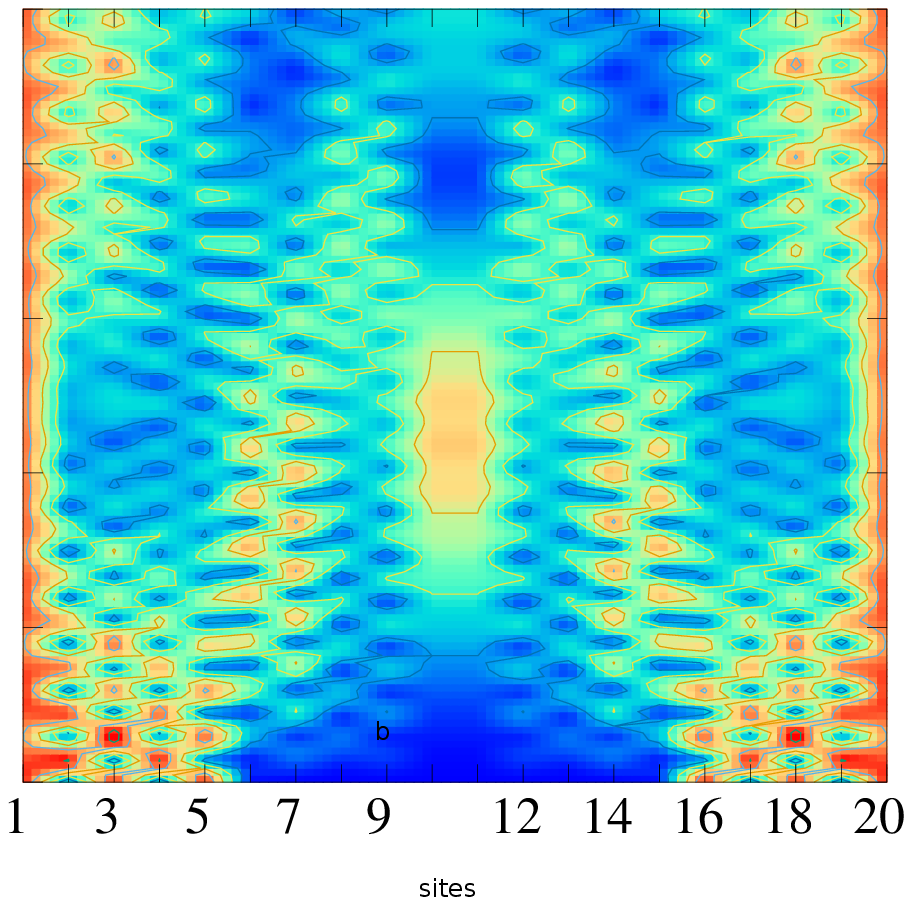}
  \end{minipage}\hskip 0.3cm
  \begin{minipage}{0.28\textwidth}
  \includegraphics[width=180pt,angle=-90]{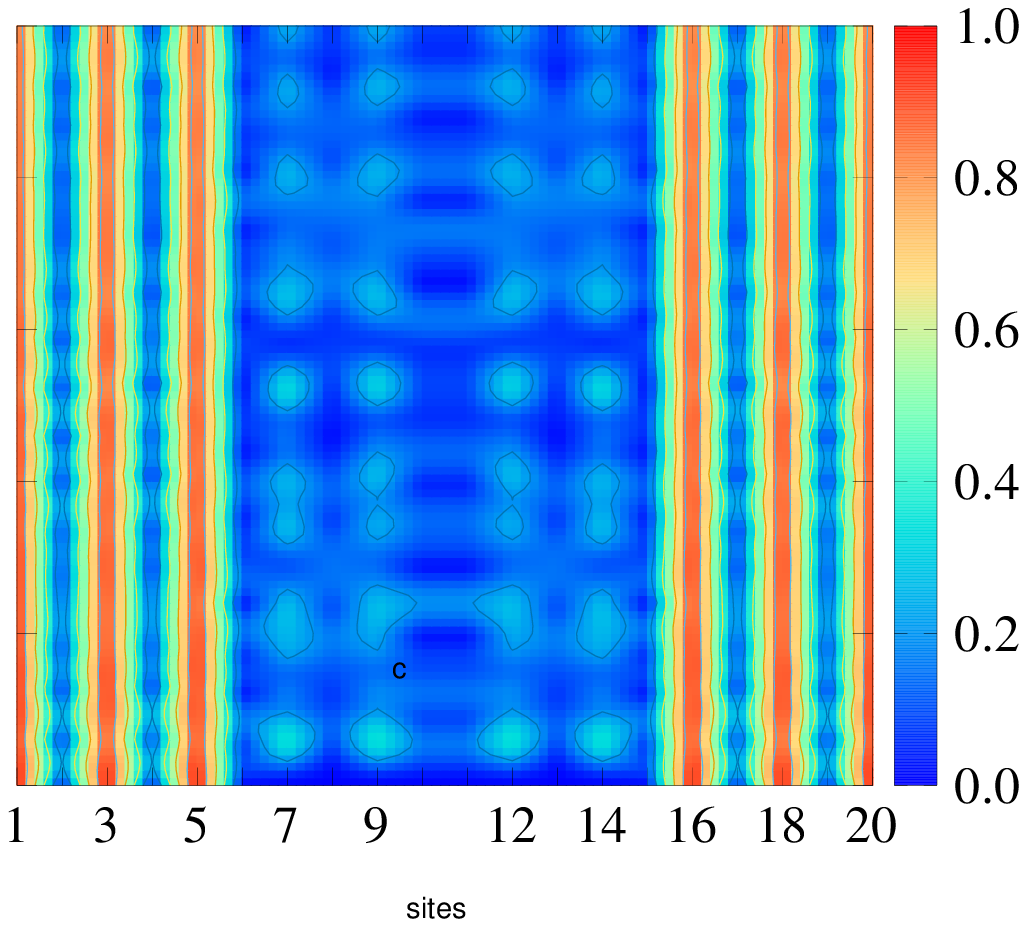}
  \end{minipage}
\caption{Quench dynamics for $H=H_{vw}+H^{\rm {B\mbox{-}A}}_{z,3}$
  when $\zeta=0.5$, (a) for $v=2.5$, $w=0.25$, $z=0.25$, 
   (b) for $v=0.25$, $w=2.5$, $z=0.25$,
  (c) for $v=0.25$, $w=0.25$, $z=2.5$.
Figures are drawn for lattice of 20 sites. }
\label{Quench-dynamics-SSH-4}
\end{figure*}

Evolution of edge states for the
nonlinear system is shown in 
Fig. \ref{Quench-dynamics-SSH-4}, 
by solving the set of Eq. \ref{coupled-equations-3},
for $L=20$ when $\zeta=0.5$.
Contour plot for the time evolution of
$|\psi_l(t)|$, is drawn for every site 
which is shown along the horizontal axis. 
Three contour plots are shown 
(a) for $v=2.5$, $w=0.25$, $z=0.25$, 
   (b) for $v=0.25$, $w=2.5$, $z=0.25$,
(c) for $v=0.25$, $w=0.25$, $z=2.5$,
where (a) indicates trivial phase as before while
(b) and (c) for the topological phases of $\nu=1$
and $\nu=3$, respectively. In this case,
initial condition is set by 
$\psi_l(0)=\delta_{l,m}$, where $m=1,3,5,16,18,20$. 
As a result, conservation
rule is modified by the equation,
$\sum_{l=1}^L|\psi_l(t)|^2=6$ for every case.

Evolution of the system is explored
for the time range $0\le t\le 20$, as shown 
along the vertical
axis. The diagram in (b) clearly indicates that
probability amplitudes for $l=1,20$,  {\em i. e.}, $|\psi_1(t)|$
and $|\psi_{20}(t)|$ survive with time. 
So the edge states bound to the topological phase
with $\nu=1$ exhibit their quenching. Obviously, no such quenching
is found for any site in the trivial phase as shown in (a).
Quenching of amplitudes of wave function
for six sites, $|\psi_l(t)|$, when 
$l=1,3,5,16,18,20$ are found in (c) which correspond to the
topological phase with $\nu=3$.
So, quenching will be found for the 
amplitude with sites $l=1,3,5,L\!-\!4,L\!-\!2,L$, 
in case of chain of length $L$. 
The number of quenched sites increases with the increase of $\nu$.
\subsection{Topological phases for $H=H_{vw}+H^{\rm {A\mbox{-}B}}_{z,3}$}
\begin{figure}[h]
\psfrag{A}{\large A}
\psfrag{B}{\large B}
\psfrag{v}{\large $v$}
\psfrag{w}{\large $w$}
\psfrag{z}{\large $z$}
\includegraphics[width=230pt]{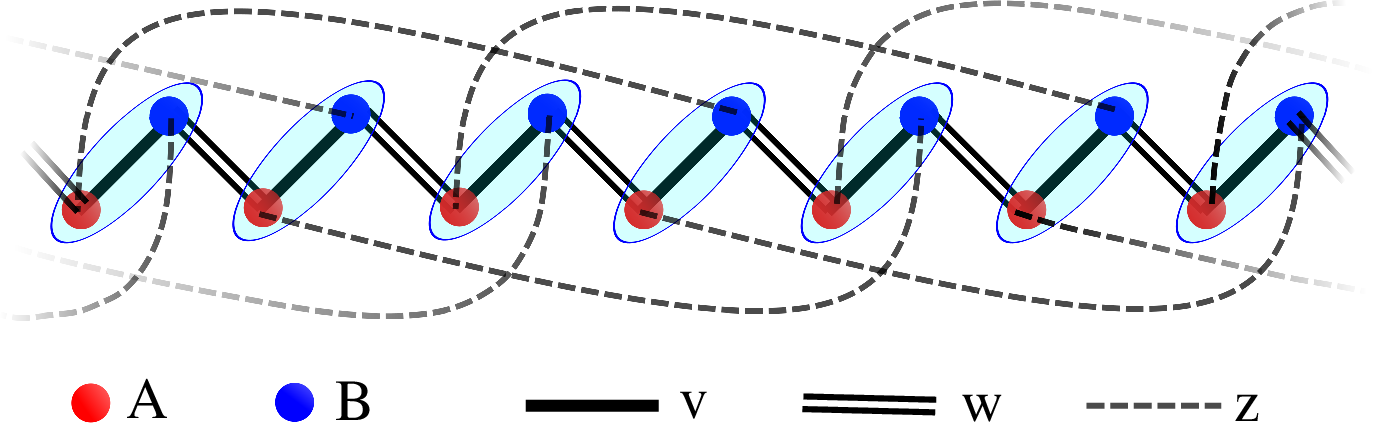}
\caption{Extended SSH model describing the hopping in
  $H=H_{vw}+H^{\rm {A\mbox{-}B}}_{z,3}$.}
\label{Extended-SSH-5}
\end{figure}

\begin{figure}[h]
\psfrag{x}{\large $v/|w+z|$}
\psfrag{y}{\large k}
\psfrag{ep}{\large $E_+({\rm k})$}
\psfrag{a}{\large (a)}
\psfrag{b}{\large (b)}
\psfrag{v}{\large $w=1$}
\psfrag{za}{\large $z=1/2$}
\psfrag{vbb}{\large $w=1/2$}
\psfrag{zb}{\large $z=1$}
\psfrag{5}{\large 5}
\psfrag{4}{\large 4}
\psfrag{3}{\large 3}
\psfrag{2}{\large 2}
\psfrag{1}{\large 1}
\psfrag{0}{\large 0}
\psfrag{0.0}{\large 0}
\psfrag{-2}{\large $-$2}
\psfrag{-1}{\large $-$1}
\psfrag{-3.1}{\large $-\pi$}
\psfrag{-1.6}{\large $-\pi/2$}
\psfrag{3.1}{\large $\pi$}
\psfrag{1.6}{\large $\pi/2$}
  \includegraphics[width=230pt]{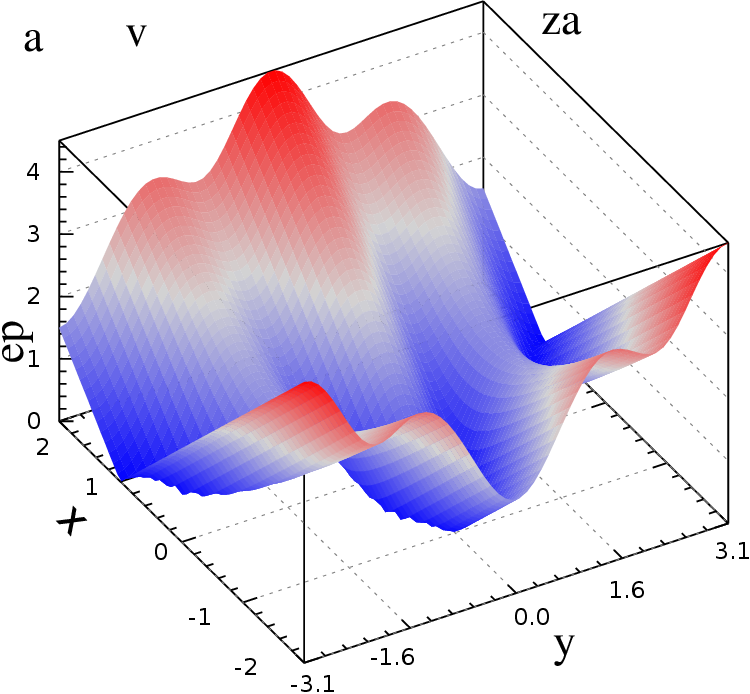}
  \vskip 0.6cm
  \includegraphics[width=230pt]{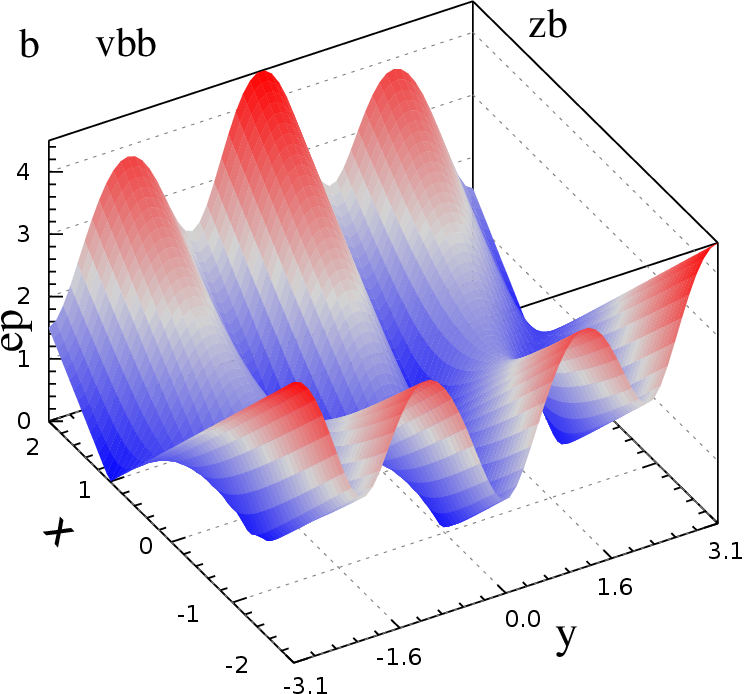}
\caption{Dispersion relation for $H=H_{vw}+H^{\rm {A\mbox{-}B}}_{z,3}$
  when $w/z>1$, (a) and $w/z<1$, (b).}
\label{Dispersion-SSH-5}
  \end{figure}

Total Hamiltonian now is 
\bea
H&=& H_{vw}+H^{\rm {A\mbox{-}B}}_{z,3},\nonumber\\ [0.4em]
H^{\rm {A\mbox{-}B}}_{z,3}&=&\sum_{j=1}^Nz\,c^\dag_{{\rm A},j}c_{{\rm B},j+3}+{\rm h.c.},
\label{HAB-z-4}
\eea
where the hopping term extends over two intermediate primitive cells,
as shown in Fig. \ref{Extended-SSH-5}.
The components of $\boldsymbol g(\rm k)$ in this case are 
\[\boldsymbol g(\rm k)\equiv\left\{\begin{array}{l}
g_x=v+w\cos{\!(\rm k)}+z\cos{\!(3\rm k)},\\[0.3em]
g_y=w\sin{\!(\rm k)}-z\sin{\!(3\rm k)},\\[0.3em]
g_z=0.
\end{array}\right.\]
The corresponding dispersion relation is
$E_\pm({\rm k})=\pm \sqrt{v^2+w^2+z^2+2[vw\cos{\!(\rm k)}
    +vz\cos{\!(3\rm k)}+wz\cos{\!(4\rm k)}]}$.
  Variation of dispersion relation, $E_+({\rm k})$,
with $v/|w+z|$ for $w=1$, $z=1/2$ and $w=1/2$, $z=1$ are shown in
Fig. \ref{Dispersion-SSH-5} (a) and (b), respectively
for the region $-2 \le v/|w+z|\le +2$.
Figures for $w>z$ and $w<z$ will be of similar shape
as shown in (a) and (b).
Dispersions exhibits three broad peaks in the regions
away from the point 
$v/|w+z|= 1$ for both the cases $w/z>1$ and $w/z<1$. 
Again the band gap vanishes at the BZ boundaries, ${\rm k}=\pm \pi$,
and ${\rm k}=0$ when $v=|w+z|$. 
As a result, $\nu$ is undefined at the point
when $v=|w+z|$.
  
This time, three different nontrivial phases
with $\nu=+1,-1$ and $-3$ appear
in the parameter space as given below which are separated by
distinct phase transition lines. 
\be
\nu=\left\{\begin{array}{ll}
    0,&  v>|w+z|,\\[0.3em]
    1,-1& v<|w+z|,\;{\rm and}\;w>z,\\[0.3em]
    -3,-1,&v<|w+z|, \;{\rm and}\;w<z.
  \end{array}\right.
\ee
Trivial phase exists when $v>|w+z|$.
A pair of distinct topological phase with
$\nu=\pm 1$ emerges when $v<|w+z|$ and $w>z$.
Another pair of nontrivial phase with
$\nu=-3$ and $-1$ appears $v<|w+z|$ when $w<z$.
It means the phase with  $\nu=-1$ emerges in
two different regions when $v<|w+z|$ but
for both $w>z$ and $w<z$.
The location of this transition point
can be determined by satisfying the
conditions, $E_\pm({\rm k})=0$, and
$\frac{dE_\pm({\rm k})}{d{\rm k}}=0$.
So, this model is capable to host a
new topological phase with $\nu=-3$.

\begin{figure}[h]
\psfrag{gxx}{\large $g_x$}
\psfrag{gyy}{\large $g_y$}
\psfrag{a}{\large (a)}
\psfrag{b}{\large (b)}
\psfrag{c}{\large (c)}
\psfrag{d}{\large (d)}
\includegraphics[width=230pt]{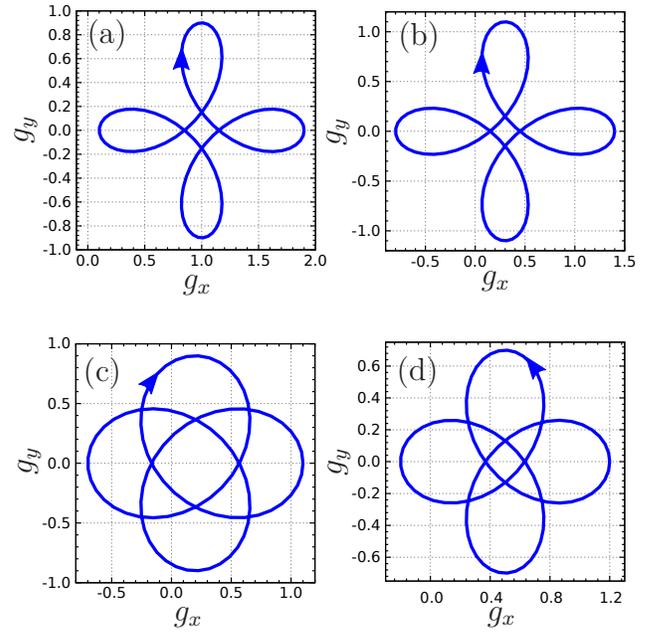}
\caption{Parametric winding diagrams in the
$g_x\mbox{-}g_y$ plane for the Hamiltonian,
  $H=H_{vw}+H^{\rm {A\mbox{-}B}}_{z,3}$. Four figures are drawn for
  (a) $v=1.0$, $w=0.5$, $z=0.4$, (b) $v=0.3$, $w=0.6$, $z=0.5$,
  (c) $v=0.2$, $w=0.3$, $z=0.6$, and (d) $v=0.5$, $w=0.4$, $z=0.3$.}
\label{parametric-windings-SSH-5}
\end{figure}

The parametric plot of winding by the tip of
the vector, $\boldsymbol g(\rm k)$ 
in the $g_x\mbox{-}g_y$ plane are shown in
Fig. \ref{parametric-windings-SSH-5}. 
Four figures are drawn for
  (a) $v=1.0$, $w=0.5$, $z=0.4$, (b) $v=0.3$, $w=0.6$, $z=0.5$,
(c) $v=0.2$, $w=0.3$, $z=0.6$, and (d) $v=0.5$, $w=0.4$, $z=0.3$.
Curves shown in (a) and (d) traverse along the 
counter clockwise direction, while those in (b) and
(c) traverse along the clockwise direction. 
As a result the winding number for (d) is positive,
but that for (b) and (c) is negative. 
Those figures can be regarded as prototype contours for the
four different regions. For example, 
the phase is trivial ($\nu=0$),
for any values of $v$, $w$ and $z$ as long as $v>|w+z|$.
On the other hand, four different regions are identified
when $v<|w+z|$, where distinct topological phases
appear. A pair of phases appear for $w>z$ and
another pair for $w<z$.
Nonzero band gap exists for all the cases.

\begin{figure}[h]
  \psfrag{v}{\large $v/|w\!+\!z|$}
  \psfrag{3}{\hskip -0.15 cm 3}
\psfrag{2}{\hskip -0.15 cm 2}
\psfrag{1}{\hskip -0.15 cm 1}
\psfrag{0}{\hskip -0.15 cm 0}
\psfrag{-3}{\hskip -0.15 cm -3}
\psfrag{-2}{\hskip -0.15 cm -2}
\psfrag{-1}{\hskip -0.15 cm -1}
\psfrag{y}{\large k}
\psfrag{E}{\hskip -0.5 cm Energy}
\psfrag{a}{\large (a)}
\psfrag{b}{\large (b)}
\hskip -3.0 cm
\begin{minipage}{0.20\textwidth}
  \includegraphics[width=130pt,angle=-90]{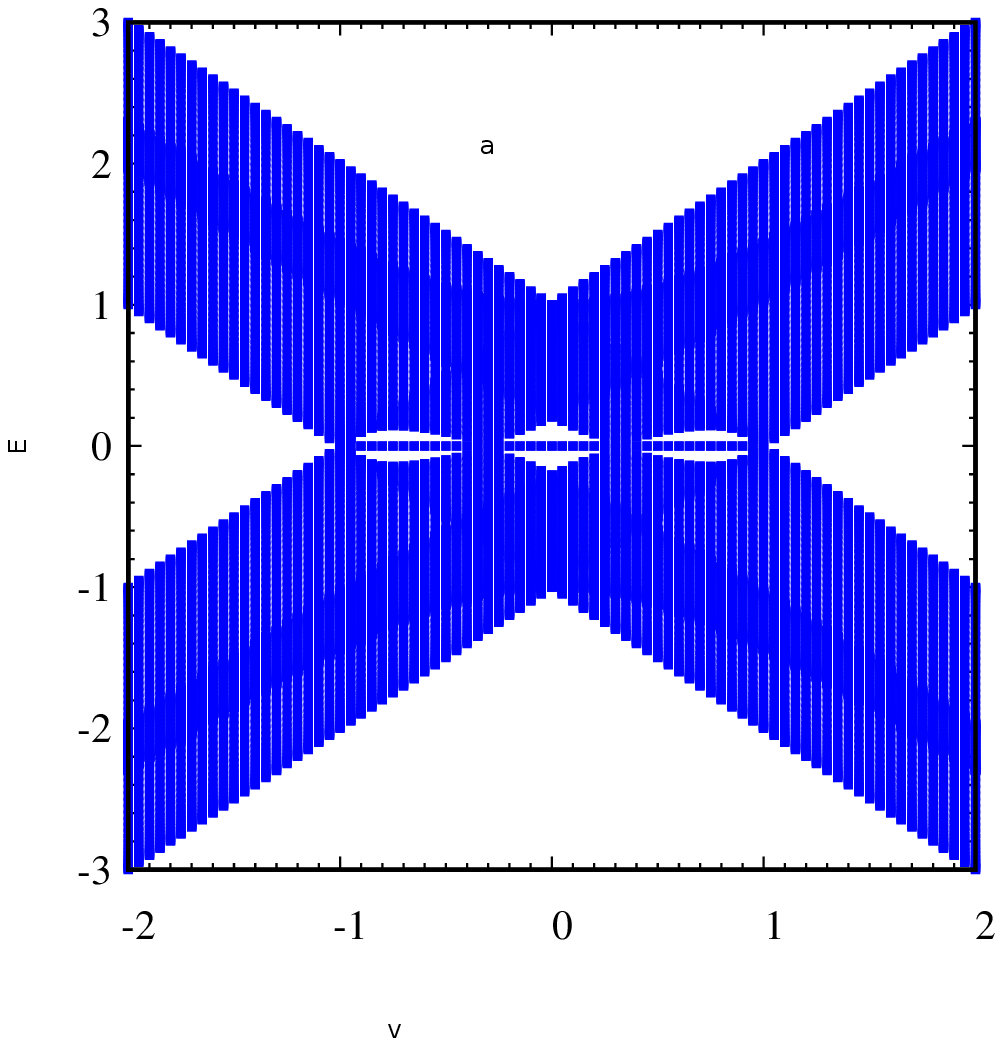}
  \end{minipage}\hskip 0.6cm
  \begin{minipage}{0.2\textwidth}
  \includegraphics[width=130pt,angle=-90]{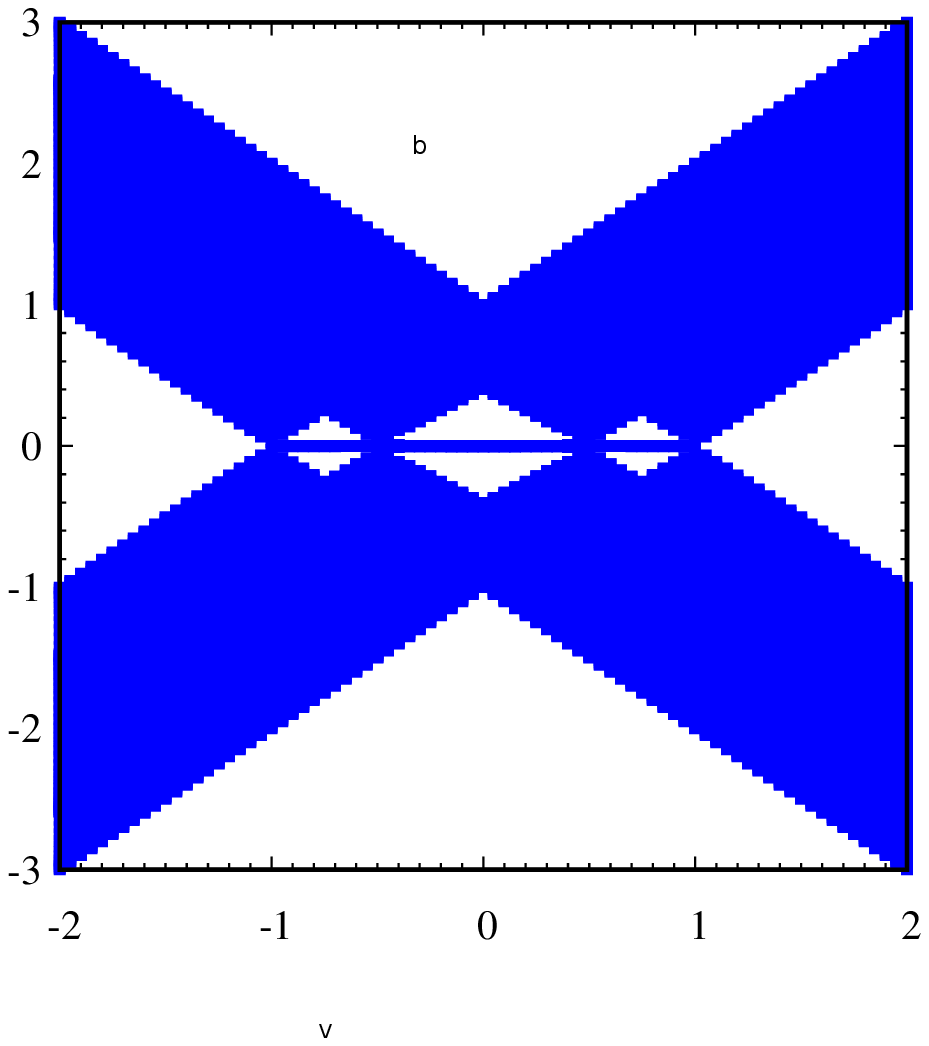}
  \end{minipage}
\caption{Edge states for $H=H_{vw}+H^{\rm {A\mbox{-}B}}_{z,3}$
  when $w/z>1$, (a) for $w=0.6$, $z=0.4$, and $w/z<1$, (b)
  for $w=0.3$, $z=0.7$.}
\label{Edge-states-SSH-5}
\end{figure}

Variation of bulk-edge state energies with respect to
$v/|w+z|$ is shown in Fig. \ref{Edge-states-SSH-5}
for the regime $-2\le (v/|w+z|)\le 2$. 
Zero energy edge states emerges as long as
$-1\le (v/|w+z|)\le 1$, which is consistent to the
previous observation.
So, no edge state is there
in this system when $v>|w+z|$. However, 
three pairs of zero energy edge states appear
in a region around the point  $v/|w+z|=0$ 
when $v<|w+z|$ and $w<z$ which is shown in
Fig. \ref{Edge-states-SSH-5} (b). This particular
region is surrounded by another topological phase
with $\nu=-1$ as long as $-1\le (v/|w+z|)\le 1$.
The figures are drawn for lattice of sites 200, and
the results conform to the bulk-edge correspondence rule 
in the topological phases.

\begin{figure}[h]
\psfrag{mps}{\hskip -0.1 cm $|\psi|^2$}
\psfrag{a}{ (a)}
\psfrag{b}{ (b)}
\psfrag{n}{sites}
\psfrag{p}{$v=0.25,\,w=2.5,\,z=0.25$}
\psfrag{q}{$v=0.25,\,w=0.25,\,z=2.5$}
\psfrag{150}{150}
\psfrag{125}{125}
\psfrag{100}{100}
\psfrag{75}{75}
\psfrag{50}{50}
\psfrag{25}{25}
\psfrag{200}{200}
\psfrag{175}{175}
\psfrag{225}{225}
\psfrag{0.5}{0.5}
\psfrag{1.0}{1.0}
\psfrag{0} {0}
\psfrag{0.0}{0.0}
\includegraphics[width=230pt]{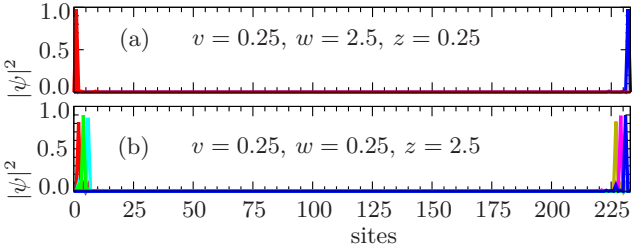}
\caption{Probability density of edge states for $H=H_{vw}+H^{\rm {A\mbox{-}B}}_{z,3}$: 
  (a) for $v=0.25$, $w=2.5$, $z=0.25$, one pair of edge states, 
  (b) for $v=0.25$, $w=0.25$, $z=2.5$, three pairs of edge states.
Figures are drawn for 232 sites. }
\label{Edge-state-probability-for-SSH-5}
\end{figure}

In order to confirm the existence of zero energy edge states,
probability density of those states are drawn in Fig. 
\ref{Edge-state-probability-for-SSH-5} for the lattice of
200 sites.
Two figures are drawn for two distinct topological phases.
In the upper panel (a), probability densities of two
distinct edge states with $E=0$ are shown when
$v=0.25$, $w=2.5$, $z=0.25$. Those values are
selected for satisfying the 
conditions, $v<|w+z|$ and $w>z$. Probability density
of one edge state exhibits sharp peak at site $m=1$
and another one at site $m=200$. This corresponds to
the topological phase with $\nu=1$.
On the other hand, for $v<|w+z|$ 
and $w<z$, probability density of six 
distinct zero energy edge states are shown
in the lower panel (b) when
$v=0.25$, $w=0.25$, $z=2.5$. 
Probability density 
of six orthonormal edge states exhibit
sharp peak at sites $m=2,4,6,195,197$ and 199. 
This result is in accordance to
the topological phase of $\nu=-3$.
Localization of zero energy states are found on
B sublattice near left edge and A sublattice near right edge. 

\begin{figure}[h]
  \psfrag{nu}{\large $\nu$}
   \psfrag{nu0}{\hskip -.cm \color{blue} \Large $\nu\!=\!0$}
   \psfrag{nu-3}{\hskip -.5cm \color{white} \Large $\nu=-3$}
   \psfrag{nu-1}{\hskip -.250cm \color{blue} \Large $\nu\!=\!-1$}
   \psfrag{nu1}{\hskip -.50cm \color{white} \Large $\nu=1$}
\psfrag{w}{\large $w$}
\psfrag{zw}{\large $w+z=2$}
\psfrag{v}{\hskip -0.4 cm \large $v/|w+z|$}
\psfrag{0.5}{\large 0.5}
\psfrag{0.0}{\large 0.0}
\psfrag{1.0}{\large 1.0}
\psfrag{1.5}{\large 1.5}
\psfrag{2.0}{\large 2.0}
\psfrag{ 1}{\large 1}
\psfrag{ 0}{\large 0}
\psfrag{-1}{\large -1}
\psfrag{-2}{\large -2}
\psfrag{-3}{\large -3}
\psfrag{-2.0}{\large -2.0}
\psfrag{-2.5}{\large -2.5}
\psfrag{-3.0}{\large -3.0}
\psfrag{-0.5}{\large -0.5}
\psfrag{-1.0}{\large -1.0}
\psfrag{-1.5}{\large -1.5}
\includegraphics[width=230pt,angle=-90]{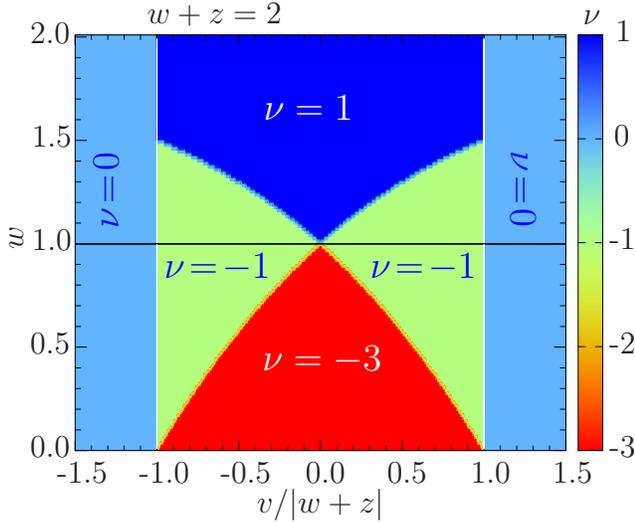}
\caption{Topological phase diagram for the Hamiltonian 
  $H=H_{vw}+H^{\rm {A\mbox{-}B}}_{z,3}$. Three distinct topological
  phases are shown by blue ($\nu=1$),
  green ($\nu=-1$) and red  ($\nu=-3$).
  The remaining portion is trivial 
  ($\nu=0$).  
  This diagram is drawn for $w+z=2.$
The horizontal line 
indicates the value $w=1$ or $w/z=1.$}
\label{topological-phase-windings-SSH-5}
\end{figure}

A rigorous phase diagram for this model is shown in Fig
\ref{topological-phase-windings-SSH-5}, where contour plot
for $\nu$ is drawn in the $w$-$v/|w+z|$ space.
Variation of the parameters is made by maintaining
the constraint $w+z=2$. The
existence of three different topological phases,
$\nu=\pm 1$ and $-3$ along with the trivial phase, $\nu=0$ is
shown in four different colors. 
The horizontal line is drawn at $w/z=1$,
which separates the topological phase with $\nu=1$
from that with $\nu=-3$. Three topolgical phases meet
at the point, $v/(w+z)=0$, $w=1$. 
All the topological phases remain within the
region bounded by the vertical lines drawn at
$v/(w+z)= \pm 1$. So the trivial phase lies beyond the 
region $-1 \le v/(w+z)\le +1$ for any value of $w$. 
The curved boundary lines
separating the nontrivial phases can be 
obtained from the solutions of the Eqs. 
$E_\pm({\rm k})=0$, and $\frac{dE_\pm({\rm k})}{d{\rm k}}=0$.
Those equations yield a cubic equation, whose solutions
along with the constraint, $w+z=2$, provides the
equation of transition lines. 

According to the formalism for quenching
of edge states as discussed before,
dynamics of the edge states in the presence of
nonlinear terms for the topological phases
of this model will be discussed. 
The set of coupled nonlinear 
equation for chain of $L$ lattice sites and for the Hamiltonian
defined in Eq. \ref{HAB-z-4} with OBC is 
given by 
\bea
i\,\frac{\partial \psi_{2j-1}}{\partial t}&=&v(\psi_{2j}-\psi_{2j-1})+
w(\psi_{2j-2}-\psi_{2j-1})\nonumber \\[0.0 em]
&&+z(\psi_{2j+6}-\psi_{2j-1})-\zeta |\psi_{2j-1}|^2\psi_{2j-1},\nonumber \\[0.2 em]
\vdots\quad\; &=&\qquad \qquad \vdots \label{coupled-equations-4}\\[0.3 em]
i\,\frac{\partial \psi_{2j}}{\partial t}&=&w(\psi_{2j+1}-\psi_{2j})+
v(\psi_{2j-1}-\psi_{2j})\nonumber \\[0.0 em]
&&+z(\psi_{2j-7}-\psi_{2j})-\zeta |\psi_{2j}|^2\psi_{2j}.\nonumber
\eea
\begin{figure*}[t]
  \psfrag{a}{\hskip -0.1 cm \large \color{white} (a)}
\psfrag{c}{\hskip -0.07 cm \large \color{white} (c)}
\psfrag{b}{\hskip -0.1 cm \large \color{white}  (b)}
\psfrag{sites}{\hskip -0.1 cm sites}
\psfrag{t}{time}
\hskip -2.2 cm
\begin{minipage}{0.277\textwidth}
  \includegraphics[width=177pt,angle=-90]{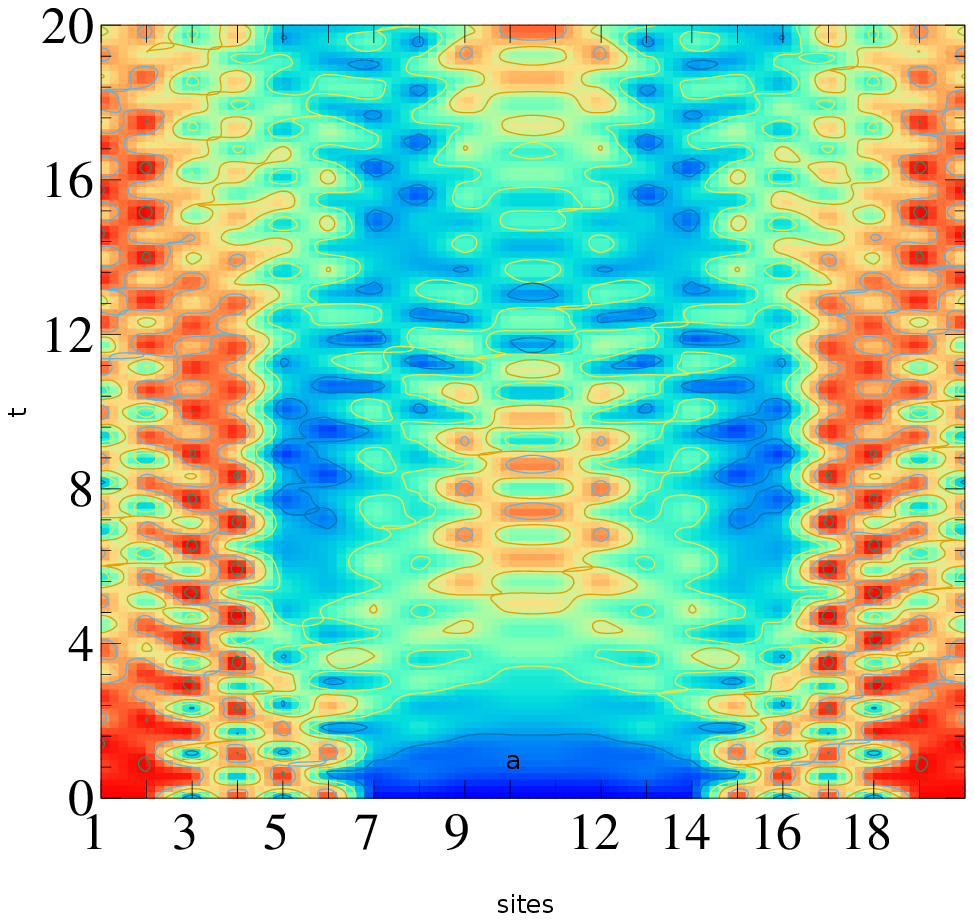}
    \end{minipage}\hskip 0.3cm
  \begin{minipage}{0.277\textwidth}
  \includegraphics[width=177pt,angle=-90]{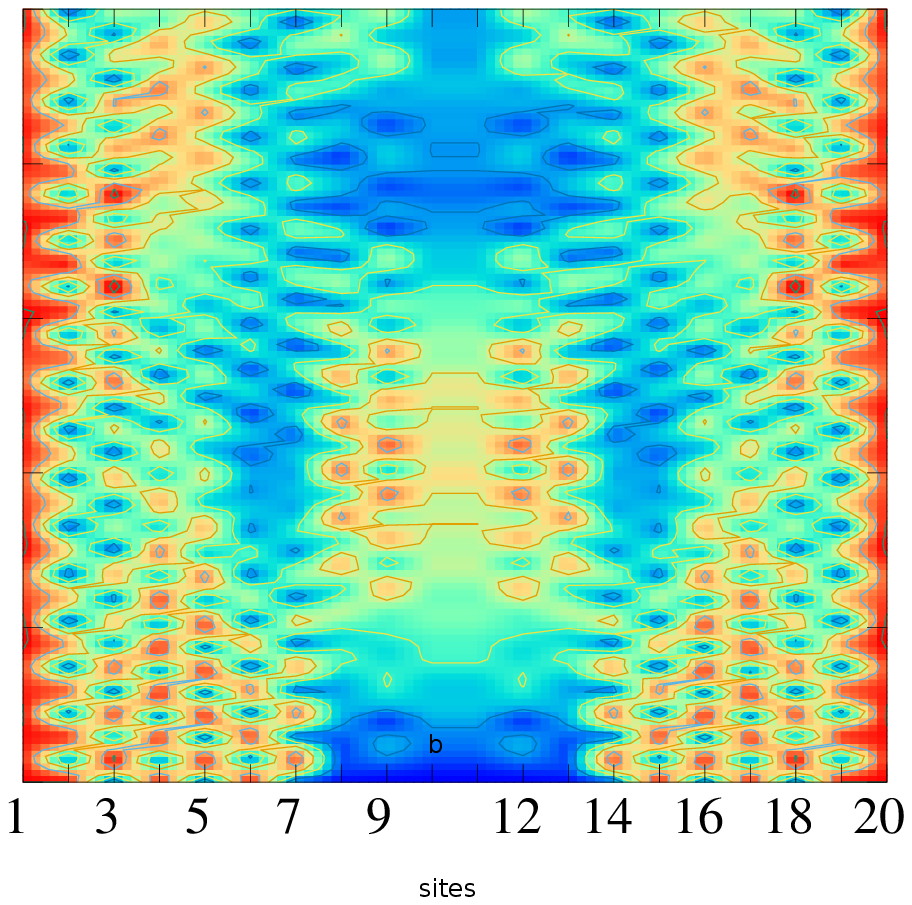}
  \end{minipage}\hskip 0.3cm
  \begin{minipage}{0.28\textwidth}
  \includegraphics[width=180pt,angle=-90]{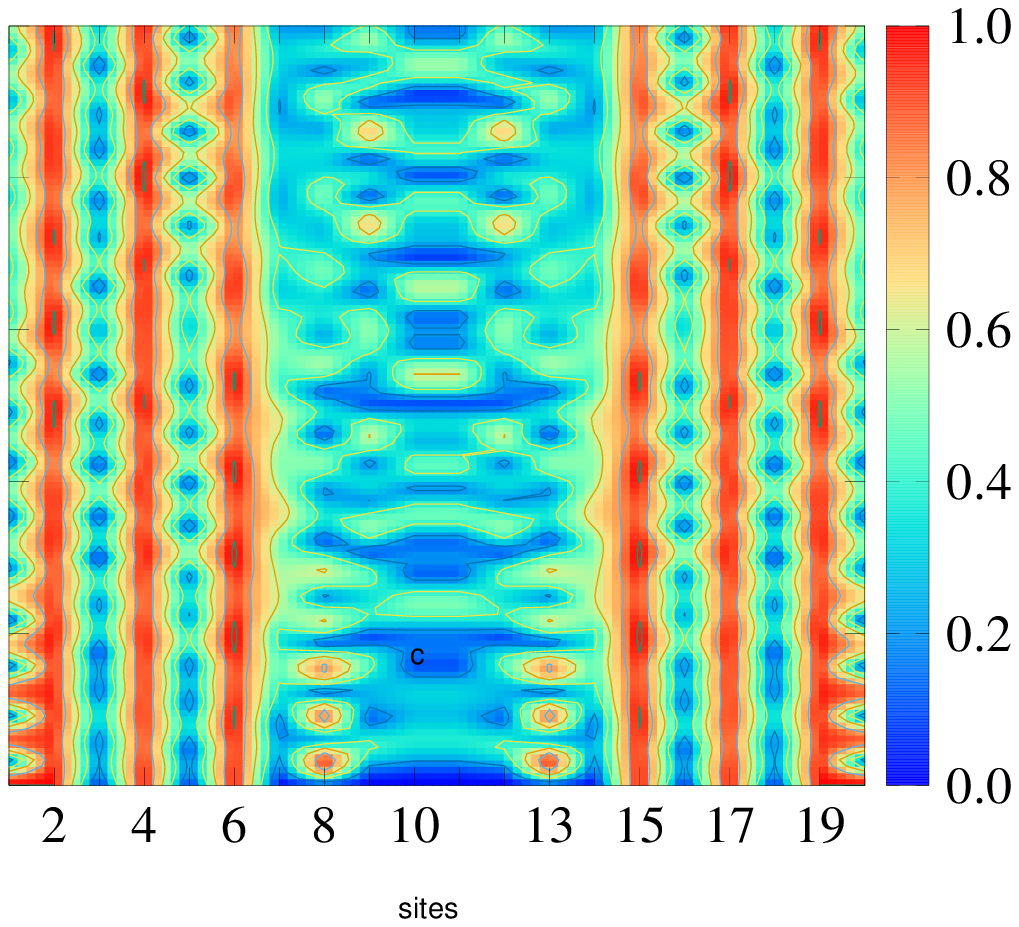}
  \end{minipage}
\caption{Quench dynamics for $H=H_{vw}+H^{\rm {A\mbox{-}B}}_{z,3}$
  when $\zeta=0.5$, (a) for $v=2.5$, $w=0.25$, $z=0.25$, 
   (b) for $v=0.25$, $w=2.5$, $z=0.25$,
  (c) for $v=0.25$, $w=0.25$, $z=2.5$.
Figures are drawn for lattice of 20 sites. }
\label{Quench-dynamics-SSH-5}
\end{figure*}

Evolution of edge states for the
nonlinear system is shown in 
Fig. \ref{Quench-dynamics-SSH-5}, 
by solving the set of Eq. \ref{coupled-equations-4},
for $L=20$ when $\zeta=0.5$.
Contour plot for the time evolution of
$|\psi_l(t)|$, is drawn for every site 
which is shown along the horizontal axis. 
Three contour plots are shown 
(a) for $v=2.5$, $w=0.25$, $z=0.25$, 
   (b) for $v=0.25$, $w=2.5$, $z=0.25$,
(c) for $v=0.25$, $w=0.25$, $z=2.5$,
where (a) indicates trivial phase as before while
(b) and (c) for the topological phases of $\nu=-1$
and $\nu=-3$, respectively. In this case,
initial condition is set by 
$\psi_l(0)=\delta_{l,m}$, where $m=2,4,6,16,18,20$. 
As a result, conservation
rule is modified by the equation,
$\sum_{l=1}^L|\psi_l(t)|^2=6$, for every case.

Evolution of the system is explored
for the time range $0\le t\le 20$, as shown 
along the vertical
axis. The diagram in (b) clearly indicates that
probability amplitudes for $l=1,20$,  {\em i. e.}, $|\psi_1(t)|$
and $|\psi_{20}(t)|$ survive with time. 
So, the edge states bound to the topological phase
with $\nu=1$ exhibit their quenching. Obviously, no such quenching
is found for any site in the trivial phase as shown in (a).
Quenching of amplitudes of wave function
for six sites, $|\psi_l(t)|$, when 
$l=2,4,6,15,17,19$ is found in (c) which correspond to the
topological phase with $\nu=-3$.
Therefore, quenching will be found in general for the
amplitudes on sites $l=2,4,6,L\!-\!5,L\!-\!3,L\!-\!1$,
if a chain of length $L$ is considered.
Again, the quenched sites for A and B sublattices interchange
the edges with respect to the last case. 
So the phase with higher values of $\nu$
can be studied where the quenching of absolute value of
the probability amplitude for higher number of sites
close to the ends of the lattice is observed.

Summarizing the results of above findings it is concluded that 
the method proposed in this work by constructing
series of Hamiltonians, $H=H_{vw}+H^{\rm {B\mbox{-}A}}_{z,m}$, and
$H=H_{vw}+H^{\rm {A\mbox{-}B}}_{z,m}$, with $m=1,2,3,\cdots$,
topological phases with  $\nu=\pm 1,\pm 2,\pm 3,\cdots,$
can be realized. Only one further neighbour hopping term
of strength $z$ is introduced within the standard SSH model,
whose extent is determined by the integer $m$. 
In this formulation, $m$ denotes the extent
of further neighbor hopping which passes over $(m-1)$
intermediate cells.
With the increase of $m$, topological phases of
higher values of $\nu$ will be realized. Hamiltonians of the type
$H=H_{vw}+H^{\rm {B\mbox{-}A}}_{z,m}$, yield phases of positive
winding number only, while that of type
$H=H_{vw}+H^{\rm {A\mbox{-}B}}_{z,m}$, yield those 
of negative winding numbers along with $\nu=0$ and $\nu=+1$.
A list of Hamiltonians along with
the winding numbers of accompanying 
phases are given in Tab. 1. 
 \begin{table}[h!]
\centering
\begin{tabular}{|c|c|c|}
\hline \;$m$\; &\;$H=H_{vw}+H^{\rm {B\mbox{-}A}}_{z,m}$ \;&$H=H_{vw}+H^{\rm {A\mbox{-}B}}_{z,m}$
 \\ \hline
 &&\\[-1.2em]
 1&$\nu=0,1$&$\nu=-1,0,1$ \\ \hline
  &&\\[-1.2em]
 2&$\nu=0,1,2$&$\nu=-2,0,1$ \\ \hline &&\\[-1.2em]
 3&$\nu=0,1,3$&$\nu=-3-1,0,1$ \\ \hline &&\\[-1.2em]
 4&$\nu=0,1,2,4$&$\nu=-4,-2,0,1$ \\ \hline &&\\[-1.2em]
 5&$\nu=0,1,3,5$&$\nu=-5,-3,-1,0,1$ \\ \hline&&\\[-1.2em]
 $\vdots$ & $\vdots$ & $\vdots$\\ \hline&&\\[-1.2em]
 $2p$&$\nu=0,1,2,\cdots\!,2p$&$\nu=-2p,-2p\!+\!2,\cdots\!,-2,0,1$ \\ \hline&&\\[-1.2em]
 $2p\!+\!1$&$\nu\!=\!0,1,3,\cdots\!,2p\!+\!1$&$\nu\!=\!-2p\!-\!1,-2p\!+\!1,\cdots\!,-1,0,1$ \\ \hline
\end{tabular}
 \label{Table}
 \caption{Distribution of winding numbers with the value of $m$
   for the Hamiltonians $H=H_{vw}+H^{\rm {B\mbox{-}A}}_{z,m}$ and
   $H=H_{vw}+H^{\rm {A\mbox{-}B}}_{z,m}$. Here $p$ is integer.}
 \end{table}
 
From this table, it is evident that, the topological phase with
$\nu=1$ and the trivial phase are present in every case. 
Apart from these common phases, an `odd-even' effect for
the values of $\nu$ is found with respect to $m$.
In addition to $\nu=0,1$ only even (odd) values of $\nu$
appear when $m$ is even (odd). This is true for both `B-A' and
`A-B' types of Hamiltonians. Also, number of nontrivial phases
increases with the value of $m$ for both types of Hamiltonians.
The result is generalized in the last two rows of the table,
which refer to even, ($m=2p$) and odd ($m=2p+1$) values of
$m$, where $p$ is an integer. 
More precisely, the number of nontrivial phase is equal to the
value of ($p+1$) and ($p+2$), respectively, for `B-A' and `A-B'
types of Hamiltonians when $m=(2p+1)$ (odd).
For even $m=2p$, both `B-A' and `A-B' types yield
($p+1$) number of topological phase.

 Another interesting finding is that, in order to realize
 the topological phase of the largest possible values of
 winding numbers from a specific Hamiltonian,
 value of $z$ is to be made larger than
 the individual values of $v$ and $w$. More elaborately, 
 it is known that maximum value of $\nu$ for the topological phase
 host by $H=H_{vw}+H^{\rm {B\mbox{-}A}}_{z,m}$ is $+m$, while that host by 
 $H=H_{vw}+H^{\rm {A\mbox{-}B}}_{z,m}$ is $-m$.
 However, this particular phase cannot be generated from the relevant
 Hamiltonian by substituting any value of $v/z$ and $w/z$.
By examining the phase diagrams as depicted in the Figs. 
\ref{topological-phase-windings-SSH-2}, \ref{topological-phase-windings-SSH-3}, 
\ref{topological-phase-windings-SSH-4}, and \ref{topological-phase-windings-SSH-5},
it can be concluded that the phase of the maximum value of $\nu$
can be achieved easily by choosing the values of the parameters in
such a way that they must satisfy the relations: 
$v/z\rightarrow 0$, and $w/z\rightarrow 0$. 
These limiting values indicate that phase with the 
maximum value of $\nu$ emerges when both the inter and intracell
hopping strength are much weaker than that of the further neighbour
cells.

Figs. (b) in \ref{Edge-state-probability-for-SSH-2},
\ref{Edge-state-probability-for-SSH-3}, 
\ref{Edge-state-probability-for-SSH-4} and 
\ref{Edge-state-probability-for-SSH-5}
indicate that localization of symmetry-protected
zero energy states are found on A and B sublattices
in a different way for phases with $+$ve and $-$ve values of 
$\nu$. For $\nu>0$, localization are found on
A sublattice for left edge and on B sublattice for right edge. 
But the localization occurs in the opposite fashion when $\nu<0$.
\section{Topological phases in terms of Chern numbers}
\label{phase-Chern-number}
In this case, further neighbour hoppings are allowed
within the same types of sublattice, means 
between A-A and B-B types of sites. 
In addition, further neighbour hoppings are limited between the
NN cells. The simplest model which exhibits topological phases
with any values of Chern numbers is defined by the
Hamiltonian,
\bea
H&=& H_{vw}+H_t,\nonumber\\ [0.4em]
H_t&=&\sum_{j=1}^N\left(t_a\,c^\dag_{{\rm A},j}c_{{\rm A},j+1}
+t_b\,c^\dag_{{\rm B},j}c_{{\rm B},j+1}\right)\!+{\rm h.c.},
\label{H-t}
\eea
\begin{figure}[h]
\psfrag{A}{\large A}
\psfrag{B}{\large B}
\psfrag{v}{\Large $v$}
\psfrag{w}{\Large $w$}
\psfrag{a}{\Large $t_a$}
\psfrag{b}{\Large $t_b$}
\includegraphics[width=230pt]{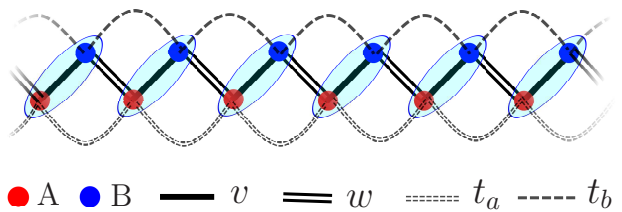}
\caption{Extended SSH model 
  describing the hopping for the Hamiltonian  
 in Eq. \ref{H-t}.}
\label{Extended-SSH-6}
\end{figure}
where $t_a$ and $t_b$ denote respectively the hopping
parameters between A-A and B-B types of sites
belonging to the NN cells. 
The model is described in the
Fig. \ref{Extended-SSH-6}. 
Assuming PBC, the Hamiltonian
in the k-space becomes
$H({\rm k})=g_{I}({\rm k}) I
+\boldsymbol g({\rm k})\cdot \boldsymbol \sigma$,  
where $I$ is the $2\times 2$ identity matrix, 
$g_{I}({\rm k})=(t_a+t_b)\cos{\!(\rm k)}$, and 
\[\boldsymbol g(\rm k)\equiv\left\{\begin{array}{l}
g_x=v+w\cos{\!(\rm k)},\\[0.3em]
g_y=w\sin{\!(\rm k)},\\[0.3em]
g_z=(t_a-t_b)\cos{\!(\rm k)}.
\end{array}\right.\]

\begin{figure}[h]
\psfrag{cp2}{$\mathcal C=+2$}
\psfrag{cm2}{$\mathcal C=-2$}
\psfrag{cp3}{$\mathcal C=+3$}
\psfrag{cm3}{$\mathcal C=-3$}
\psfrag{cp4}{$\mathcal C=+4$}
\psfrag{cm4}{$\mathcal C=-4$}
\psfrag{cp5}{$\mathcal C=+5$}
\psfrag{cm5}{$\mathcal C=-5$}
\psfrag{c0}{$\mathcal C=0$}
\psfrag{nu0}{\hskip -.cm \color{blue} \Large $\nu\!=\!0$}
\psfrag{nu-3}{\hskip -.5cm \color{white} \Large $\nu=-3$}
\psfrag{nu-1}{\hskip -.250cm \color{blue} \Large $\nu\!=\!-1$}
\psfrag{nu1}{\hskip -.50cm \color{white} \Large $\nu=1$}
\psfrag{f}{ $f$}
\psfrag{p1}{$-\pi$}
\psfrag{p5}{\hskip -0.2 cm$-\pi/2$}
\psfrag{p2}{$\pi$}
\psfrag{p3}{$\pi/2$}
\psfrag{p4}{\hskip -0.0 cm$\pi$}
\includegraphics[width=230pt,angle=0]{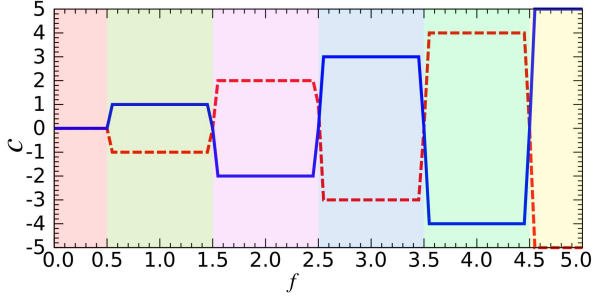}
\caption{Variation of $\mathcal C$ with $f$ for the Hamiltonian
defined in Eq. \ref{H-t} with the parameters in Eq. \ref{parameters-2}
is shown. Phases with higher values of $\mathcal C$ will
  appear with the increase of $f$. Solid blue line for
  $\phi=(n+\frac{1}{2})\pi$,
and dashed red line for  $\phi=(n-\frac{1}{2})\pi$.}
\label{topological-phase-chern-number}
\end{figure}

\begin{figure*}[h]
  \psfrag{mps}{\hskip -0.1 cm $|\psi|^2$}
  \psfrag{sites}{\hskip -0.1 cm sites}
  \psfrag{0.5}{\scriptsize 0.5}
  \psfrag{1.0}{\scriptsize 1.0}
  \psfrag{100}{\hskip -0.15 cm 100}
\psfrag{80}{80}
\psfrag{60}{60}
\psfrag{40}{40}
\psfrag{20}{20}
\psfrag{cp2}{\color{white} $\mathcal C=+2$}
\psfrag{cm2}{\color{white} $\mathcal C=-2$}
\psfrag{cp3}{\color{white} $\mathcal C=-3$}
\psfrag{cm3}{\color{white} $\mathcal C=+3$}
\psfrag{cp4}{\color{white} $\mathcal C=+4$}
\psfrag{cm4}{\color{white} $\mathcal C=-4$}
\psfrag{n2}{ $n=2$}
\psfrag{n3}{ $n=3$}
\psfrag{n4}{ $n=4$}  
\psfrag{a}{ (a)}
\psfrag{b}{(b)}
\psfrag{c}{(c)}
\psfrag{d}{(d)}
\psfrag{e}{(e)}
\psfrag{f}{(f)}
\psfrag{Energy}{Energy}
\psfrag{th}{$\theta$}
\psfrag{ 2}{\hskip -0.0 cm 2}
\psfrag{ 1}{\hskip -0.0 cm 1}
\psfrag{ 0}{\hskip -0.0 cm 0}
\psfrag{0}{\hskip 0.01 cm 0}
\psfrag{0.0}{\scriptsize \hskip -0.0 cm 0.0}
\psfrag{-2}{\hskip -0.2 cm $-2$}
\psfrag{-1}{\hskip -0.2 cm $-1$}
\psfrag{p1}{$\frac{\pi}{2}$}
\psfrag{p2}{$\pi$}
\psfrag{p3}{$\frac{3\pi}{2}$}
\psfrag{p4}{\hskip -0.2 cm$2\pi$}
\hskip -1.2 cm
\begin{minipage}{0.28\textwidth}
  \includegraphics[width=180pt,angle=0]{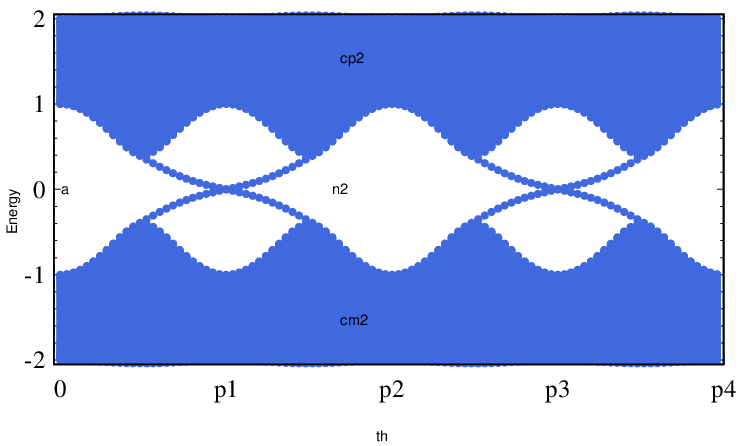}
    \end{minipage}\hskip 0.9cm
  \begin{minipage}{0.28\textwidth}
  \includegraphics[width=180pt,angle=0]{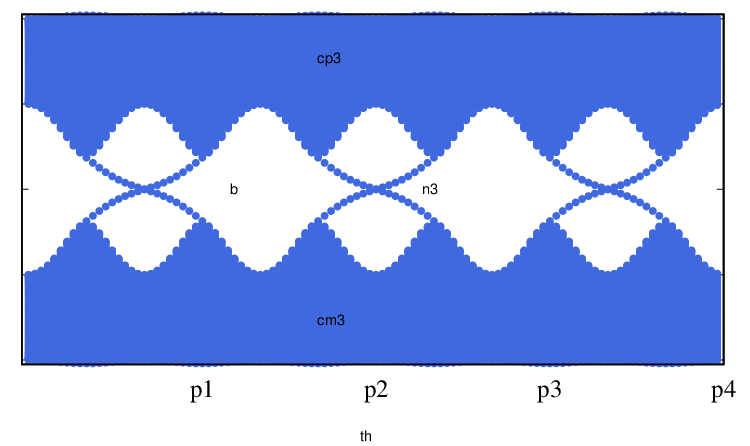}
  \end{minipage}\hskip 0.9cm
  \begin{minipage}{0.28\textwidth}
  \includegraphics[width=180pt,angle=0]{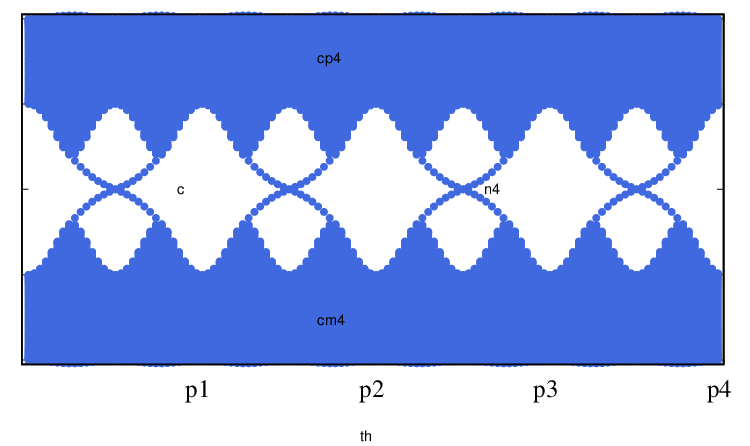}
  \end{minipage}\vskip 0.02 cm
  \hskip -1.2 cm
  \begin{minipage}{0.28\textwidth}
    \includegraphics[width=182pt,angle=0]{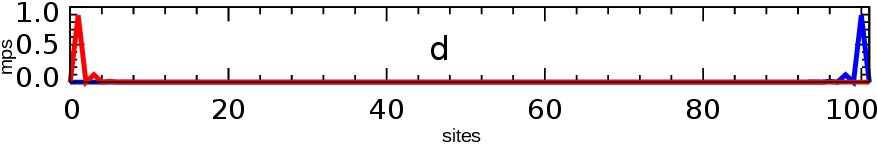}
    \end{minipage}  \hskip 1.25cm
   \begin{minipage}{0.27\textwidth}
     \includegraphics[width=166pt,angle=0]{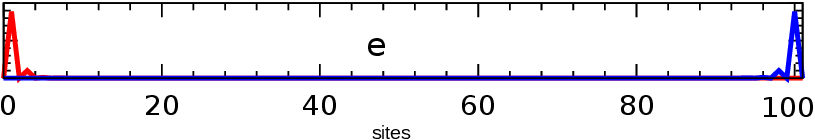}
   \end{minipage}\hskip 0.95cm
   \begin{minipage}{0.27\textwidth}
     \includegraphics[width=166pt,angle=0]{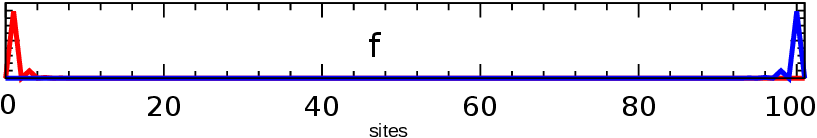}
   \end{minipage}
   \caption{Variation of energy with $\theta$ for the Hamiltonian
     defined in Eq. \ref{H-t} and the
     parametrization defined in Eq. \ref{parameters-2}
     are drawn in the upper panel. 
     Figures are drawn for $t=1$, $\delta=0.5$, $h=0.2$,
     $\phi=(n+\frac{1}{2})\pi$ 
     and $n=2$ in (a), $n=3$ in (b), $n=4$ in (c).
     Figures are drawn for lattice of 100 sites.
     Probability density of the corresponding edge states
     are shown in the lower panel.
     $\theta=\pi/2$ when $n=2$ in (d), $\theta=(\pi-1)/2$ when $n=3$ in (e), 
     and $\theta=\pi/4$ when $n=4$ in (f).}
\label{Energy-theta-higher-Chern}
\end{figure*}

\begin{figure*}[h]
\psfrag{n2}{ $n=2$}
\psfrag{n3}{\hskip 0.13 cm $n\!=\!3$}
\psfrag{n4}{\hskip 0.05 cm $n\!=\!4$}  
\psfrag{a}{ (a)}
\psfrag{b}{(b)}
\psfrag{c}{(c)}
\psfrag{Energy}{Energy}
\psfrag{th}{$\theta$}
\psfrag{ 2}{\hskip -0.0 cm 2}
\psfrag{ 1}{\hskip -0.0 cm 1}
\psfrag{ 0}{\hskip -0.0 cm 0}
\psfrag{0.0}{\hskip -0.0 cm 0}
\psfrag{-2}{\hskip -0.2 cm $-2$}
\psfrag{-1}{\hskip -0.2 cm $-1$}
\psfrag{p1}{$\frac{\pi}{2}$}
\psfrag{p2}{$\pi$}
\psfrag{p3}{$\frac{3\pi}{2}$}
\psfrag{p4}{\hskip -0.2 cm$2\pi$}
  \hskip -1.2 cm
\begin{minipage}{0.27\textwidth}
  \includegraphics[width=170pt,angle=0]{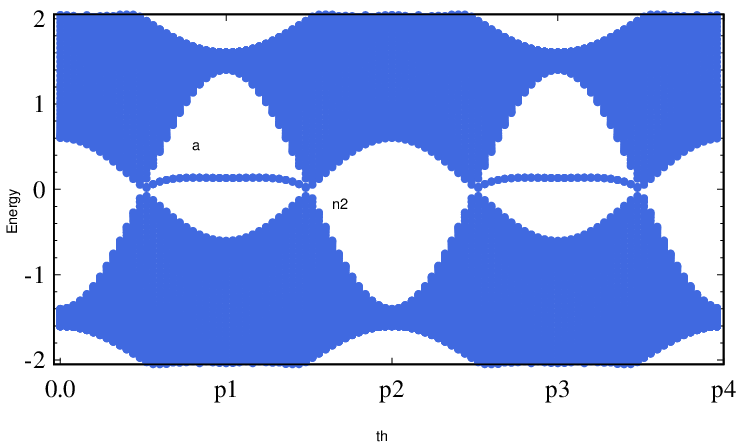}
    \end{minipage}\hskip 0.9cm
  \begin{minipage}{0.27\textwidth}
  \includegraphics[width=170pt,angle=0]{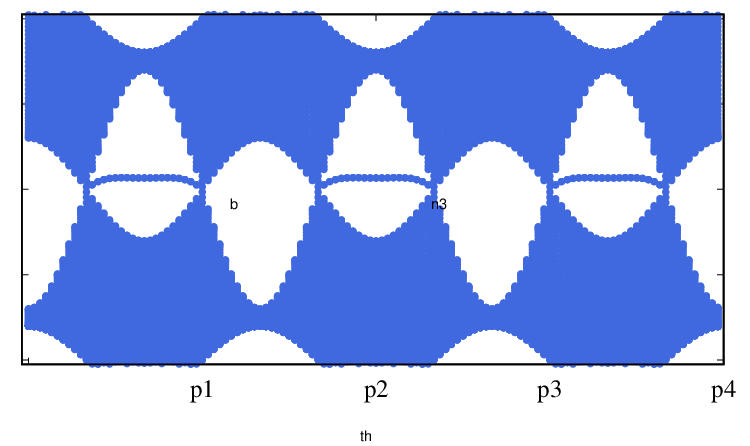}
  \end{minipage}\hskip 0.9cm
  \begin{minipage}{0.27\textwidth}
  \includegraphics[width=170pt,angle=0]{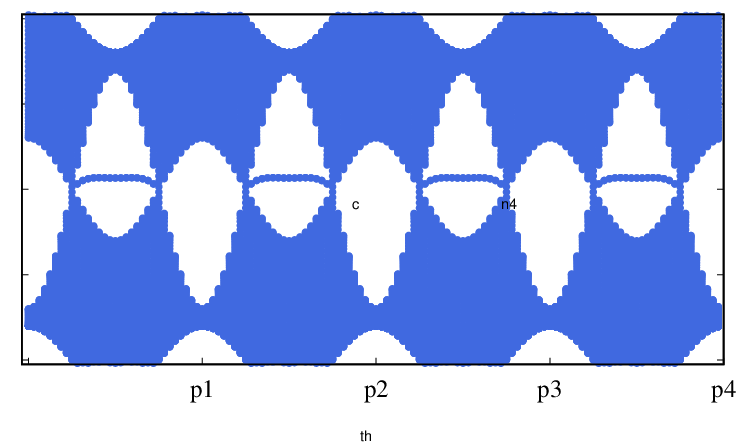}
  \end{minipage}
   \caption{Variation of energy with $\theta$ for the Hamiltonian
     defined in Eq. \ref{H-t} and the
     parametrization defined in Eq. \ref{parameters-2}
     are drawn when $\phi=0$. 
     Figures are drawn for $t=1$, $\delta=0.5$, $h=0.2$, 
     and $n=2$ in (a), $n=3$ in (b), $n=4$ in (c).
     Figures are drawn for lattice of 100 sites.}
\label{edge-states-with-phi-eq-0}
\end{figure*}

\begin{figure*}[h]
   \psfrag{mps}{\hskip 0.0 cm $|\psi|^2$}
  \psfrag{sites}{\hskip -0.1 cm sites}
  \psfrag{0.5}{\scriptsize 0.5}
  \psfrag{1.0}{\scriptsize 1.0}
  \psfrag{0.0}{\scriptsize \hskip -0.0 cm 0.0}
  \psfrag{100}{\hskip -0.15 cm 100}
\psfrag{80}{80}
\psfrag{60}{60}
\psfrag{40}{40}
\psfrag{20}{20}
\psfrag{cp1}{\color{white} $\mathcal C=+1$}
\psfrag{cm1}{\color{white} $\mathcal C=-1$}
\psfrag{n2}{ $n=2$}
\psfrag{n3}{ $n=3$}
\psfrag{n4}{ $n=4$}  
\psfrag{a}{ (a)}
\psfrag{b}{(b)}
\psfrag{c}{(c)}
\psfrag{d}{(d)}
\psfrag{e}{(e)}
\psfrag{f}{(f)}
\psfrag{Energy}{Energy}
\psfrag{th}{$\theta$}
\psfrag{ 2}{\hskip -0.0 cm 2}
\psfrag{ 1}{\hskip -0.0 cm 1}
\psfrag{ 0}{\hskip -0.0 cm 0}
\psfrag{0}{\hskip -0.0 cm 0}
\psfrag{-2}{\hskip -0.2 cm $-2$}
\psfrag{-1}{\hskip -0.2 cm $-1$}
\psfrag{p1}{$\frac{\pi}{2}$}
\psfrag{p2}{$\pi$}
\psfrag{p3}{$\frac{3\pi}{2}$}
\psfrag{p4}{\hskip -0.2 cm$2\pi$}
  \hskip -1.2 cm
\begin{minipage}{0.27\textwidth}
  \includegraphics[width=170pt,angle=0]{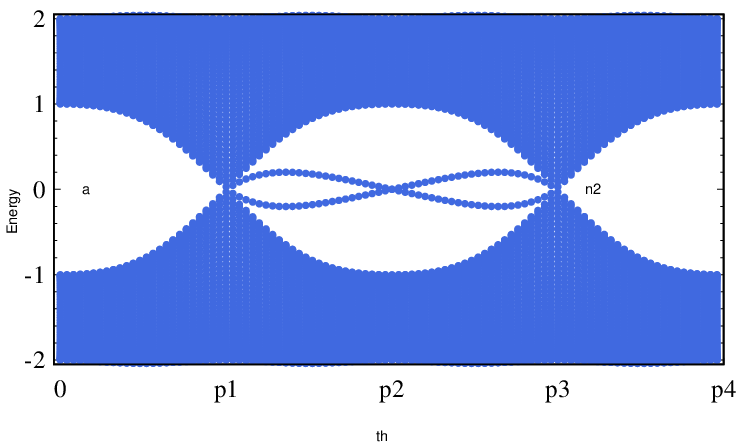}
    \end{minipage}\hskip 0.9cm
  \begin{minipage}{0.27\textwidth}
  \includegraphics[width=170pt,angle=0]{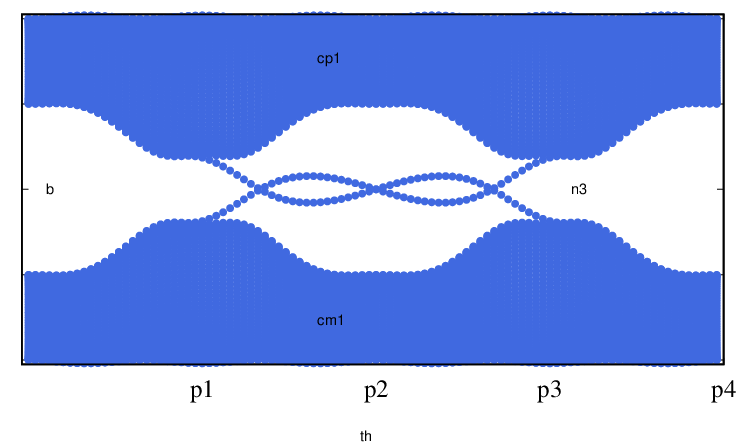}
  \end{minipage}\hskip 0.9cm
  \begin{minipage}{0.27\textwidth}
  \includegraphics[width=170pt,angle=0]{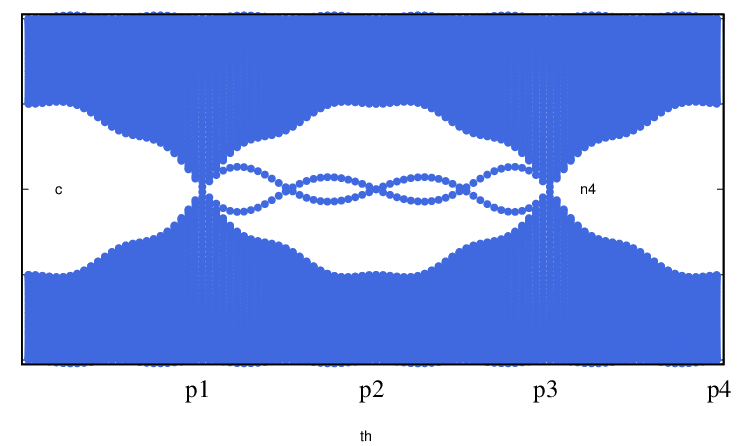}
  \end{minipage}\vskip 0.02 cm
  \hskip -1.2 cm
  \begin{minipage}{0.28\textwidth}
    \includegraphics[width=175pt,angle=0]{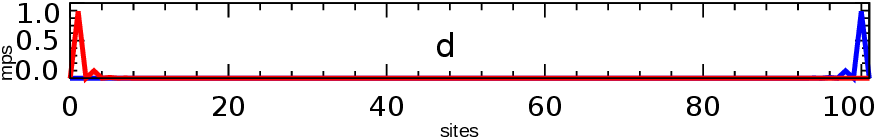}
    \end{minipage}  \hskip 1.1 cm
   \begin{minipage}{0.27\textwidth}
     \includegraphics[width=166pt,angle=0]{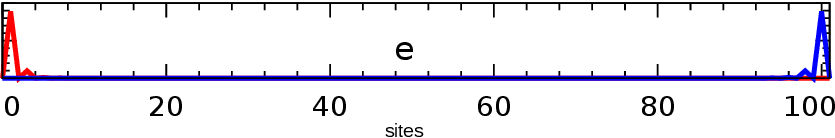}
   \end{minipage}\hskip 1.1 cm
   \begin{minipage}{0.27\textwidth}
     \includegraphics[width=166pt,angle=0]{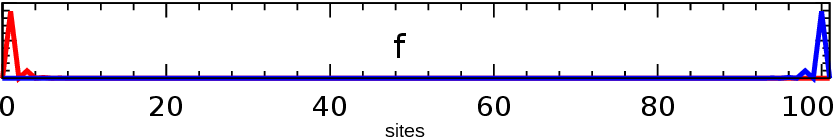}
   \end{minipage}
    \caption{Variation of energy with $\theta$ for the Hamiltonian
     defined in Eq. \ref{H-t} and the
     parametrization defined in Eq. \ref{parameters-3}
     are drawn in the upper panel. 
     Figures are drawn for $t=1$, $\delta=0.5$, $h=0.2$, $\phi=\pi/2$
     and $n=2$ in (a), $n=3$ in (b), $n=4$ in (c).
     Figures are drawn for lattice of 100 sites.
     Probability density of the corresponding edge states
     when $\theta=\pi$ 
     are shown in the lower panel for $n=2$ in (d),  $n=3$ in (e), 
     and $n=4$ in (f).}
\label{edge-states-with-multiple-crossing}
\end{figure*}

Since $\boldsymbol g(\rm k)$ is a three-component
vector, chiral symmetry is not preserved for this model.
In addition particle-hole and inversion
symmetries are not preserved. The standard forms of those
symmetries for 1D system now satisfy
\[\left\{\begin{array}{l}
\mathcal P H({\rm k}) \mathcal P^{-1}\ne -H(-{\rm k}),\\ [0.4em]
 \sigma_x H({\rm k}) \sigma_x\ne H(-{\rm k}),\\ [0.4em]
 \sigma_z H({\rm k}) \sigma_z\ne -H({\rm k}).\end{array}\right. \]
However, those symmetries in the ${\rm k}$-$\theta$ space
can be restored by choosing the hopping terms accordingly.  
 No topological phase in terms of nonzero
 winding number is present in this case.
However, this model is capable to behave like an
effective 2D model in the virtual momentum space
if the amplitude of hopping parameters are
modulated cyclically in terms of two additional
angular parameters $\theta$ and $\phi$.
Li {\em et al} introduced this model and reported the
emergence of topological phases characterized by
the Chern numbers of Haldane like two-band 2D system \cite{Li2}. 
Parametrization behind the realization of
this phase is 
\be
\left\{\begin{array}{l}
v=t[1+\delta \cos{\!(\theta)}],\\[0.3em]
w=t[1-\delta \cos{\!(\theta)}],\\[0.3em]
t_a=h\,\cos{\!(\theta+\phi)},\\[0.3em]
t_b=h\,\cos{\!(\theta-\phi)}.
\end{array}\right.
\label{parameters-1}
\ee
Investigation on this model has been carried out 
under periodic drive in order to find
Floquet topological phase\cite{Agarwal}. 

The Hamiltonian with this parametrization 
preserve the inversion symmetry
in terms of two variables ${\rm k}$ and $\theta$, since
\[\sigma_x H({\rm k},\theta) \sigma_x=H(-{\rm k},-\theta),\]
for any values of $\phi$, as long as $\theta \ne 0$. 
In addition, Hamiltonian preserves the
mirror symmetry with respect to $\theta$, 
when $\phi=0,\pm m\pi$, where $m$ is integer.
Let $\mathcal M_\theta$ be the operator for mirror symmetry 
and it acts as, $\mathcal M_\theta g(\theta)=g(-\theta)$,
where $g(\theta)$ is an arbitrary function of
$\theta$. Hamiltonian obeys the relation,
\[\mathcal M_\theta H({\rm k},\theta) \mathcal M_\theta^{-1}=H({\rm k},\theta),\]
only for $\phi=0,\pm m\pi$. Anyway, this symmetry is not
relevant to the topological properties 
in this case. 
However,  Hamiltonian preserves the
mirror symmetry with respect to both $\theta$, and $\phi$.
It means, if $\mathcal M_{\theta,\phi}$ be the operator of that symmetry,
Hamiltonian holds the relation:
\[\mathcal M_{\theta,\phi} H({\rm k},\theta) \mathcal M_{\theta,\phi}^{-1}=H({\rm k},\theta).\]
It occurs due to the fact that hopping parameters,
$v,\,w,\,t_a$ and $t_b$ do not change sign upon simultaneous sign reversal of
angular variables $\theta$ and $\phi$.

Interestingly,
topological phase of 2D system is realized in this 1D model
when $\theta$ is allowed to vary from $-\pi$ to $\pi$, 
for a specific value of another angular variable $\phi$.
For example, $\mathcal C=\pm 1$ is realized when
$0<\phi<\pi$, while $\mathcal C=\mp 1$ when
$-\pi<\phi <0$. Mirror symmetry with respect to
$\mathcal M_\theta$ is broken in this entire
regime, but that with respect to
$\mathcal M_{\theta,\phi}$ is preserved.
Band gap vanishes for $\phi=0$, as well as 
band inversion occurs around this point. Hence these
two distinct topological phases are realized in this model.
No other phase is realized if the further neighbour
hopping extends beyond the NN primitive cells
for the parametrization defined in Eq. \ref{parameters-1}.
However, band inversion is found to take place if further
neighbor hopping between NN cells is replaced by NNN
cells and this phenomenon occurs recursively if the
further neighbor hopping terms extend
beyond NNN cells successively in addition.

Chern number can be defined in this virtual reciprocal
space as
\be \mathcal C=\frac{1}{2\pi}\iint_{\rm BZ} d{\rm k}d\theta\,
(\partial_{\theta}A_{\rm k}-\partial_{\rm k}A_{\theta}),
\label{Chern}
\ee
where the Berry phase,
$A_\nu=i\langle {\rm k}\theta|\partial_{\nu}|{\rm k}\theta \rangle$,
with $\nu={\rm k},\,\theta$ and $|{\rm k}\theta \rangle$
is the Bloch state. Integration is performed over the
BZ in the 2D reciprocal space. 
The reciprocal space is called virtual in a sense that
no parameter in the real space can be connected to
the angular variable $\theta$, as on contrary the wave number
${\rm k}$ corresponds to the reciprocal of the
lattice parameter for the real space.
In order to find the Chern number the integral in Eq \ref{Chern}
is numerically evaluated \cite{Suzuki}.

In this study, new topological phases other than
$\mathcal C=\pm 1$ and $\mp 1$
have been obtained in a very simple way in which 
the angular variable $\theta$ is replaced by ($f\theta$) 
where $f$ may assume any values. Another cyclic parameter
$\phi$ also depends on $f$ as shown below. 
Higher values of
$f$ lead to phases with higher values of $\mathcal C$. 
In other words for the realization of phases with
$\mathcal C=\pm n$ and $\mp n$,
with $n=1,2,3,\cdots,$ sequentially, 
the following parametrization is implemented,
\be\left\{\begin{array}{l}
v=t[1+\delta \cos{\!(f\theta)}],\\[0.3em]
w=t[1-\delta \cos{\!(f\theta)}],\\[0.3em]
t_a=h\,\cos{\!(f\theta+\phi)},\\[0.3em]
t_b=h\,\cos{\!(f\theta-\phi)},
\end{array}\right.
\label{parameters-2}
\ee
where $\phi=(f\pm\frac{1}{2})\pi$.
The mirror symmetry with respect to the operator,
$\mathcal M_\theta$ is preserved only when
$f=\pm \frac{1}{2},\pm \frac{3}{2},\pm \frac{5}{2},\cdots$.
Because at those points all the hopping parameters,
$v,\,w,\,t_a$ and $t_b$ do not change sign upon the sign reversal of
angular variable $\theta$. So, the Hamiltonian remains invariant 
under the transformation. 
In contrast, the particle-hole symmetry is preserved in the
${\rm k}$-$\theta$ space only when $f$ is integer, since the relation,
\[\mathcal P H({\rm k},\theta) \mathcal P^{-1}=-H(-{\rm k},-\theta),\]
is satisfied at those points. But the chiral symmetry is not preserved anymore.

The system is found to host nontrivial phases with 
\[\mathcal C=\left\{\begin{array}{lll}
\pm n,&(n\!-\!\frac{1}{2})\!<\!f\!<\!(n\!+\!\frac{1}{2}),&{\rm when}\;n=1,3,5,\cdots,\\[0.3em]
\mp n,&(n\!-\!\frac{1}{2})\!<\!f\!<\!(n\!+\!\frac{1}{2}),&{\rm when}\;n=2,4,6,\cdots,
\end{array}\right.\]
when $\phi=(f+\frac{1}{2})\pi$. On the other hand, it yields 
\[\mathcal C=\left\{\begin{array}{lll}
\mp n,&(n\!-\!\frac{1}{2})\!<\!f\!<\!(n\!+\!\frac{1}{2}),&{\rm when}\;n=1,3,5,\cdots,\\[0.3em]
\pm n,&(n\!-\!\frac{1}{2})\!<\!f\!<\!(n\!+\!\frac{1}{2}),&{\rm when}\;n=2,4,6,\cdots,
\end{array}\right.\]
when $\phi=(f-\frac{1}{2})\pi$.
$\mathcal C$ is undefined when $f=\frac{1}{2},\frac{3}{2},\frac{5}{2},\cdots,$
as band gap closes at those points when $\phi=(f\pm\frac{1}{2})\pi$. 
At these points the system preserves the mirror symmetry, $\mathcal M_\theta$. 
Which means topological phase emerges
when this mirror symmetry is broken.

Appearance of topological phases with increasing Chern numbers
with the increase of $f$ is shown in Fig. \ref{topological-phase-chern-number}.
Topological phases $\mathcal C=\pm 1,\pm 2, \pm 3,
\pm 4$ and $\pm 5$ are shown here. New phases with higher values of
$\mathcal C$s may appear with the increase of $f$.
Chern number is undefined at the
intermediate points when $f=\frac{1}{2},\frac{3}{2},\frac{5}{2},\cdots$.
This phase diagram is independent of the value of $t$, $\delta$ and
$h$. 

In order to visualize the edge states, variation of
energy with $\theta$ is shown in the upper panel
of Fig. \ref{Energy-theta-higher-Chern}. A chain
of 100 sites is considered. 
Figures are drawn for $t=1$, $\delta=0.5$, $h=0.2$, $\phi=(n+\frac{1}{2})\pi$
and $n=2$ in (a), $n=3$ in (b), $n=4$ in (c). $n$ pairs of
edge state lines are found for each case when $\mathcal C=\pm n$. 
Those results are consistent with the `bulk-boundary 
correspondence' rule which states that:
chern number is equal to the 
number of pair of edge states in the gap
for the two-band model \cite{Hatsugai1,Hatsugai2,Mook}.
Probability density of a specific pair edge states
for a definite value of $\theta$ is shown in the
respective lower panels.
For example,  $\theta=\pi/2$ when $n=2$ in (d),
$\theta=(\pi-1)/2$ when $n=3$ in (e), 
and $\theta=\pi/4$ when $n=4$ in (f).
Variation of energy with $\theta$ is shown Fig. 
\ref{edge-states-with-phi-eq-0} when $\phi=0$
for $t=1$, $\delta=0.5$, $h=0.2$. 
Value of $\mathcal C$ is undefined as there is no
band gap when $\phi=0$. Particle-hole symmetry is not
preserved in this case.

For another type of parametrization, where 
\be\left\{\begin{array}{l}
v=t[1+\delta \cos{\!(\theta)}],\\[0.3em]
w=t[1-\delta \cos{\!(\theta)}],\\[0.3em]
t_a=h\,\cos{\!(n\theta+\phi)},\\[0.3em]
t_b=h\,\cos{\!(n\theta-\phi)},
\end{array}\right.
\label{parameters-3}\ee
a single topological phase with $\mathcal C=\pm 1$
always appear when $n=1,3,5,\cdots$. $\mathcal C$ is undefined
when $n=2,4,6,\cdots$ as band gap closes for the
even $n$. Interestingly edge states appear for the
trivial phases instead. Not only that, 
multiple crossing of the edge state energy lines are found
for both trivial and topological phases
where the number of crossing points are
exactly equal to the value of $n$.
Bulk-edge energy dispersion and probability density of the edge states for this
parametrization are shown in Fig. \ref{edge-states-with-multiple-crossing}.
In the upper panel bulk-edge energy variation with $\theta$
is shown for $n=2$ in (a), $n=3$ in (b),  and $n=4$ in (c).
Values of the fixed parameters are $t=1$, $\delta=0.5$, $h=0.2$, and
$\phi=\pi/2$.
Lower panel shows the probability densities of
edge states in (d), (e) and (f) for the respective cases.
Figures are drawn for lattice of 100 sites.

\begin{figure}[h]
\psfrag{C}{$\mathcal C$}
\psfrag{cp1}{$\mathcal C=+1$}
\psfrag{cm1}{$\mathcal C=-1$}
\psfrag{cp2}{$\mathcal C=+2$}
\psfrag{cm2}{$\mathcal C=-2$}
\psfrag{cp3}{$\mathcal C=+3$}
\psfrag{cm3}{$\mathcal C=-3$}
\psfrag{cp4}{$\mathcal C=+4$}
\psfrag{cm4}{$\mathcal C=-4$}
\psfrag{cp5}{$\mathcal C=+5$}
\psfrag{cm5}{$\mathcal C=-5$}
\psfrag{c0}{$\mathcal C=0$}
\psfrag{nu0}{\hskip -.cm \color{blue} \Large $\nu\!=\!0$}
\psfrag{nu-3}{\hskip -.5cm \color{white} \Large $\nu=-3$}
\psfrag{nu-1}{\hskip -.250cm \color{blue} \Large $\nu\!=\!-1$}
\psfrag{nu1}{\hskip -.50cm \color{white} \Large $\nu=1$}
\psfrag{f}{ $f$}
\psfrag{v}{\hskip -0. cm $\phi$}
\psfrag{0.5}{0.5}
\psfrag{0.0}{0.0}
\psfrag{1.0}{1.0}
\psfrag{1.5}{1.5}
\psfrag{2.0}{2.0}
\psfrag{2}{2}
\psfrag{3}{3}
\psfrag{4}{4}
\psfrag{5}{5}
\psfrag{1}{1}
\psfrag{0}{0}
\psfrag{-1}{-1}
\psfrag{-2}{-2}
\psfrag{-3}{-3}
\psfrag{-4}{-4}
\psfrag{-5}{-5}
\psfrag{2.5}{2.5}
\psfrag{3.5}{3.5}
\psfrag{3.0}{3.0}
\psfrag{4.5}{4.5}
\psfrag{4.0}{4.0}
\psfrag{5.0}{5.0}
\psfrag{p1}{$-\pi$}
\psfrag{p5}{\hskip -0.2 cm$-\pi/2$}
\psfrag{p2}{$\pi$}
\psfrag{p3}{$\pi/2$}
\psfrag{p4}{\hskip -0.0 cm$\pi$}
\includegraphics[width=230pt,angle=0]{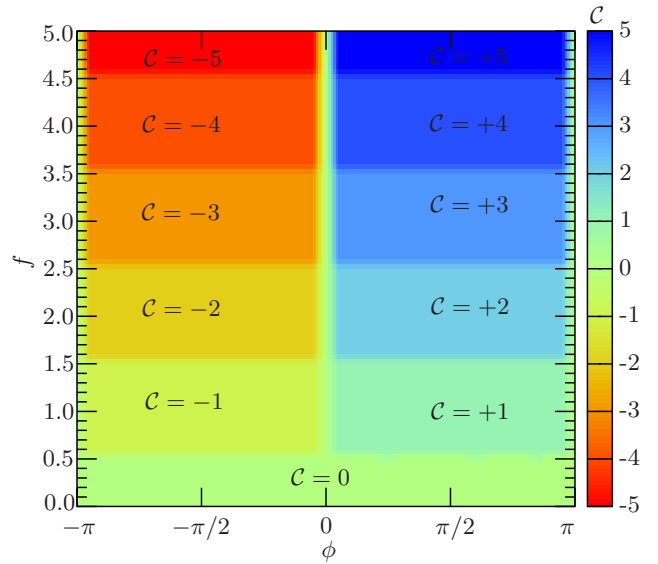}
\caption{Variation of $\mathcal C$ in the $f$-$\phi$ phase space
  for the Hamiltonian
  defined in Eq. \ref{H-t}
  with the parameters in Eq. \ref{parameters-2}
  is shown but when $-\pi<\phi<\pi$.}
\label{nontopological-phase-chern-number}
\end{figure}

Variation of $\mathcal C$ in the $f$-$\phi$ phase space
  for the Hamiltonian
  defined in Eq. \ref{H-t}
  with the parameters in Eq. \ref{parameters-2}
  is shown in Fig. \ref{nontopological-phase-chern-number}
  but when $-\pi<\phi<\pi$. In this particular case
  $\phi$ does not depend on the parameter $f$. 
  Phases with different values of
  $\mathcal C$ will
  appear with the increase of $f$ along the 
  vertical axis. However, in this case 
  phases with different $\mathcal C$s are 
  not separated by band gap closing. The Hamiltonian remains 
  invariant under the transformation of $\mathcal M_{\theta,\phi}$ 
  for any value of $f$. In contast, symmetry with respect to 
  $\mathcal M_\theta$, is preserved only when  
  $f=0,\,\pm \pi$. The states shown in Fig.
  \ref{nontopological-phase-chern-number} are no more topological
  in nature since they are not separated by
  vanishing band gap.
  In contrast, phases across the line $\phi=0$ are separated by
  zero band gap.
\section{Discussion}
\label{Discussion}
In this investigation, emergence of two different series of
topological phases with
$\nu=\pm 1,\pm 2,\pm 3,\cdots,$ and $\mathcal C=\pm 1,\pm 2,\pm 3,\cdots,$
has been successfully demonstrated using the 1D eSSH models.
The manuscript is composed of two main parts, say Sec \ref{phase-Chern-number}
and \ref{phase-winding-number}, where the major
results are presented. In the first case, a single further neighbor
hopping term beyond NN is found enough for the realization of the series of new
phase where particle-hole and inversion symmetries are preserved.
Those phases can be realized by adding multiple further neighbor
terms as well. 
In the second case, a single pair of NNN 
hopping terms is found sufficient for the realization of the series of new
phases where the standard forms of particle-hole and inversion symmetries
for 1D are not preserved.

Four different eSSH models in the first case have been considered
in which extend of further neighbor hopping is limited by $m=2,3$, and
their properties have been studied rigorously. Topological properties have
been characterized in terms of winding number, edge states and quench dynamics
in the presence of an additional nonlinear term. Finally, the results
are generalized for $m>2$, where phases with higher values of $\nu$
appear. Comprehensive phase diagrams are drawn, where the equations of
phase transition lines are obtained. It is also expected that
a pair of further neighbor staggered hopping terms
with varying extent can yield
series of topological phases with different $\nu$, however,  
this case is not addressed here.

It is known that topological interface states emerge when two lattices with
different topological phases are joined.
Nowadays, study on these interface states in phononic crystals have been
initiated \cite{Li3}. So, the results obtained for these tight-binding 
models will become helpful for constructing the phononic model
in order to realize the topological phase of any desired winding number,
as well as to study the properties of interface states. 
In addition, the most simple route for the demonstration of topological
phase of higher winding numbers using systems of ultracold atoms in
optical lattice can be obtained by mimicking the structure of 
these tight-binding models. 

In the second case, emergence of topological phase with any
value of $\mathcal C$ is discussed when the NN and NNN hopping terms
in the two-band eSSH model are expressed in terms
of two angular variables, $\theta$ and $\phi$.
Particle-hole and inversion symmetries are preserved in the
${\rm k}$-$\theta$ space.  
Three different types of parametrization are employed
when $\theta$ and $\phi$ depend on another
parameter $f$ in three different ways. 
A specific parametrization gives rise to a nontrivial phase of
any value of $\mathcal C$ which is controlled by the parameter $f$.
Phase transition points are
marked on the phase diagram where the band gap vanishes.
Experimental realization of these phases using systems of ultracold atoms in
optical lattice is possible, as discussed here\cite{Li2}.

The system hosts topological phases with
$\mathcal C=\pm 1, \mp 1$, for another parameterization, 
but with peculiar types of edge states.
Here, multiple crossing between the edge states lines 
is found within the band gap. This particular phenomenon
of multiply-crossed edge states as shown in
Fig \ref{edge-states-with-multiple-crossing} is not reported before. 
In addition, fake topological phases with series of
different $\mathcal C$s appear in the third kind of parametrization.
They are spurious because of the fact that 
the band gap does not vanish at the transition points in this case. 
All the results obtained in this study are
insensitive to the external magnetic field.
This is because of the fact that there is no spin dependent term
in the Hamiltonians. Magnetic field will be registered here as
an additional constant in the diagonal element of ($2\times 2$) 
$H({\rm k})$ matrix in every case. 
As a result, symmetry of this matrix does not change in
the presence of magnetic field and all the
results remain valid.
\section{ACKNOWLEDGMENTS}
  RKM acknowledges the DST/INSPIRE Fellowship/2019/IF190085.
  Authors acknowledge the fruitful discussion with Oindrila Deb.

\end{document}